\documentclass[twocolumn]{aastex63}
\usepackage{amsmath,multirow,graphicx}

\accepted{\today}

\submitjournal{ApJ, accepted for publication}
\shortauthors{Saikia et al. 2025}
\graphicspath{{./}{figures/}}
\begin{document}


\title{Peering into the heart of darkness with VLBA : \\Radio Quiet AGN in the JWST North Ecliptic Pole Time-Domain Field
}


\correspondingauthor{Payaswini Saikia}
\email{ps164@nyu.edu}

\author[0000-0002-5319-6620]{Payaswini Saikia}
\affiliation{Center for Astrophysics and Space Science (CASS), New York University Abu Dhabi, PO Box 129188, Abu Dhabi, UAE }

\author[0000-0002-2200-0592]{Ramon Wrzosek}
\affiliation{Center for Astrophysics and Space Science (CASS), New York University Abu Dhabi, PO Box 129188, Abu Dhabi, UAE }

\author[0000-0003-4679-1058]{Joseph Gelfand}
\affiliation{Center for Astrophysics and Space Science (CASS), New York University Abu Dhabi, PO Box 129188, Abu Dhabi, UAE }

\author{Walter Brisken}
\affiliation{National Radio Astronomy Observatory, 520 Edgemont Road, Charlottesville, VA 22903, USA}

\author[0000-0001-7363-6489]{William Cotton }
\affiliation{National Radio Astronomy Observatory, 520 Edgemont Road, Charlottesville, VA 22903, USA}

\author[0000-0002-9895-5758]{S. P. Willner}
\affiliation{Center for Astrophysics, Harvard \& Smithsonian, 60 Garden Street, Cambridge, MA 02138, USA}

\author[0000-0003-1436-7658]{Hansung B. Gim}
\affiliation{Department of Physics, Montana State University, P.O. Box 173840, Bozeman, MT 59717, USA}

\author[0000-0001-8156-6281]{Rogier A. Windhorst}
\affiliation{School of Earth and Space Exploration, Arizona State University,
Tempe, AZ 85287-1404, USA}

\author[0000-0001-8489-2349]{Vicente Estrada-Carpenter}
\affiliation{Institute for Computational Astrophysics and Department of Astronomy \& Physics, Saint Mary’s University, 923 Robie Street, Halifax, Nova Scotia B3H 3C3, Canada}

\author[0000-0002-6425-6879]{Ivan Yu. Katkov}
\affiliation{Center for Astrophysics and Space Science (CASS), New York University Abu Dhabi, PO Box 129188, Abu Dhabi, UAE }
\affiliation{Sternberg Astronomical Institute, M.V. Lomonosov Moscow State University, 13 Universitetsky prospect, Moscow, 119991, Russia}

\author[0000-0002-5208-1426]{Ingyin Zaw}
\affiliation{Center for Astrophysics and Space Science (CASS), New York University Abu Dhabi, PO Box 129188, Abu Dhabi, UAE }

\author[0000-0003-3910-6446]{Michael Rosenthal}
\affiliation{Department of Astronomy, University of Wisconsin–Madison, 475 North Charter Street, Madison, WI 53706, USA }

\author[0000-0001-9012-4020]{Hanaan Shafi}
\affiliation{Center for Astrophysics and Space Science (CASS), New York University Abu Dhabi, PO Box 129188, Abu Dhabi, UAE }

\author[0000-0002-0093-4917]{Kenneth Kellermann}
\affiliation{National Radio Astronomy Observatory, 520 Edgemont Road, Charlottesville, VA 22903, USA}

\author[0000-0003-4724-1939]{James Condon}
\affiliation{ Unaffiliated, 2571 Old Lynchburg Road, North Garden, VA 22959, USA}

\author[0000-0002-6610-2048]{Anton M. Koekemoer} 
\affiliation{Space Telescope Science Institute,
3700 San Martin Drive, Baltimore, MD 21218, USA}

\author[0000-0003-1949-7638]{Christopher J. Conselice} 
\affiliation{Jodrell Bank Centre for Astrophysics, Alan Turing Building,
University of Manchester, Oxford Road, Manchester M13 9PL, UK}

\author[0000-0002-6150-833X]{Rafael {Ortiz~III}} 
\affiliation{School of Earth and Space Exploration, Arizona State University,
Tempe, AZ 85287-1404, USA}

\author[0000-0001-9262-9997]{Christopher N. A. Willmer} 
\affiliation{Steward Observatory, University of Arizona, 933 N Cherry Ave, Tucson, AZ, 85721-0009, USA}

\author[0000-0003-1625-8009]{Brenda Frye} 
\affiliation{Department of Astronomy/Steward Observatory, University of Arizona, 933 N Cherry Ave,
Tucson, AZ, 85721-0009, USA}

\author[0000-0001-9440-8872]{Norman A. Grogin} 
\affiliation{Space Telescope Science Institute,
3700 San Martin Drive, Baltimore, MD 21218, USA}

\author[0000-0001-8751-3463]{Heidi B.~Hammel} 
\affiliation{Association of Universities for Research in Astronomy, 1331 Pennsylvania 
Avenue NW, Suite 1475, Washington, DC 20005, USA}

\author[0000-0003-3329-1337]{Seth H. Cohen} 
\affiliation{School of Earth and Space Exploration, Arizona State University,
Tempe, AZ 85287-1404, USA}

\author[0000-0003-1268-5230]{Rolf A. Jansen} 
\affiliation{School of Earth and Space Exploration, Arizona State University,
Tempe, AZ 85287-1404, USA}

\author[0000-0002-7265-7920]{Jake Summers} 
\affiliation{School of Earth and Space Exploration, Arizona State University,
Tempe, AZ 85287-1404, USA}

\author[0000-0002-9816-1931]{Jordan C. J. D'Silva} 
\affiliation{International Centre for Radio Astronomy Research (ICRAR) and the
International Space Centre (ISC), The University of Western Australia, M468,
35 Stirling Highway, Crawley, WA 6009, Australia}
\affiliation{ARC Centre of Excellence for All Sky Astrophysics in 3 Dimensions
(ASTRO 3D), Australia}

\author[0000-0001-9491-7327]{Simon P. Driver} 
\affiliation{International Centre for Radio Astronomy Research (ICRAR) and the International Space Centre (ISC), The University of Western Australia, M468, 35 Stirling Highway, Crawley, WA 6009, Australia}

\author[0000-0003-3382-5941]{Nor Pirzkal} 
\affiliation{Space Telescope Science Institute,
3700 San Martin Drive, Baltimore, MD 21218, USA}

\author[0000-0001-7592-7714]{Haojing Yan}
\affiliation{Department of Physics and Astronomy, University of Missouri, Columbia, MO 65211, USA}

\author[0000-0001-7095-7543]{Min S. Yun} 
\affiliation{Department of Astronomy, University of Massachusetts, Amherst, MA 01003, USA}


\begin{abstract}

We present initial results from the 4.8 GHz Very Long Baseline Array (VLBA) survey of the JWST North Ecliptic Pole Time-Domain Field (TDF). From 106 radio sources found in the Karl G. Jansky Very Large Array observations in the TDF, we detected 12 sources ($\sim$11\% detection rate) at $\sim3.3 \, \mu$Jy rms sensitivity and $\sim$4 mas resolution. Most detections exhibit pc-scale emission (less than 40 pc) with high VLBA/VLA flux density ratios and brightness temperatures exceeding $10^5$ K, confirming non-thermal AGN activity. Spectral indices $\alpha \gtrsim -0.5$ correlate with higher VLBA/VLA flux ratios, consistent with synchrotron emission from AGN coronae or jets. In the majority of our sources star formation contributes less than 50\% of the total VLBA radio emission, with a few cases where the emission is almost entirely AGN-driven. Although the radio emission from radio quiet AGN is thought to be primarily driven by star formation, our VLBA observations confirm that there is also often a contribution at various levels from black hole driven AGN. Eight VLBA detections have JWST/NIRCam counterparts, predominantly early-type, bulge-dominated galaxies, which we use to get an estimate of the redshift and star formation rate (SFR). WISE colors indicate that VLBA detections are either AGN or intermediate-disk-dominated systems, while VLBA non-detections correspond to extended, star-forming galaxies. We compare SFRs derived from previous SCUBA-2 850 $\mu$m observations with new JWST-based estimates, and discuss the observed discrepancies, highlighting JWST’s improved capability to disentangle AGN activity from star formation.
\\ 

\vspace{1.0 cm}
\end{abstract}

\section{Introduction} \label{sec:1}

\begin{table*}[tbh]
     \caption{Expected properties of proposed AGN emission mechanisms, and correlations with {\bf(1)} spectral indices, {\bf(2)} emission mechanisms, {\bf(3)} radio morphology on VLBI scales, {\bf(4)} Eddington ratios, {\bf(5)} the variability observed, and {\bf(6)} physical mechanisms responsible for the observed radio emission \cite{Panessa}.} 
     \label{tab:mechanisms}
     \centering
     \begin{tabular}{cccccc}
     \hline
     \hline
     Spectral Index & Emission & VLBI Radio & Eddington & \multirow{2}{*}{Variability} & Physical \\
     $\alpha$ & mechanism & morphology & ratio & & mechanism \\
     (1) & (2) & (3) & (4) & (5) & (6) \\
     \hline
     $\lesssim -0.5$ & \multirow{2}{*}{Opt. thin synchrotron} & Highly elongated & \multirow{2}{*}{$>0.3$ (High)} & Y & Jet\\
     (steep) & & Aspherical & & N & AGN-driven Wind \\
     \hline
     $\gtrsim -0.5$ & Opt. thick synchrotron & Point-like & \multirow{2}{*}{$<0.3$ (Low)} & Y & Coronal Emission \\ 
     (flat) & Opt. thin bremsstrahlung & Diffuse & & N & Broad-line Region\\
     \hline
     \hline
     \end{tabular}
\end{table*}

Growing evidence suggests that galaxies and their supermassive black holes (SMBHs) strongly influence each other’s properties and evolution. Feedback from SMBH ($\textup{M}_{\rm BH} \gtrsim 10^6 {M}_{\odot}$) in active galactic nuclei (AGN) also affects stellar populations, as indicated by the linear correlation between galactic bulge and SMBH masses \citep{Gebhardt}, and the quenching of star formation (SF) in AGN host galaxies \citep{Hopkins}. Further interconnection is observed in an order of magnitude decline since cosmic noon (z $\sim$ 2) of the comoving star formation rate (SFR) density, the mean specific SFR (sSFR $\equiv$ SFR/$\textup{M}_{*}$) of galaxies, and SMBH growth rates \citep{Madau,jordan}. Radio-continuum milliarcsecond (mas) resolution studies have found that $>$50\% of the emission of compact radio sources, that are unresolved by the NSF's Karl G.\ Jansky Very Large Array (VLA)\footnote{The National Radio Astronomy Observatory is a facility of the National Science Foundation operated under cooperative agreement by Associated Universities, Inc.}, is produced by AGN accretion processes, while the rest might be traced to SF processes \citep{Maini,Herrera,Radcliffe}. Notably, for the fainter source population, SF dominates on kiloparsec (kpc) scales or larger, which are often marginally resolved or entirely unresolved by the VLA \citep{Cotton2018}.
The origin of this radio emission in faint sources is still uncertain. Deep-field and multi-wavelength studies are essential to study Radio-Quiet (RQ) AGN, which represent more than 90\% of the detected AGN \citep{Padovani17}. The high sensitivity of deep-field radio surveys makes them the most effective way to study AGN emission mechanisms, and their effects on the surrounding media can be analyzed with multi-wavelength follow-ups \citep{Smolcic,Maini,saikia}. However, in order to distinguish between the SF radio emission and the AGN radio emission, fainter RQ AGN samples must be targeted to eliminate the Very Long Baseline Interferometry (VLBI) bias towards AGN-dominated sources \citep{Padovani}.

In the case of RQ AGN, different correlations (see also Table \ref{tab:mechanisms}) can help distinguish between different emission mechanisms  \citep{Panessa}. In most cases, RQ AGN emission has been observed to be synchrotron radiation, emitted by relativistic electrons that are accelerated by moving shocks in organized AGN outflows \citep{2014MNRAS.442..784Z}. However, various mechanisms, such as SF, winds, and/or low-power jets driven by an AGN, free-free emission from photoionized gas, and coronal activity in the innermost accretion disk, could also contribute to radio emission in these sources \citep{Panessa}. A comprehensive study of 144 bright quasars (PG quasar sample) led \cite{Laor} to discover that the radio spectral index ($\alpha$; Flux density $S_{\nu} \propto \nu^\alpha)$, the variability of the AGN's radio emission, and the Eddington ratios ($\frac{L}{L_{\rm Edd}}$) \footnote{Ratio of the bolometric luminosity $L$ and the Eddington luminosity $L_{\rm Edd}$ of the AGN. $L_{\rm Edd}$ is the maximum luminosity that the AGN can achieve when its radiation pressure and gravitational force are balanced.} are correlated with the emission mechanisms of the AGN. As can be seen in Table \ref{tab:mechanisms}, RQ AGN with higher Eddington ratios ($>0.3$) have a ``steeper spectra” than those with lower Eddington ratios. This suggests that the origin of the AGN's radio emission strongly depends on the accretion modes onto the SMBH (see also Table \ref{tab:mechanisms}). \\

Deep multi-wavelength surveys of the extragalactic sky can be used to determine the evolution of galaxies and AGNs earlier in the history of the cosmos. The James Webb Space Telescope (JWST) North Ecliptic Pole (NEP) Time-Domain Field (TDF) is a $\sim 14'$ diameter field that was observed by the JWST Guaranteed Time Observations program (JWST-GTO-2738; PI: R. A. Windhorst). The JWST NEP TDF (hereafter TDF) has been chosen to study the nature of AGN given its optimal properties for time domain observations at multiple wavelengths \citep{2018PASP..130l4001J}. The TDF is notable for its continuous accessibility by the JWST, low Galactic extinction, low zodiacal foreground light, and absence of bright galactic stars (free from sources brighter than $m_{AB} \sim 16$). The field has a wealth of multi-wavelength ancillary data obtained with both ground-based and space-based surveys, as tabulated by R. Jansen\footnote{http://lambda.la.asu.edu/jwst/neptdf/} and \cite{nustarxmm}; which has several dedicated surveys in optical \citep[56 narrowband optical filters plus $u, g, r$, and $i$ filters, as part of the Javalambre-Physics of the Accelerating Universe Astrophysical Survey;][]{jnep}, including a time-domain optical study \citep[at visible wavelengths using TREASUREHUNT HST data;][]{treasurehunt}, near-infrared \citep[$Y, J, H$, and $K$ imaging obtained using the MMT-Magellan Infrared Imager and Spectrometer (MMIRS) on the MMT;][]{pearlnir}, X-rays \citep[3$-$24 keV and 0.5$-$10 keV fluxes with NuSTAR and XMM-Newton;][]{nustarxmm}, radio \citep[3 GHz observations using the VLA;][]{scuba}, and sub-mm \citep[850 $\mu$m observations with the Submillimetre Common-user Bolometer Array 2 (SCUBA-2) of the James Clerk Maxwell Telescope;][]{scuba}. The initial JWST survey of the field, conducted as part of the Prime Extragalactic Areas for Reionization and Lensing Science (PEARLS) TDF program \citep[using NIRCam infrared observations to search for radio counterparts;][]{pearlsjwst}, has already been completed, along with a detailed visual search for galaxies exhibiting central point-like features in the TDF \citep{ortiz}.\\

A high-resolution radio survey of the field is necessary to study how different accretion modes affect AGN emission, to understand how AGN affects galaxy formation and evolution, and to explore the origin of low-luminosity AGN emission, which is poorly understood especially in the case of RQ AGN. The observations in this paper are part of a radio program using the VLA and the Very Long Baseline Array (VLBA) to generate a list of extragalactic radio sources using the VLA and then identify which ones contain a significant AGN component using the VLBA. 

The purpose of the VLBA TDF deep survey is to identify and determine the nature of RQ AGN and to image the sources with milli-arcsecond (mas) resolution and microJansky ($\mu$Jy) rms sensitivity. Similar deep surveys with the VLBA and other VLBI telescopes have already been conducted in different fields, as outlined in Table \ref{tab:surveys}. These surveys have proved that the methodology is successful, obtaining several detections at $\sim$1.4 GHz. Data have been published from the Hubble Deep Field \citep[HDF;][]{2001AA...366L...5G}, the NOAO Boötes Field \citep{2005ApJ...619..105G}, the Lockman Hole/XMM \citep{Middelberg}, Hubble DeepF ield North (HDF-N) and Flanking Fields \citep[HFF;][]{2013AA...550A..68C}, the COSMOS Field \citep{Herrera,2018AA...616A.128H}, the northern SKA PAthfinder Radio Continuum Surveys (SPARCS) reference field \citep{sparcs}  and the GOODS-N Field \citep{Radcliffe}. Compared with these deep surveys, the data presented in this paper are among the most sensitive (only one other field is comparable), while also being at a significantly higher frequency (4.8 GHz, while all the other fields are at 1.4--1.6 GHz), making the detected sources less susceptible to synchrotron self-absorption. \\

\begin{deluxetable*}{ccccccc}
    \tablecaption{VLBI Deep Surveys \label{tab:surveys}}
    \tablewidth{0pt}
    \tablehead{
    \colhead{Field/s} & \colhead{Telescopes} & \colhead{Central $\nu$} & \colhead{Detections} & \colhead{Detection} & \colhead{Sensitivity} & \colhead{Reference} \\
    \colhead{} & \colhead{} & \colhead{(GHz)} & \colhead{} & \colhead{fraction} & \colhead{($\mu$Jy/beam)} & \colhead{}
    }
    \startdata
     HDF  & EVN &  1.6 & 2 & $40\%$ &  $33$ & \cite{2001AA...366L...5G}\\    
    NOAO Boötes  & VLBA+GBT & 1.4 &9 & $15\%$ & $9$ & \cite{2005ApJ...619..105G}\\
    Lockman Hole & VLBA & 1.4 & 65 & $30\%$ & $20$ & \cite{Middelberg}\\
    HDF-N \& HFF & Global VLBI & 1.4 & 21 & $23\%$ & $7.3$ & \cite{2013AA...550A..68C}\\
    COSMOS & VLBA & 1.54 & 468 & $20\%$ & $10$ &  \cite{Herrera}\\ 
    COSMOS & VLBA+GBT & 1.54 & 35 & $20\%$ & $3.5$ & \cite{2018AA...616A.128H}\\
    GOODS-N & EVN & 1.6 & 31 & $10\%$ & $9$  & \cite{Radcliffe}\\ 
    SPARCS & EVN+e-Merlin & 1.6 & 11 & $21\%$ & $6-10$ &  \cite{sparcs}\\ 
    NEP & VLBA & 4.8 & 12 & $20\%$ & $3.3$ & Present work\\
    \enddata
    \tablecomments{
The table presents a compilation of deep radio surveys conducted in various fields using VLBI techniques. It includes the field name, telescope and central frequency, number of detections, detection fraction relative to the total observed sample, achieved sensitivity, and relevant references.
}
\end{deluxetable*}

This paper presents initial findings from the 4.8 GHz VLBA TDF deep field survey, based on 55.3 effective hours of VLBA observations with an rms sensitivity of $\sim3.3$ $\mu \rm{Jy/beam}$ and a resolution of 4 mas. The paper is organized as follows: In Section \ref{sec:2}, we discuss the VLBA observations of the TDF, including source selection and data reduction methods. Section \ref{sec:mwc} discusses the multi-wavelength counterparts of our sample, particularly focusing on optical (SDSS), millimeter (SCUBA-2), and infrared (WISE, JWST) wavelengths. Section \ref{sec:res} presents the radio wavelength analysis from the VLBA and previous multi-wavelength detections. In Section \ref{sec:dis} we discuss our findings and in Section \ref{sec:con} we provide a summary of our results and present the conclusions of the study. Finally, Appendix A includes a catalog of VLBA non-detections and a mosaic of their JWST counterparts.\\

Throughout this paper, we adopt the $\Lambda$CDM cosmology parameters with Hubble constant $H_0$ = 67.66 km s$^{-1}$Mpc$^{-1}$, matter density parameter $\Omega_M$ = 0.3111, and dark-energy density parameter $\Omega_{\Lambda}$ = 0.6889 \citep{planck}.


\section{Observations and Data reductions} \label{sec:2}

\subsection{Sample}
We used the 3 GHz VLA observations of the TDF consisting of 588 objects \citep{scuba}, to identify  point-like sources. 
The parent sample comprised of unresolved objects at 3 GHz, with a single phase calibrator, achieving an rms sensitivity of 1 $\mu$Jy beam$^{-1}$ at a resolution of 0\farcs7.

A central portion of the TDF was observed with the VLBA under project codes BB388 and BB397 (PI : W. Brisken). The field was centered on a VLBA calibrator source J1723+6547. The chosen field contains no other strong radio source. For the VLBA observations, we employed a single-pointing strategy centered on the same position as the VLA 3 GHz observations, using the same phase calibrator for consistency and phase referencing. All VLA-detected radio sources that were unresolved and exhibited flux densities $\geq5\sigma$ above the local rms noise were selected as candidates for VLBA follow-up. Due to the smaller primary beam of the VLBA at 4.8 GHz, the effective field of view was more limited, resulting in a reduced number of detectable sources. The initial VLBA correlation was performed with the pointing center on the phase calibrator, after which multiple re-correlations were carried out at different phase centers (PCs), targeting the brightest VLA sources. Target selection for re-correlation was constrained to sources within 6 arcminutes of the pointing center, where the primary beam response falls to 0.25. This resulted in a total sample of 106 sources within 1.5\arcsec radius of the VLA position.


\subsection{Observations}
The VLBA observations were made at a central observing frequency of 4832 MHz and bandwidth of 256 MHz using both right and left circular polarizations. The specific observing setup was chosen to maximize instantaneous sensitivity and minimize the effects of radio frequency interference. At the observing frequency, the central calibrator source had a flux density of $\sim160$ mJy, sufficient to allow self-calibration. The circumpolar VLBA calibrator J2005+7752 with $\sim1.0$ Jy flux density at 5 GHz was observed approximately once every 90 minutes to serve as a bandpass calibrator.

The observations were well suited as ``filler" observations at the VLBA. The TDF can be observed with the VLBA for about 18 hours each day. A series of template observing files were provided to VLBA Operations, which scheduled observations whenever conditions were poor for other observations in the dynamic queue. This paper presents an analysis based on the first 137 hours of observing time allocated at the VLBA for this program, including calibration and overheads, corresponding to 55.3 effective on-source hours after accounting for periods when the full array was not available. A subsequent publication will follow with a complete data reduction. The ultimate ambition of the VLBA deep field imaging is to amass 500 hours of VLBA observing time with a goal of reaching approximately $\sim$1.3 $\mu\rm{Jy/beam}$ image sensitivity at the center of the field.

\subsection{Correlation}
We correlated the observations with the DiFX correlator \citep{2011PASP..123..275D}.  The multiphase-center mode was used to allow formation of separate correlated data sets for the central calibrator source and all 106 additional fields. In this mode, the data were correlated with 31.25 kHz spectral resolution and 40 ms time resolution, sufficient to avoid bandwidth and time smearing anywhere in the VLBA primary beam.  After each of these {\em subintegrations}, visibility spectra were appropriately phase shifted to correspond to each of the phase centers correlated.  Time and spectral averaging was then performed separately for each phase center to achieve final resolution of 0.5 MHz. The correlation parameters led to a usable field of view for each phase center of about 4 arcsecond.


\begin{figure}[t]
    \epsscale{1.1}
    \plotone{"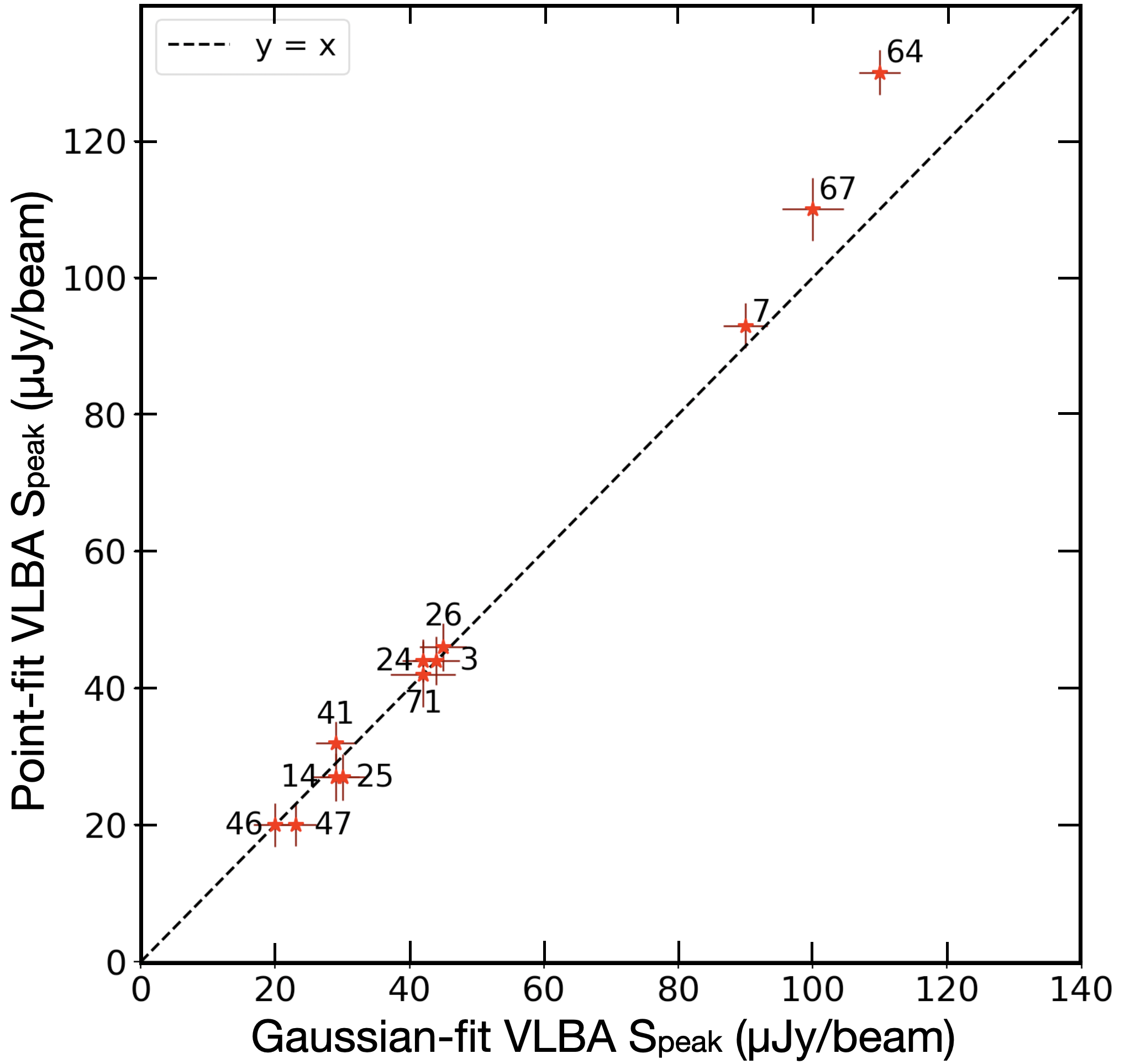"}
    \caption{Point-model vs. Gaussian-model peak flux densities (natural weighting  before primary beam correction). The sources are labeled with their VLBA PC numbers (see Figure \ref{fig:sources} for the complete VLBA snapshots, and Table \ref{tab:detections} for the coordinates). The dashed line shows equality. All sources except PC 64 (VLA ID 528) lie close to the identity line. PC 64 (VLA ID 528) was the only one resolved with the current beam settings, while PC 67 (VLA ID 554) shows hints of being marginally resolved.}
    \label{rob5_point_peak_vs_gauss_peak}
\end{figure}

\begin{figure*}[h]
\begin{center}
    \includegraphics[width=0.34\textwidth]{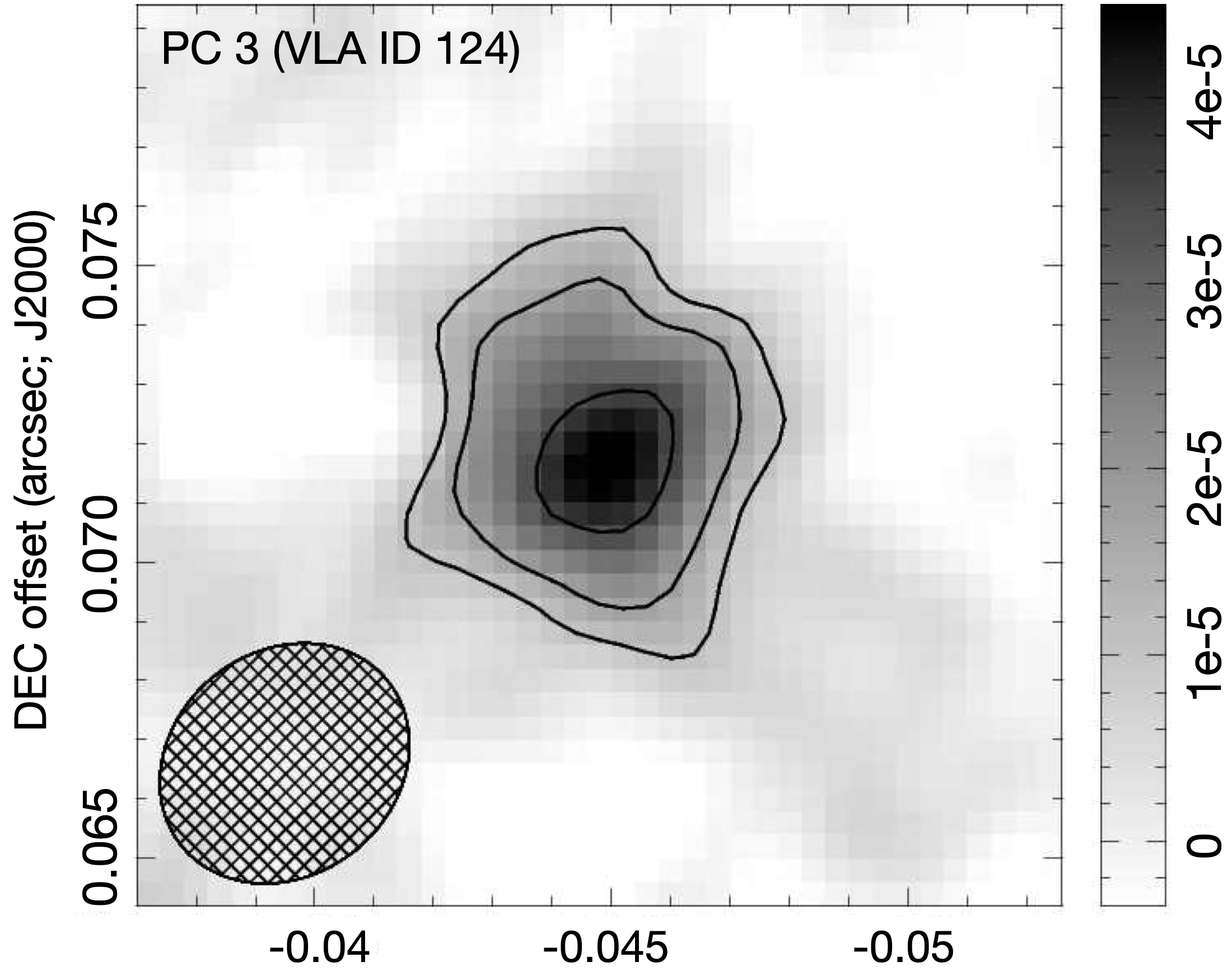}
    \includegraphics[width=0.31\textwidth]{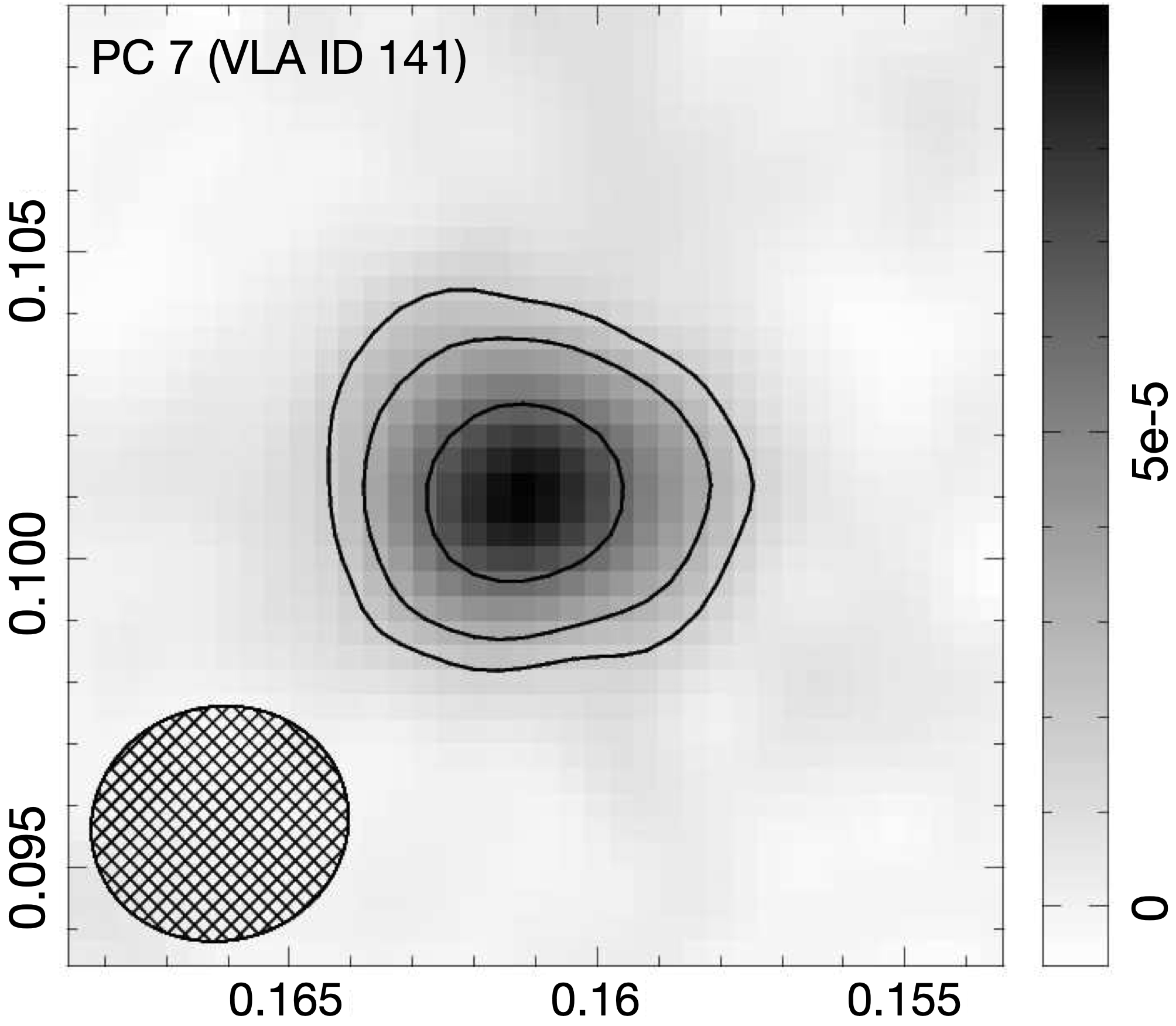} 
    \includegraphics[width=0.31\textwidth]{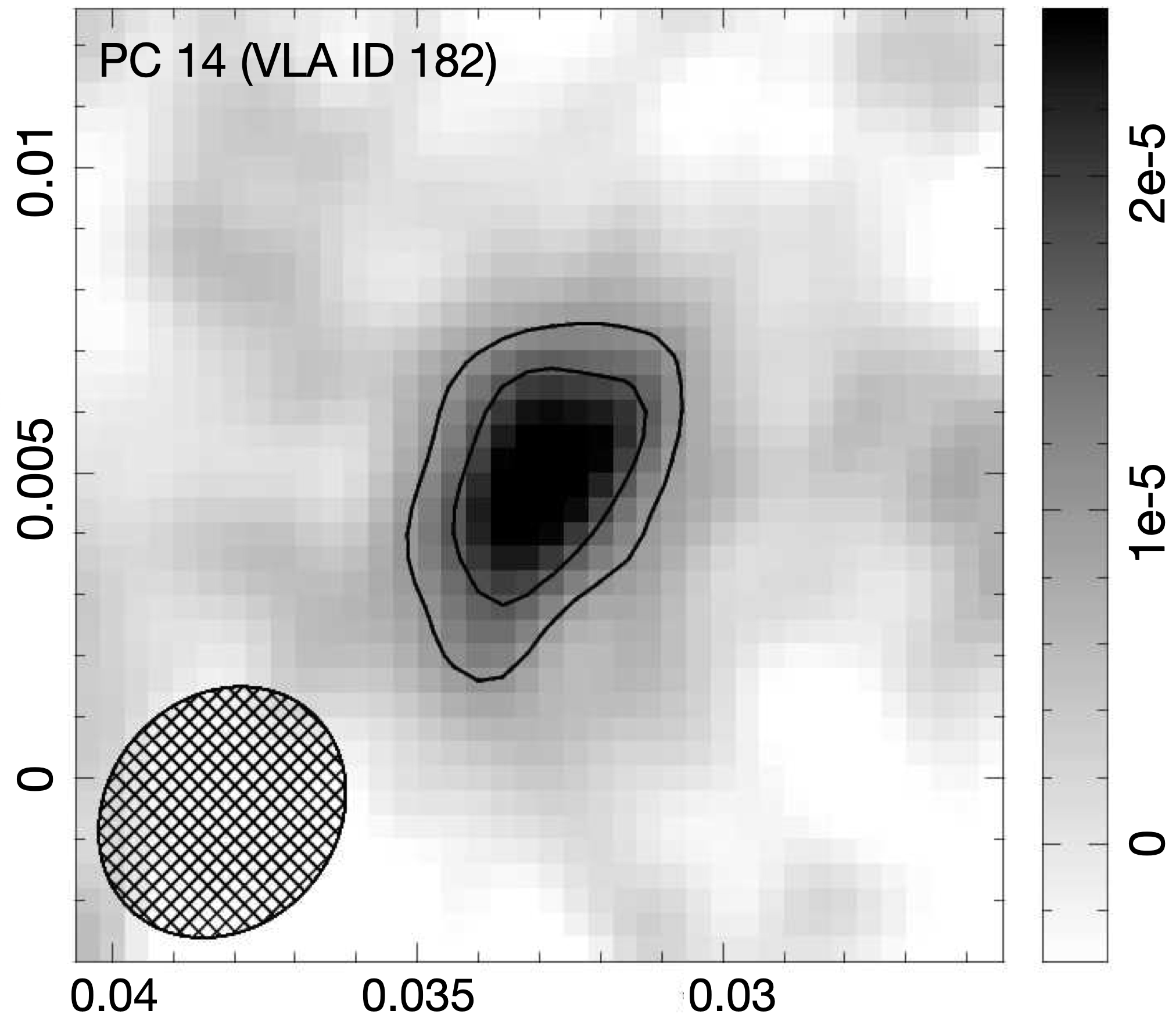} 
    \includegraphics[width=0.34\textwidth]{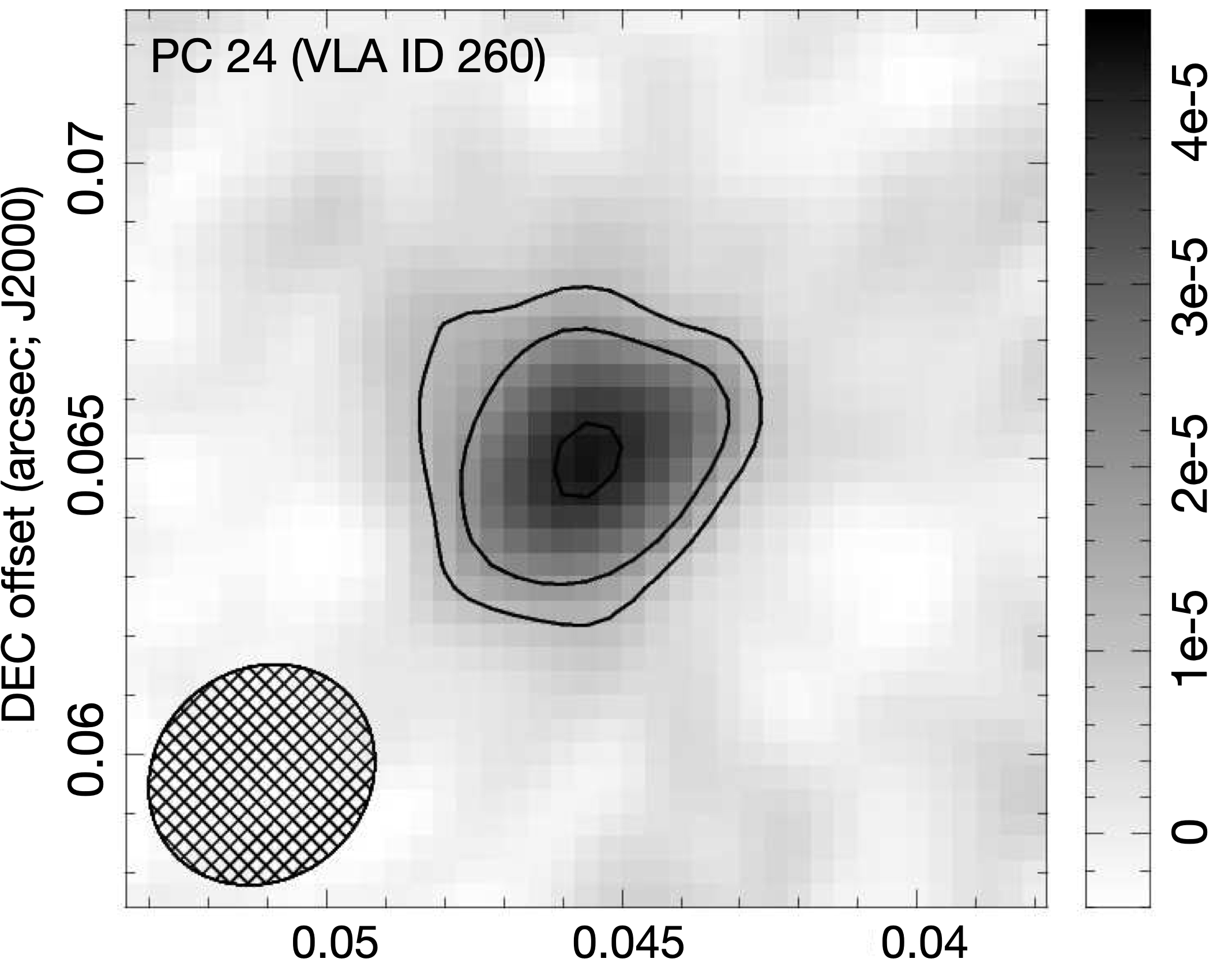} 
    \includegraphics[width=0.30\textwidth]{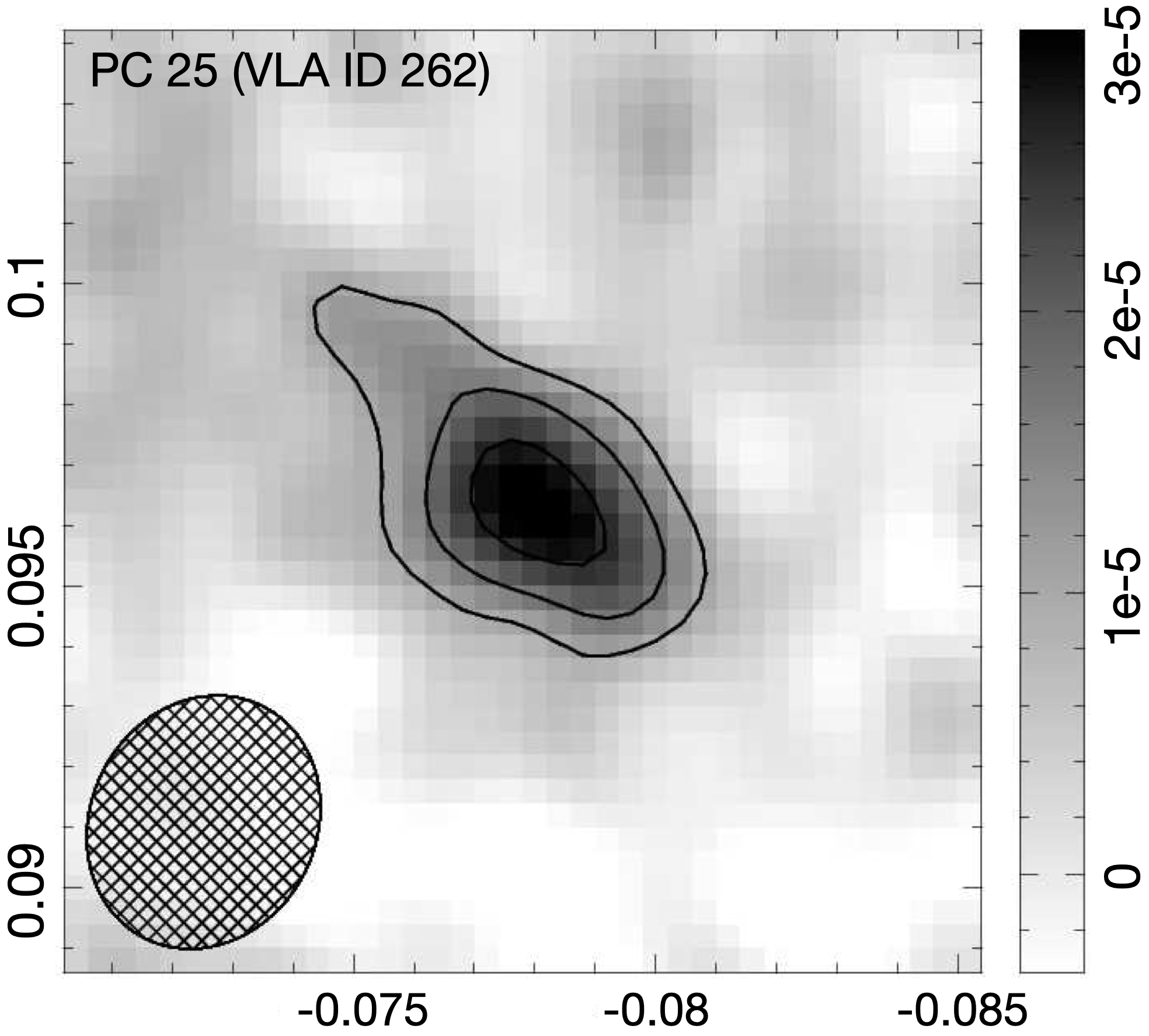}
    \includegraphics[width=0.31\textwidth]{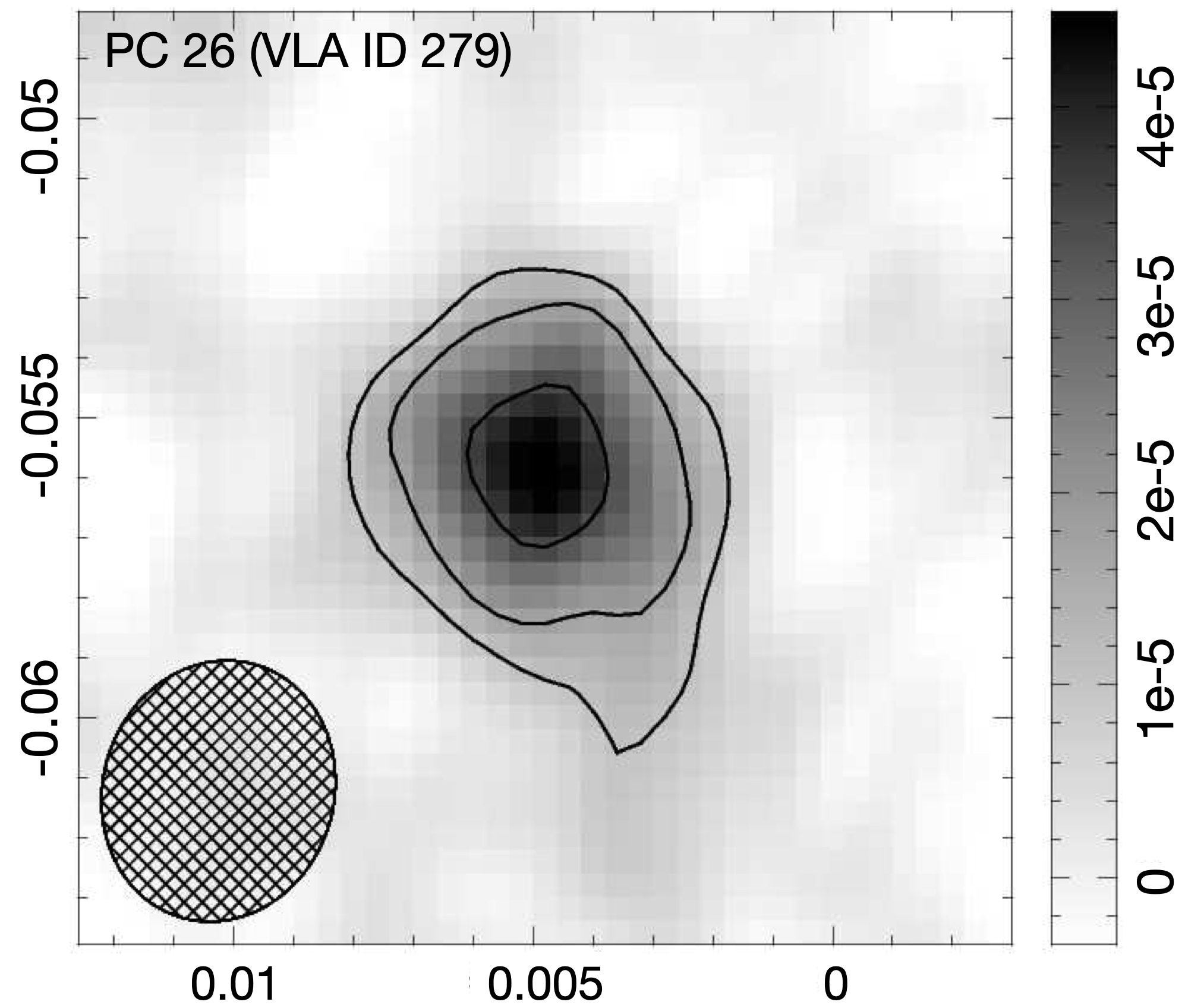}
    \includegraphics[width=0.33\textwidth]{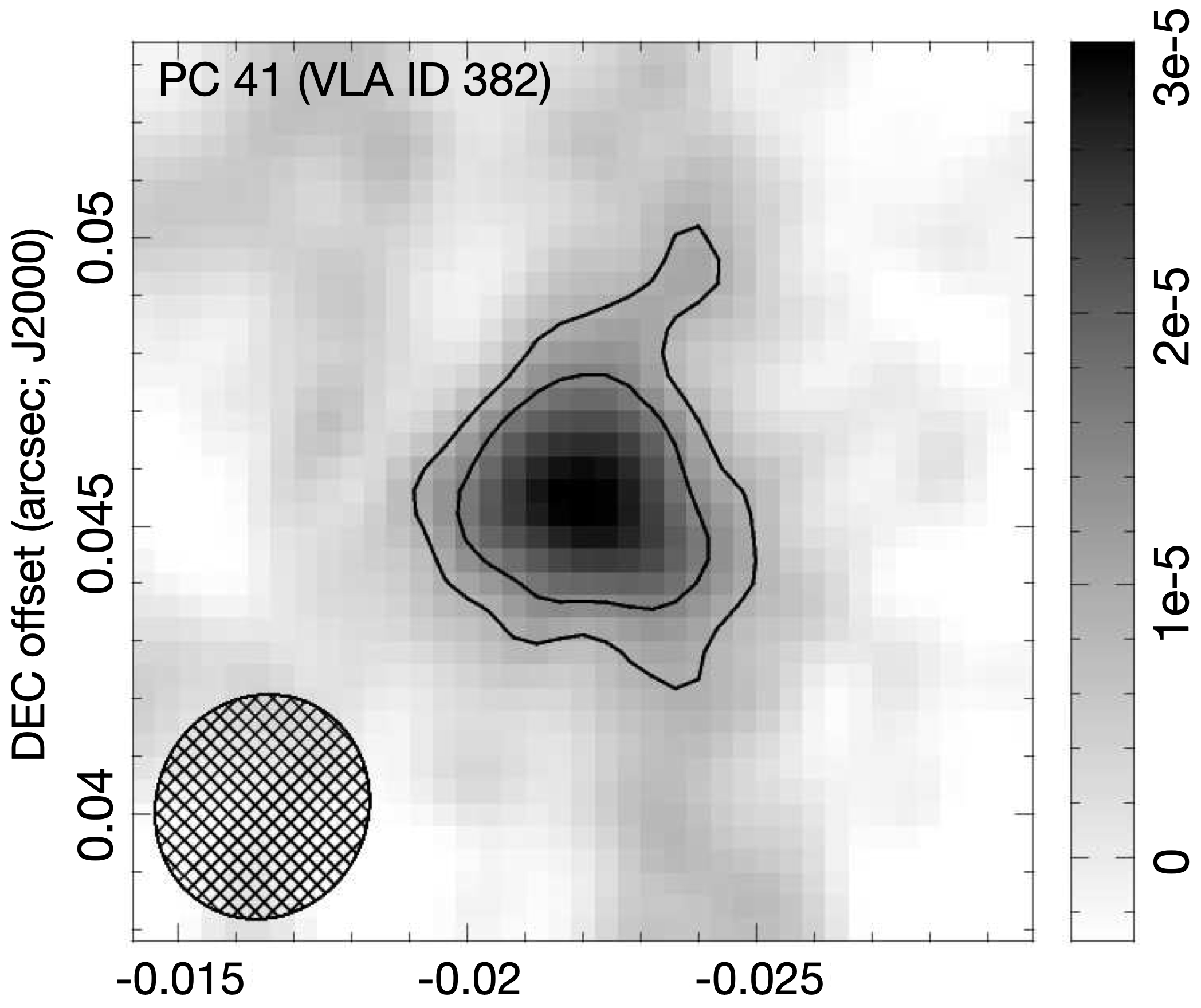}
    \includegraphics[width=0.31\textwidth]{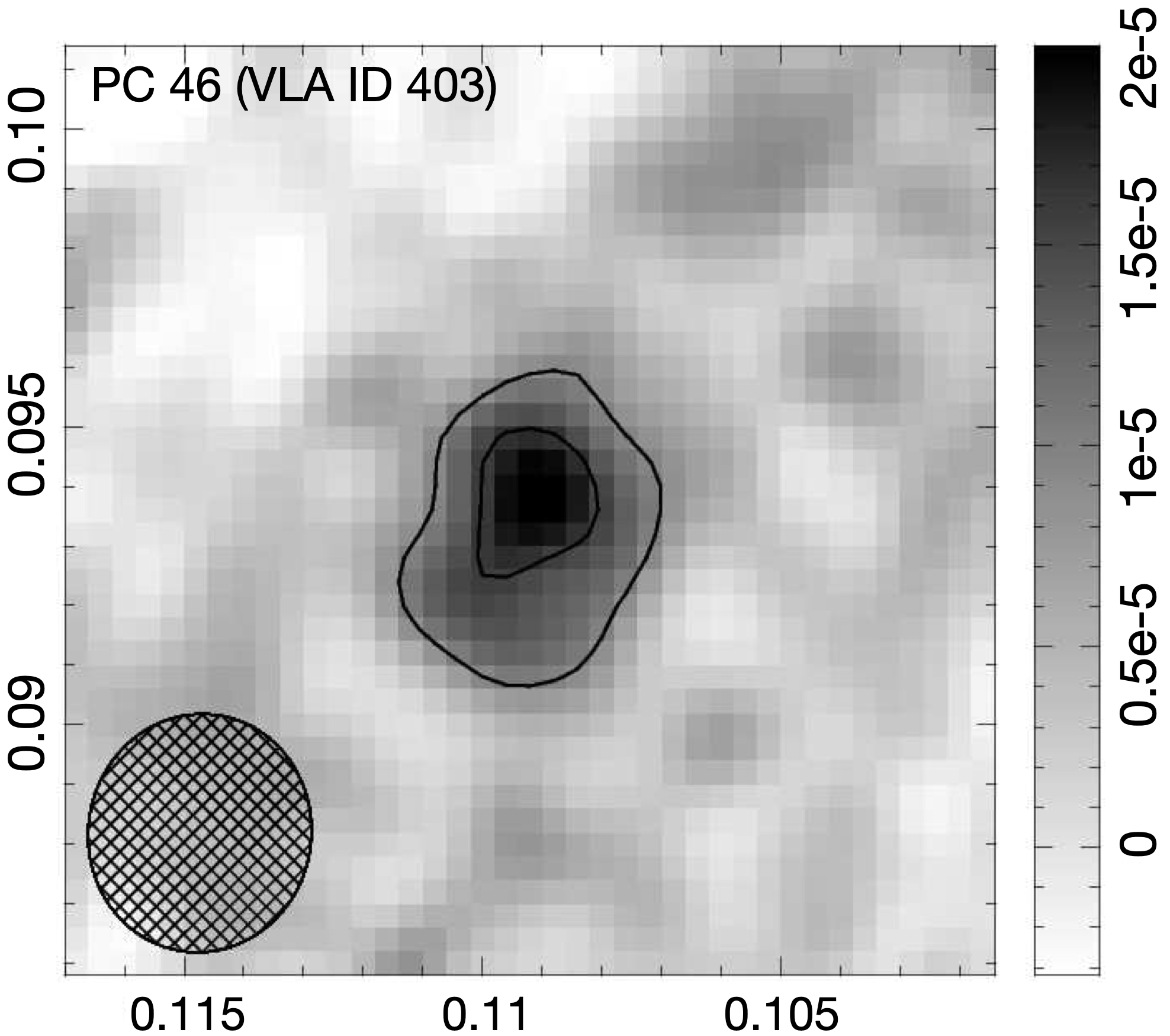}
    \includegraphics[width=0.31\textwidth]{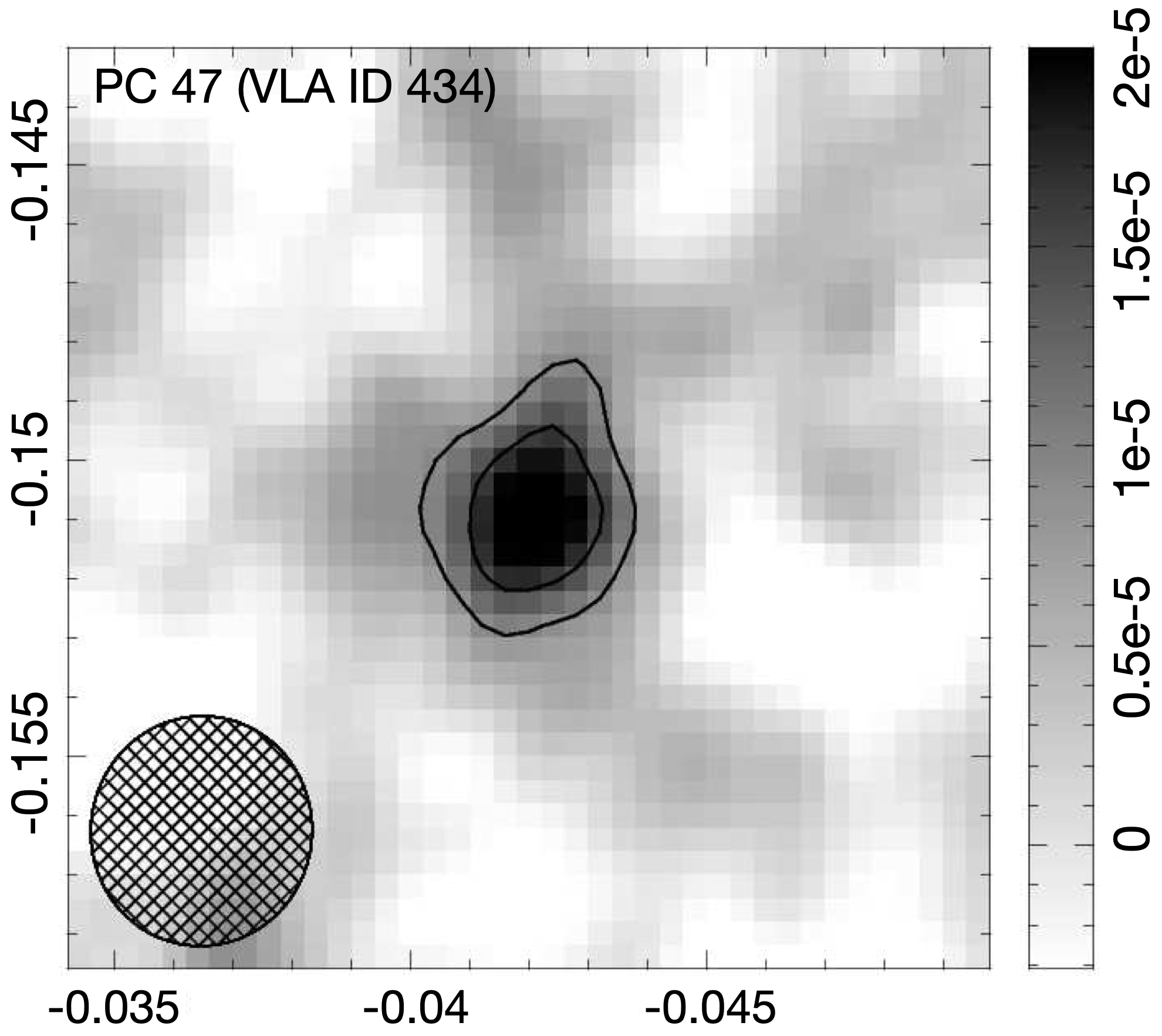}
    \includegraphics[width=0.33\textwidth]{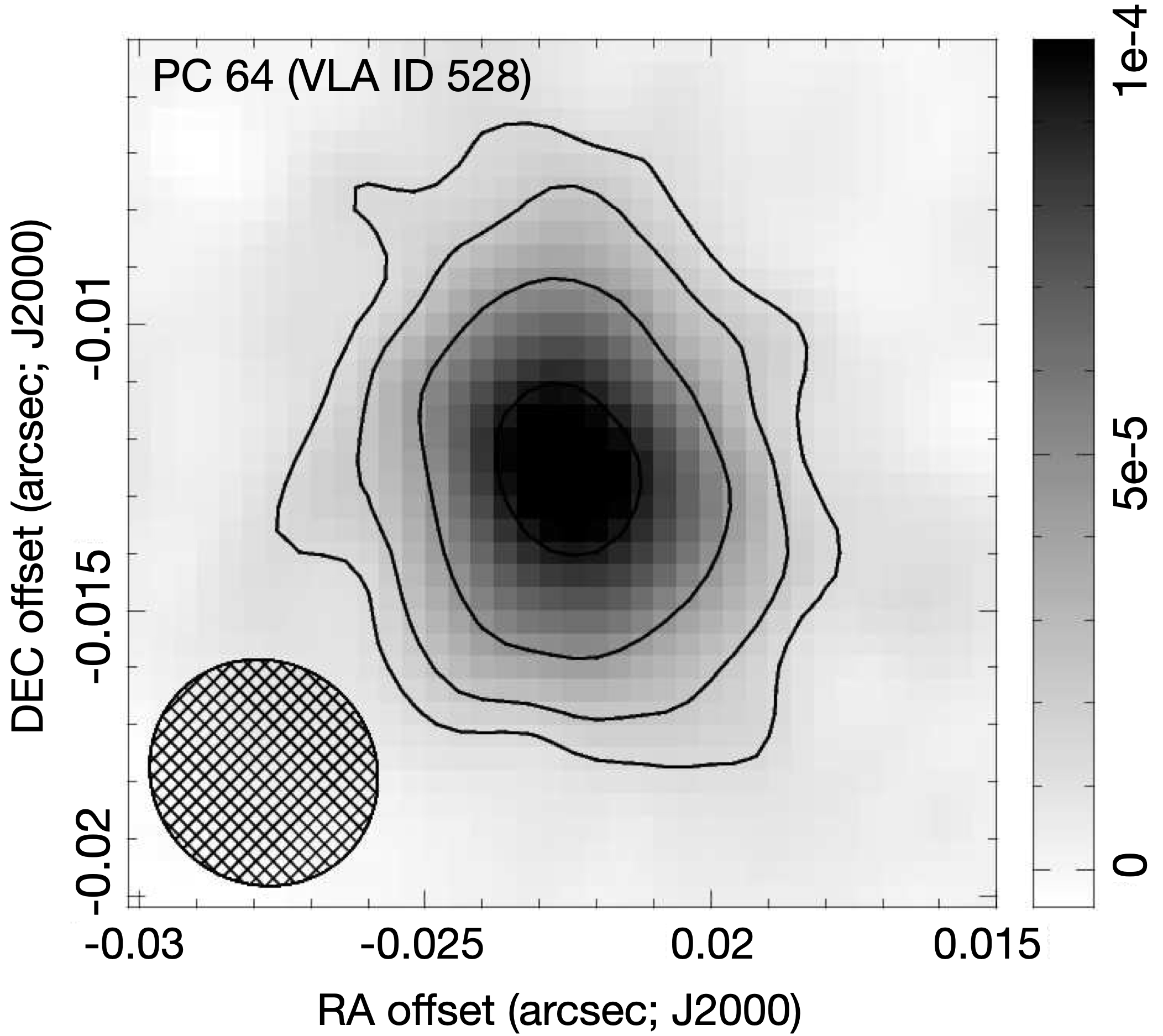}
    \includegraphics[width=0.32\textwidth]{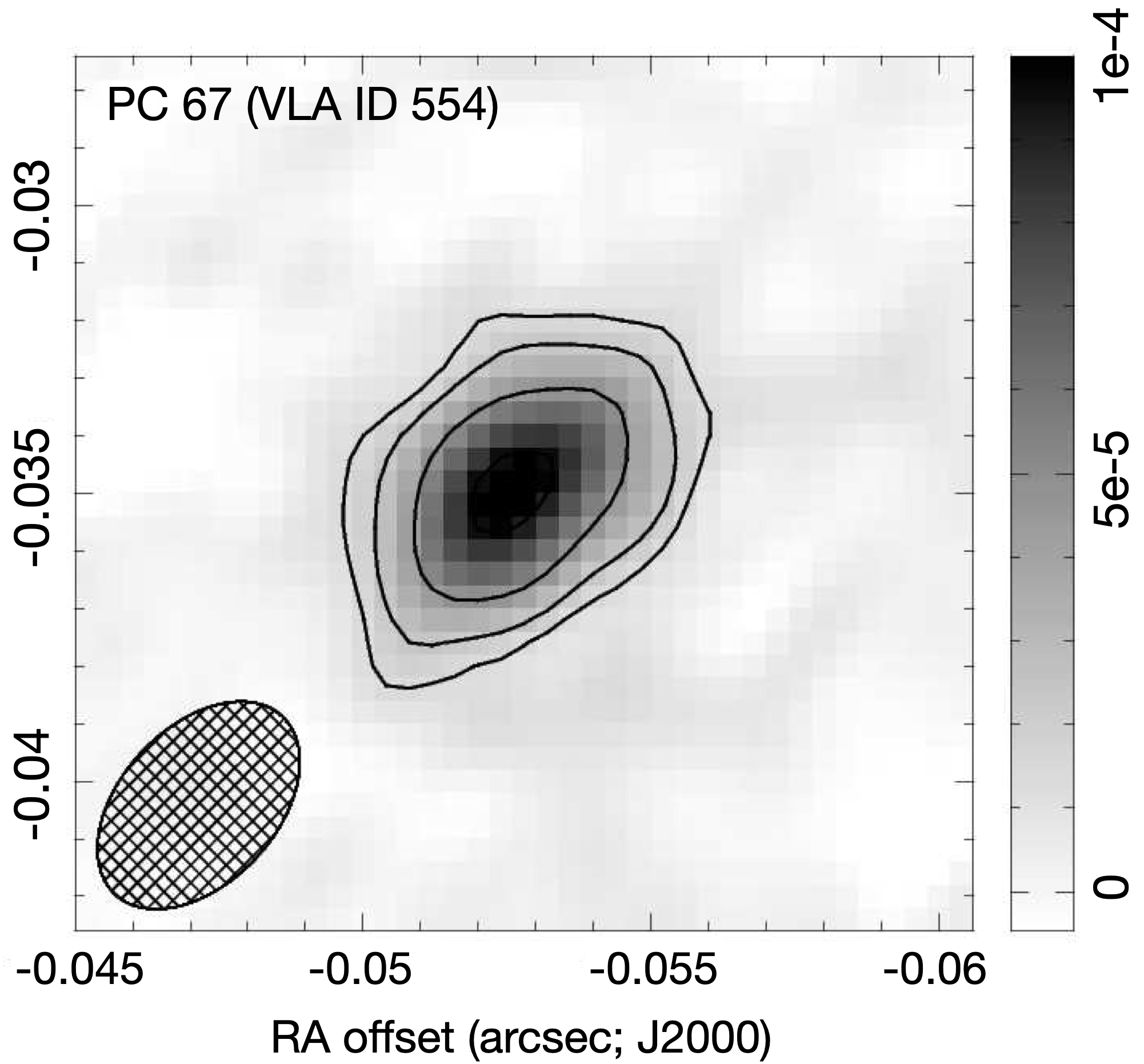}
    \includegraphics[width=0.31\textwidth]{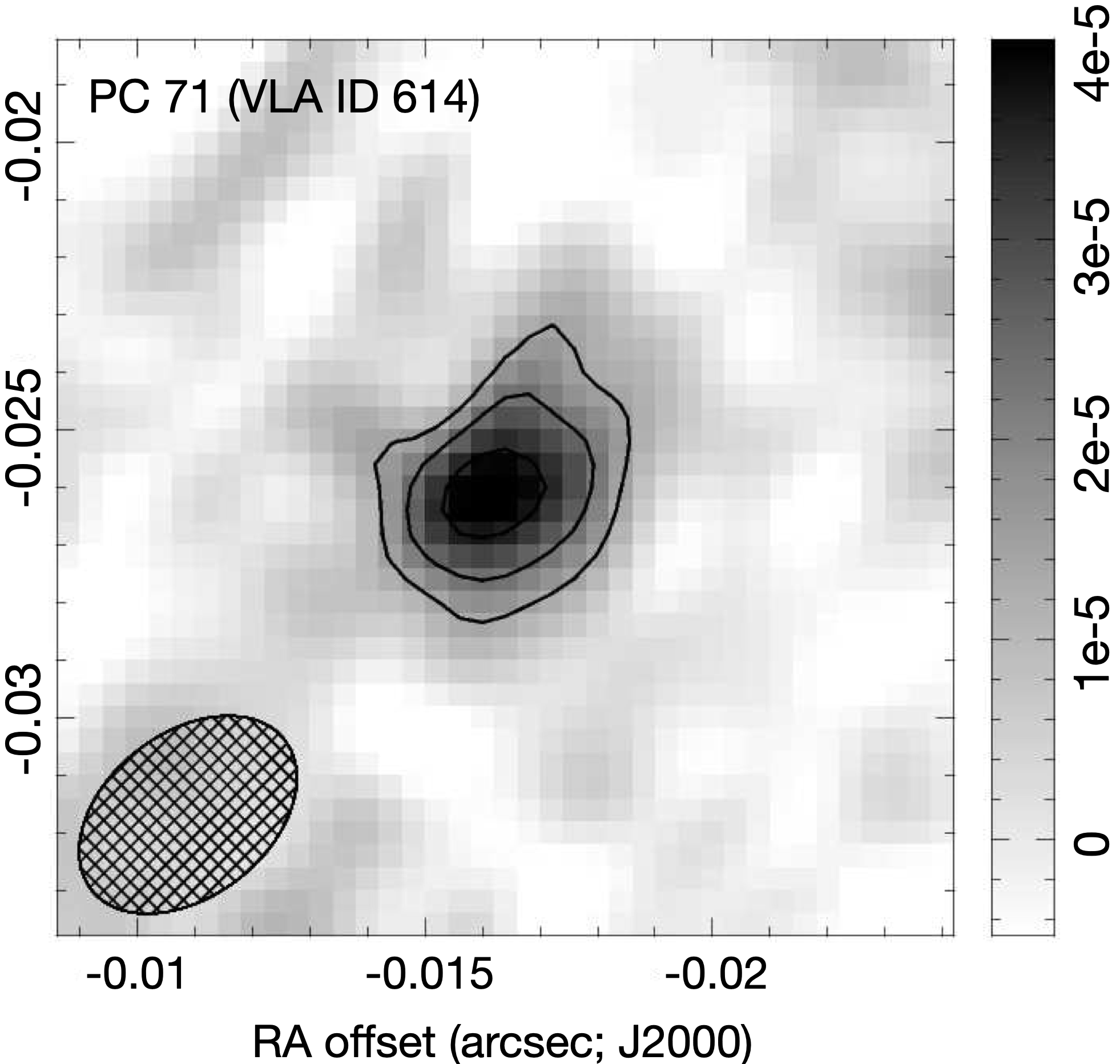}
\end{center}
    \caption{Natural weighting (\textsc{robust = 5}) 4.8 GHz VLBA images. Sources are identified in each panel. Image rms and contour levels are : {\bf a) PC~3} (image rms = 3.5 $\mu$Jy/beam; contours = $3,5,10 \sigma$), {\bf b) PC~7} (6.2 $\mu$Jy/beam; $3,5,10,15\sigma$), {\bf c) PC~14} (3.5 $\mu$Jy/beam; $3,5,7.5\sigma$), {\bf d) PC~24} (4.0 $\mu$Jy/beam; $3,5,10,15\sigma$), {\bf e) PC~25} (3.5 $\mu$Jy/beam; $3,5,7.5\sigma$), {\bf f) PC~26} (3.5 $\mu$Jy/beam; $3,5,10\sigma$), {\bf g) PC~41} (3.1 $\mu$Jy/beam; $3,5,10\sigma$), {\bf h) PC~46} (3.2 $\mu$Jy/beam; $3,5,7.5\sigma$), {\bf i) PC~47} (3.1 $\mu$Jy/beam; $3,5,7.5\sigma$), {\bf j) PC~64} (4.6 $\mu$Jy/beam; $3,5,10,20\sigma$), {\bf k) PC~67} (4.7 $\mu$Jy/beam; $3,5,10,20\sigma$), and {\bf l) PC~71} (4.8 $\mu$Jy/beam; $3,5,7.5\sigma$).  Image rms and color bar indicate the surface brightness uncorrected for primary beam attenuation, and the cross-hatched ellipses shows the restored beams.}
\label{fig:sources}
\end{figure*}


\subsection{Calibration}

Calibration and imaging were performed in the Astronomical Image Processing System (AIPS) data reduction package \citep{AIPS}. A data reduction pipeline was used to process all of the data consistently.  There were two independent steps in the calibration: determination and application.  Rather than developing a pipeline with complex logic in the `POPS' language that AIPS uses, Python language programs were used to generate explicit scripts based on a project status file which contained relevant information for each observation.  \\

The calibration determination was performed on a per-observation basis. Data validity flags from the VLBA online system were applied.  In a small number of cases, an antenna was excluded from analysis due to poor sensitivity not captured by the on-line system. Furthermore, some low-altitude data were discarded when the phases were changing rapidly. The calibration routine followed the method suggested by \cite{VLBA_Scientific_Memo_37}. The bright calibrator J2005+7752 was used for determining the residual delays and complex bandpass. The central source J1723+6547 was used for determination of delays and complex gains (using AIPS tasks \textsc{fring} and \textsc{calib}, respectively).  A high-fidelity source model for J1723+6547, constructed from the combined 137 hours of observing, was used in the complex gain determination.  When determining the gains, constraints ensured that only antennas that experienced the best weather could affect the overall gain scaling.  This was in lieu of determining corrections due to atmospheric opacity, which are typically not important at 5 GHz observing frequency.  The product of this pipeline step consisted of three tables: an AIPS `FG' table containing the final set of data validity flags, an AIPS `BP' table containing the complex bandpass table, and an AIPS `CL' task containing the delays and complex gains.

The calibration was applied on a per-phase-center basis, where data from all relevant observations for a given phase center were combined to create a single calibrated dataset. We loaded the correlated data for each observation into AIPS, with each observation occupying its own catalog entry. Subsequently, the FG', BP', and CL' tables from the prior calibration step for the corresponding observation were loaded. To transfer the calibration from J1723+6547 to the target phase center, a source renumbering was required in the CL' table. This non-standard operation was executed using the AIPS task \textsc{tabed} with \textsc{optype}=‘REPL’.

The AIPS task \textsc{split} was used to apply the calibration values, creating new databases with corrected visibilities.  Finally the AIPS task \textsc{dbapp} was used to append all of the calibrated observations into a single database ready for imaging.  This `in-beam' calibration scheme works very well for two reasons: 1.\ the calibration was determined contemporaneously, so no interpolation was required; 2.\ the angle between the calibrator source and the target field was less than 6', leaving only very small angular gradients uncorrected.


\begin{figure*}[h]
     \epsscale{1.2}     \plotone{"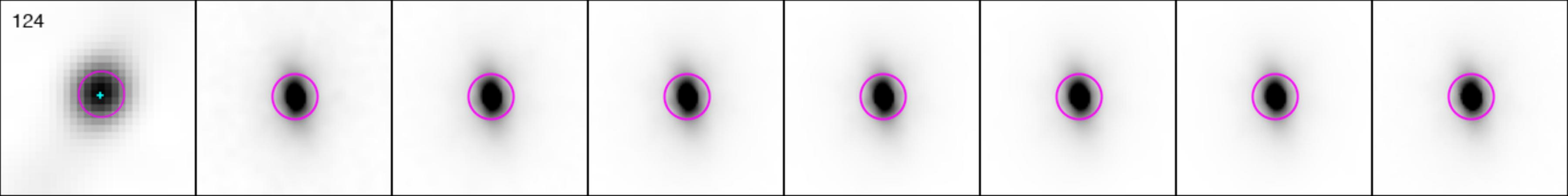"}     \plotone{"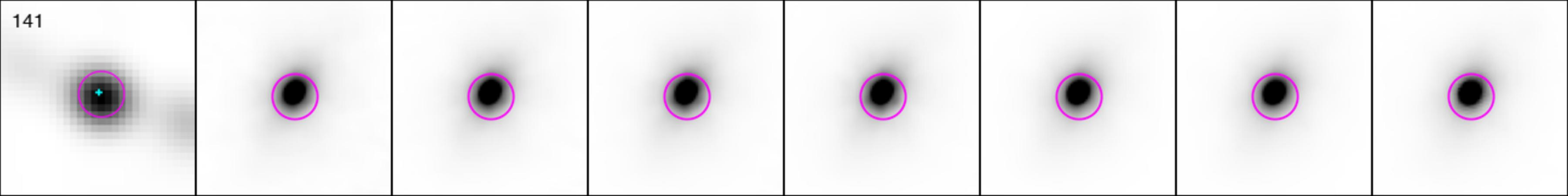"}    \plotone{"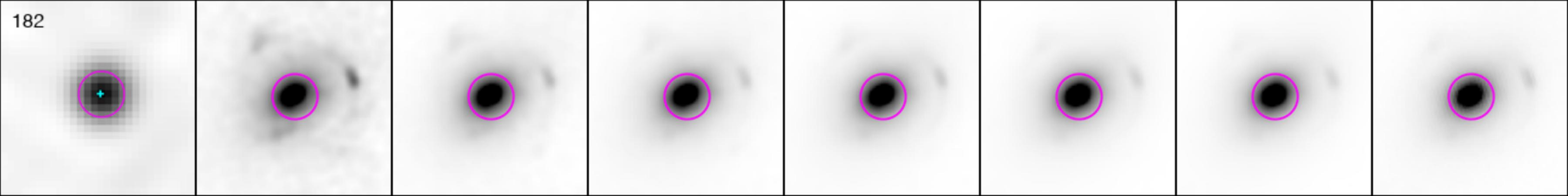"}     \plotone{"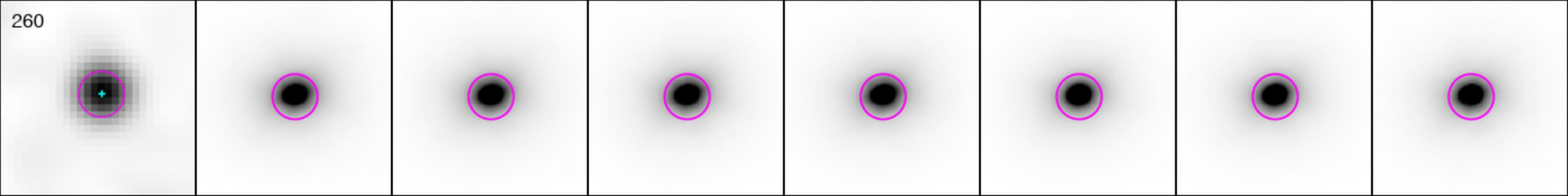"}     \plotone{"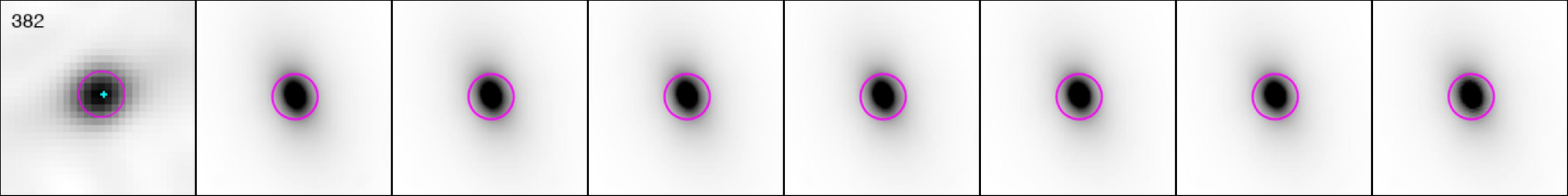"}    \plotone{"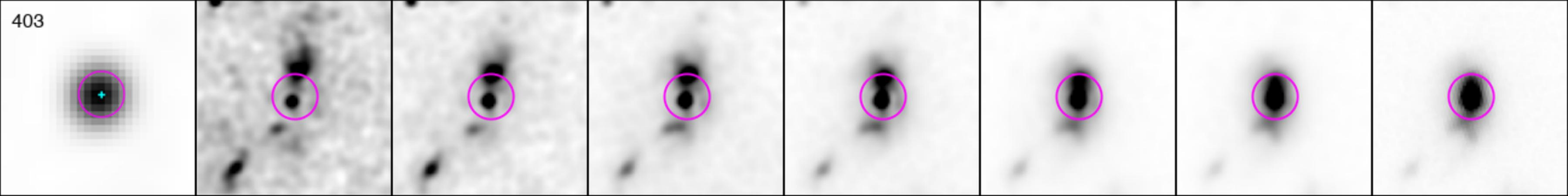"}     \plotone{"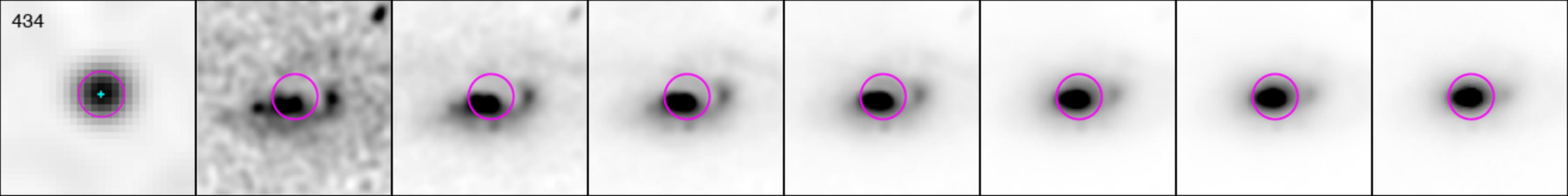"}    \plotone{"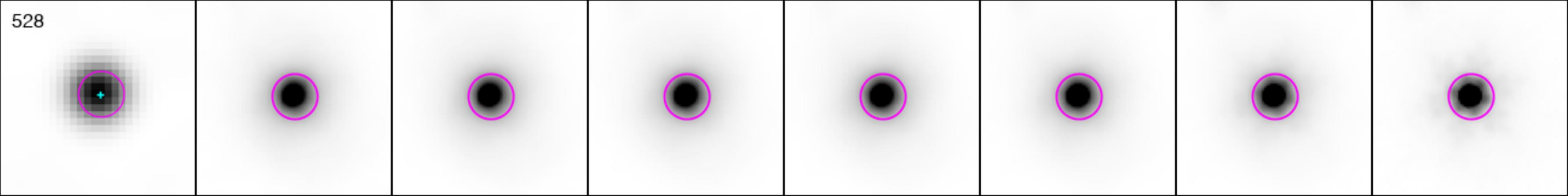"}
     \caption{Negative images of the eight VLBA-detections in the JWST/NIRCam area. The leftmost panels show the 3 GHz VLA radio image with the source ID included \citep{scuba}. Other panels show the NIRCam images in the filters of F090W, F115W, F150W, F200W, F277W, F356W, and F444W. Each panel is 3\arcsec x 3\arcsec. The green '+' sign on the first postage stamp indicates the VLBA positions. The magenta circle indicates the VLA beam size of 0.7\arcsec, centered around the VLA position. A logarithmic scale has been implemented for visualization. For the VLA images, the minimum value is fixed at -3 $\mu$Jy and the maximum is set to the 99.9999th percentile of the data. For the JWST images, the display range spans from the 0.01st percentile to the 99th percentile of the data.
\\  
     }
     \label{fig:vlajwst}
\end{figure*}


\subsection{Imaging}

We Fourier-transformed, cleaned, and imaged the source fields (UV data) with \textsc{imagr}, averaging together all 256 frequency channels (\textsc{stokes = i}). We used natural weighting (\textsc{robust = 5}) to minimize the rms noise. The central $1\farcs6\times1\farcs6$ around each phase center was imaged using 4096 pixels along each axis, and the average synthesized beam in each image was $4 \times 3.5$ mas. The typical sensitivity loss from delay and time smearing at 6" offset is $\sim 48 \%$.

Source detection required the cleaned image to contain a source with a peak flux $>5\sigma$ above the background noise. After initial imaging, we analyzed fields with clear detections using AIPS' \textsc{jmfit}.  We applied primary beam corrections (PBC) to the peak flux density ($S_{\rm peak}$) and integrated flux density ($S_{\rm int}$) of the source. To check for possible coordinate shifting we re-imaged the fields with no detections with 6\arcsec sides and $12000$ pixels in each axis. No additional detections were found.


\subsection{Source Properties\label{subsec:VLBAproperties}}

Of the sample of 106 radio VLA sources, 12 gave VLBA detections. We measured the source properties from the images by fitting them with Gaussian and point models vis AIPS \textsc{jmfit}, the point model being a Gaussian with size fixed to that of the synthesized beam. Figure \ref{rob5_point_peak_vs_gauss_peak} presents a comparison between the flux densities obtained from the point model and the Gaussian model. Higher resolution imaging was also used to check for potential structures in the detected sources. However, this increase of angular resolution comes at the cost of higher image noise. Only PC 64 appears resolved, but no specific structures were detected. PC 67 also exhibits a slight indication of being resolved, but the significance is low. For the other fields we applied a point-model fit (\textsc{dowidth = -1}). The resultant images are plotted in Figure \ref{fig:sources}.


\section{Multi-wavelength counterparts} \label{sec:mwc}

\subsection{VLA Counterparts}\label{subsec:VLAc}

The VLA survey of the TDF served as the parent sample for selecting our VLBA phase centers \citep{scuba}. The VLA sample was observed at 3 GHz with an rms sensitivity of 1 $\mu$Jy beam$^{-1}$ at a resolution of 0\farcs7. To account for positional uncertainties in the VLA, we searched for VLBA detections within a 1\farcs5 radius around each VLA coordinate. The VLA counterparts for eight of the VLBA detections are shown in Figure \ref{fig:vlajwst} as illustrative examples. 

To estimate the radio spectral index from the VLA 3 GHz data \citep{scuba}, we subdivided the 2 GHz bandwidth into 25 sub-bands, each with a 3\% fractional bandwidth. We generated images for each sub-band independently but performed joint CLEANing to improve signal quality while accounting for frequency-dependent variations in sky brightness and antenna gain. For each source in our catalog, we interpolated the flux densities from these sub-band images, applying frequency-dependent gain corrections, then fit a spectrum for each source to yield an estimated spectral index. We adopt the convention $S_{\nu} \propto \nu^{\alpha}$, where $S_{\nu}$ represents the integrated radio flux density, and $\alpha$ denotes the intrinsic spectral index of the source. Due to limitations in precise primary-beam correction with frequency, these spectral indices are uncertain, but they indicate the general spectral behavior of the sources.


\begin{figure}[ht]
     \epsscale{0.93}
     \plotone{"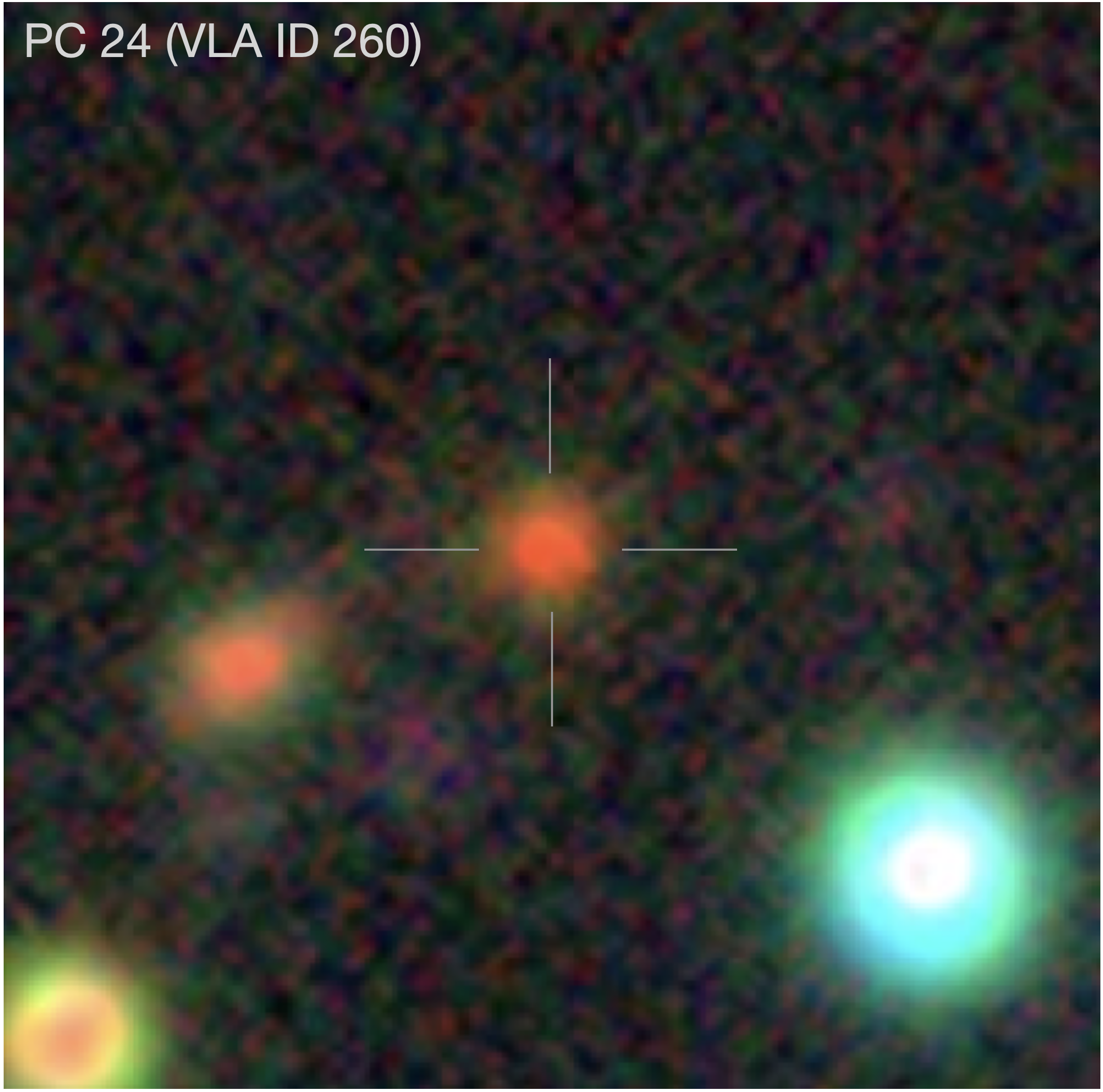"} 
     \plotone{"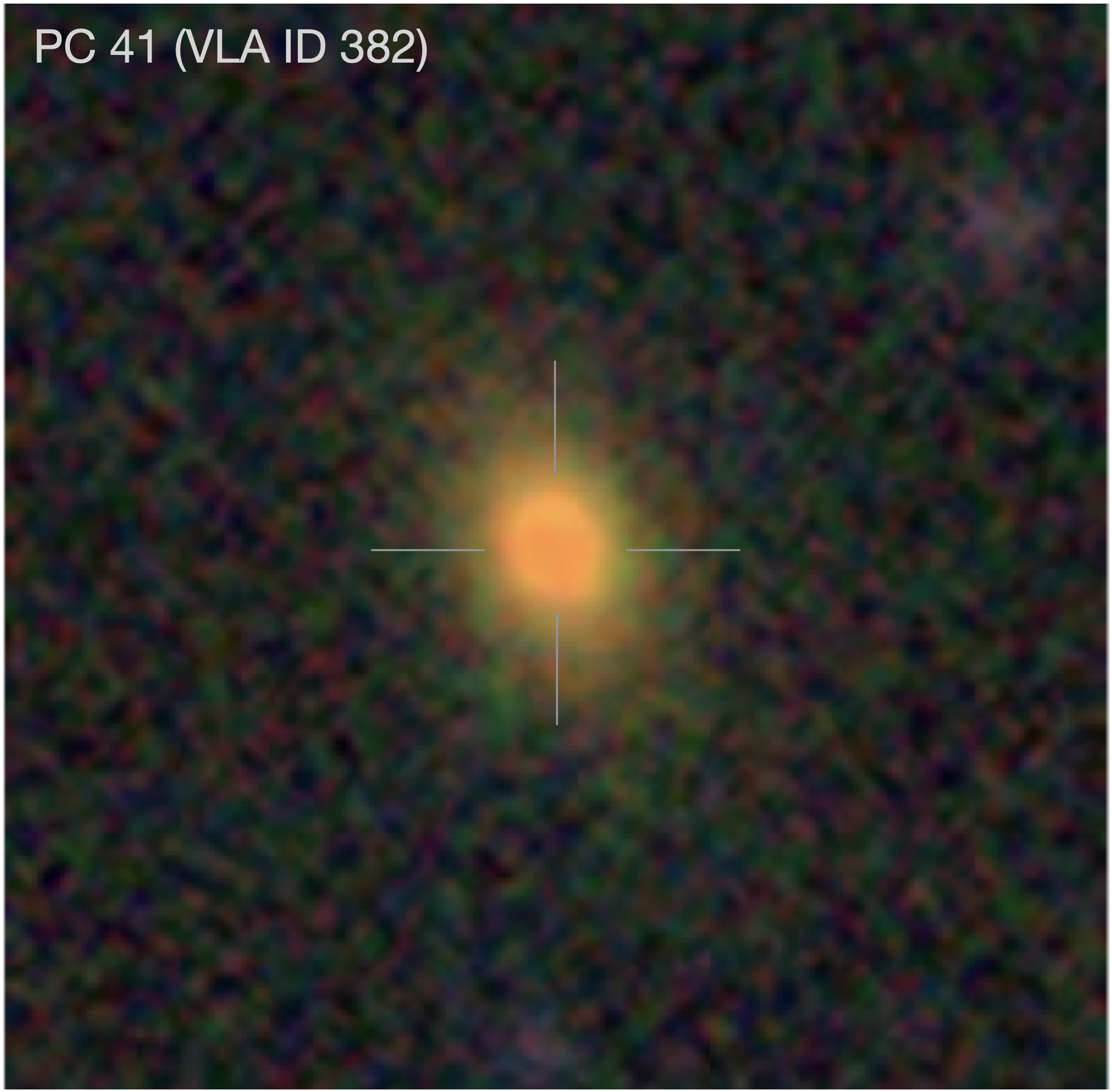"} 
     \plotone{"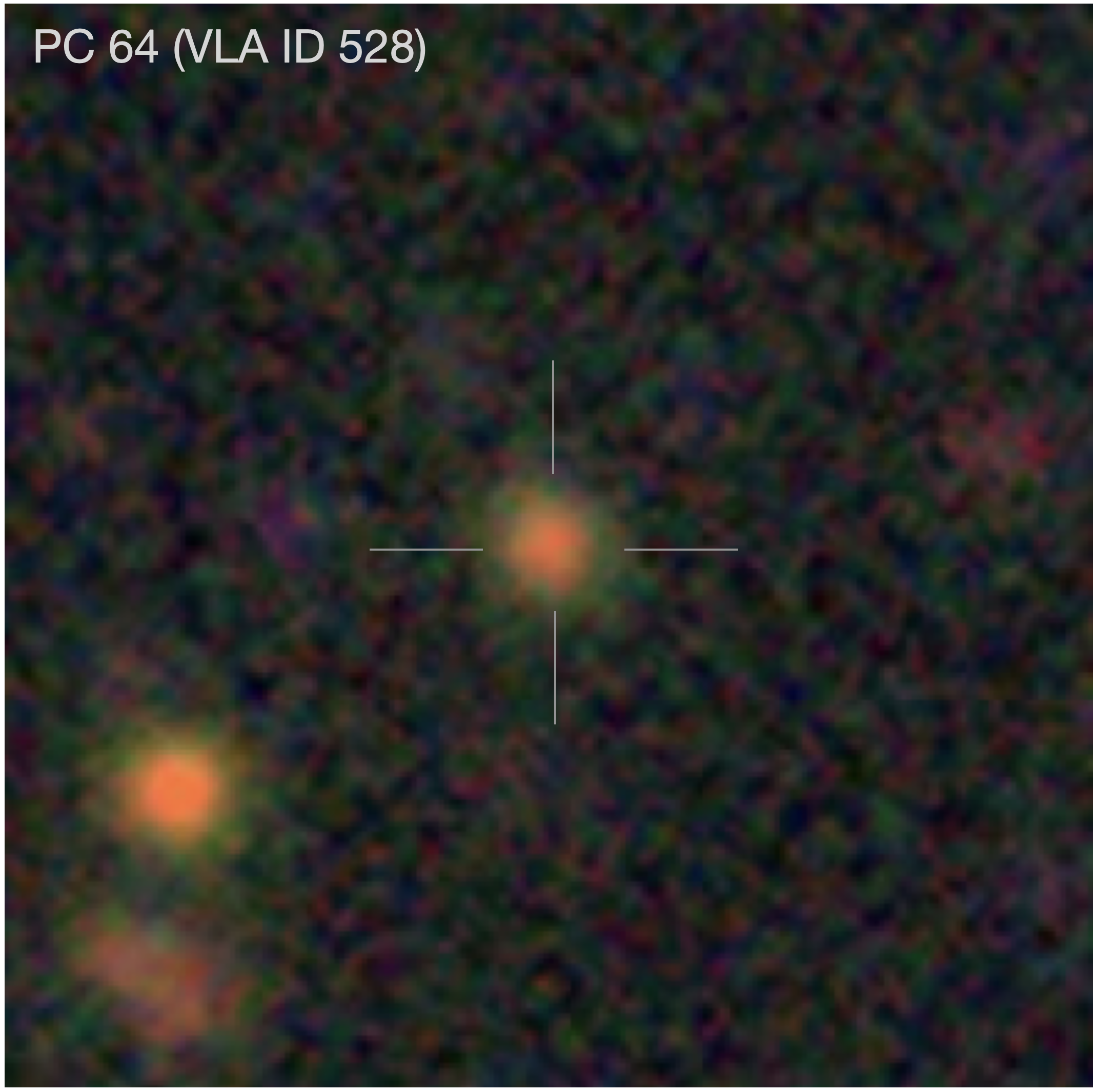"} 
     \caption{Legacy survey cutouts of the three optical counterparts of our VLBA detections (PC 24, 41 and 64). These are RGB images made from the $g$, $r$, and $z$ bands, with a size of 200 pixels, and a pixscale of 0\farcs12 per pixel.  In all the cutouts, North is oriented at the top and East to the left. 
     }
     \label{fig:sdss}
\end{figure}


\subsection{WISE counterparts}
\label{subsec:wise}

We conducted a search for mid-infrared counterparts of our VLBA sources in the WISE mission catalogue \citep{ww}, across four wavelength bands: 3.4 $\mu$m (W1), 4.6 $\mu$m (W2), 12 $\mu$m (W3), and 22 $\mu$m (W4). Using a 10\arcsec cone search radius of our VLBA positions, we find that all of our VLBA detections, except for PC 25, have WISE counterparts.


\subsection{SCUBA counterparts}

We also cross-matched the radio sample with the 850 $\mu$m observations of the TDF with the Submillimeter Common-user Bolometer Array 2 (SCUBA-2) of the James Clerk Maxwell Telescope \citep{scuba}. The survey was carried out over an area of approximately 0.087 deg$^2$ in the TDF, achieving a mean 1$\sigma$ sensitivity of 1.0 mJy beam$^{-1}$. For the discussion on the SCUBA counterparts, we use the SCUBA$-$VLA associations identified by \cite{scuba}. Currently, the SCUBA$-$VLA cross-match yielded a total of 85 sources, but only 17 of these sources matched with our VLBA phase centers, with three of them being detections (PC 26, 46, and 71) and the remaining 14 being non-detections. However, because of the relatively poor angular resolution of SCUBA observations and the resulting positional uncertainties, we remain cautious about confidently associating them with specific VLA sources. More precise submillimeter positions are needed to fully leverage the potential of SCUBA-2 data.

\subsection{JWST/NIRCam counterparts}

To further investigate potential infrared counterparts of the VLBA-detected sources, we cross-matched our sample with JWST PEARLS observations from the literature. \citet{pearlsjwst} described the association between the
\citet{scuba} VLA sources and the first spoke of the JWST/NIRCam
imaging of the TDF\null. NIRCam has now observed the remaining three
spokes, and Willner et al.\ (in prep.) will describe the results. In
brief, nearly all of the 211 VLA radio sources within the NIRCam area
have a NIRCam counterpart, and more than 80\% of the counterparts have
magnitudes $\rm [F444W]<22$~AB\null. All eight of the VLBA-detected
sources within the NIRCam area have counterparts, the faintest of which
has $\rm [F444W]\approx21.4$~AB\null. Figure \ref{fig:vlajwst} presents for the first time, postage--stamp images of these VLBA detections across seven JWST NIRCam bands (F090W, F115W, F150W, F200W, F277W, F356W, and F444W) using \textsc{astrocut}\footnote{https://astrocut.readthedocs.io/en/latest/} \citep{astrocut}. Four counterparts have
spectroscopic redshifts, and the remainder have photometric redshifts
(Table~\ref{tab:jwsttable}).\\

We also cross-matched our observations with the 66 galaxies observed by JWST/NIRCam in \cite{ortiz} that looked for galaxies with clear point-source centres indicative of AGN, and found that three of our VLBA detections (PC 3, 7, and 64) are included in their sample. We further expand on the optical-NIR properties of our detections in Section \ref{subsec:optir}.\\

The JWST postage stamps associated with the non-detections of our sample are presented in Figure \ref{fig:jwstnd}. Each image highlights the VLBA position (indicated by a central red circle with a diameter of 0\farcs15) within a 3\arcsec × 3\arcsec field of view.


\subsection{SDSS counterparts}

A search for optical counterparts within 1\arcsec of our VLBA sample in the Sloan Digital Sky Survey (SDSS) archive, identified 34 matches. Among these, three sources were detected with the VLBA (PC 24, 41, and 64). All optical counterparts are within 0\farcs1 of their corresponding radio positions and have photometric redshifts $z_{ph} \sim 0.3$. The non-detected sources, on the contrary, exhibit a broad range of distances ($0.1< z_{ph}< 0.8$). This spread in redshift may suggest diverse evolutionary stages or environments among these sources, potentially contributing to their non-detection in the higher-resolution VLBA observations.\\

Legacy survey cutouts \citep{desi} of the three optical counterparts to our VLBA detections are shown in Figure~\ref{fig:sdss}, where we present RGB images constructed from the $g$, $r$, and $z$ bands, each with a size of 200 pixels and a pixel scale of 0\farcs12 per pixel.\\


\begin{deluxetable*}{lc|cccrr|rr}
    \tablecaption{VLBA Detections \label{tab:detections}}
    \tablewidth{0pt}
    \tablehead{
     \multicolumn{2}{c|}{Source}    & \multicolumn{5}{c|}{VLBA} & \multicolumn{2}{c}{VLA} \\
    \hline
    \colhead{PC}  & \multicolumn{1}{c|}{VLA ID}  & \colhead{$\alpha_{\rm J2000}$} & \colhead{$\delta_{\rm J2000}$} &  \colhead{$\Delta$} &
    \colhead{$S_{\rm peak}$}  & \multicolumn{1}{c|}{S/N} &\colhead{$S_{\rm 3 GHz}$}   &\colhead{Spec. Index}
    \\
    \colhead{} & \multicolumn{1}{c|}{} & \colhead{(h m s)} & \colhead{($^\circ~^\prime~^{\prime \prime}$)} & \colhead{(mas)} &
    \colhead{($\mu \rm{Jy/bm}$)} & \multicolumn{1}{c|}{} &  \colhead{($\mu$Jy)} &  \colhead{($\alpha$)}
    \\
    \colhead{(1)} & \multicolumn{1}{c|}{(2)} & \colhead{(3)} & \colhead{(4)} &
    \colhead{(5)} & \colhead{(6)} & \multicolumn{1}{c|}{(7)} & \colhead{(8)} &
    \colhead{(9)} 
    }
    \startdata
    3 & 124 & 17 22 30.3766	&+65 51 07.9098 &  0.1 & $143 \pm11$ &	33.1   &   $472.0\pm12.3$ & $-$0.37$\pm$0.20	\\
    7 & 141 & 17 22 33.3964	&+65 47 57.7862 &  0.1 & $174 \pm 6$  & 38.5 &  $1071.0\pm25.7$ & $-$0.51$\pm$0.20	\\
    14 & 182 & 17 22 38.9746	&+65 51 43.0270 &  0.1 & $78  \pm10$ &	16.3 &   $74.3\pm2.6$  & $-$0.01$\pm$0.20   \\
    24 & 260 & 17 22 53.1466	&+65 48 47.3870 &  0.1 & $53  \pm 4$ & 14.3 &    $57.4\pm2.0$   & $-$0.17$\pm$0.32 	\\
    25 & 262 & 17 22 53.7332	&+65 52 17.7410 &  0.1 & $67  \pm 8$ & 16.4 &   	$ 117.0\pm3.7$   & $-$0.41$\pm$0.23	\\
    26 & 279 & 17 22 55.6433	&+65 53 01.3809 &  0.1 & $148 \pm11$ & 32.5 &   	$ 223.0\pm6.8$   & $-$0.50$\pm$0.21	\\
    41 & 382 & 17 23 11.9299	&+65 50 14.3367 &  0.2 & $40  \pm 4$ &	9.1   &   	$ 100.3\pm7.1$  & $-$0.09$\pm$0.31	\\
    46 & 403 & 17 23 16.9409	&+65 50 45.4654 &  0.3 & $27  \pm 4$ & 7.5 &   $ 181.0\pm6.5$ 	  & $-$0.90$\pm$0.22   \\
    47 & 434 & 17 23 22.5065	&+65 49 45.6346 &  0.3 & $23  \pm 4$ & 6.8 &   	$ 50.5\pm1.8$    & $-$0.41$\pm$0.33   \\
    64 & 528 & 17 23 40.6700	&+65 49 52.8134 &  0.1 & $159 \pm 5$  & 49.6 &	$ 512.0\pm15.4$& $-$1.00$\pm$0.20  	\\
    67 & 554 & 17 23 45.4225	&+65 43 57.5330 &  0.1 & $274 \pm11$ &  46.0  & 	$ 286.0\pm8.7$ & 0.51$\pm$0.21	\\
    71 & 614 & 17 23 59.7510	&+65 45 48.3340 &  0.1 & $109 \pm12$  & 21.6 &   	$195.0\pm6.0$   & $-$0.43$\pm$0.22	\\
    \enddata
    \tablecomments{
    {\bf(1)}: Phase Center (PC) number, used as reference for the VLBA sources in this project; {\bf(2)}: VLA ID \citep{scuba};
    {\bf(3, 4)}: Right Ascension ($\alpha$) and Declination ($\delta$) (J2000) of the source, measured with the VLBA (natural weighting);
    {\bf(5)}: Position uncertainty ($\Delta$) for the source;
    {\bf(6)}: Peak ($S_{\rm peak}$) VLBA flux density of the source (4.8 GHz) after primary beam correction;  
    {\bf(7)}: Signal-to-noise ratio (S/N) from the ratio of uncorrected $S_{\rm peak}$ over image rms;
    {\bf(8)}: VLA 3 GHz flux density, corrected for the primary beam, and its uncertainty
    \citep{scuba};
    {\bf(9)}: Spectral index $\alpha$ of the VLA 3 GHz counterpart.
     }
\end{deluxetable*}

\section{Results} \label{sec:res}


\begin{deluxetable*}{cccccc}
\tablecaption{Resolved source data : PC 64 (VLA ID 528) \label{tab:SRC64 Properties}}
\tablewidth{0pt}
\tablehead{
\colhead{Component} & \colhead{Maj/min axes}& \colhead{Physical Size}& \colhead{$\rm S_{int}$} & \colhead{$\rm L_{4.8~GHz}$}  & \colhead{$ \nu \cdot \rm L$}\\
\colhead{} & \colhead{(mas)}& \colhead{(pc)} & \colhead{($\rm erg~s^{-1} cm^{-2} Hz^{-1}$)} & \colhead{($\rm W Hz^{-1}$)}& \colhead{($\rm W$)}
}
\startdata
\multirow{2}{*}{1\textsuperscript{st} Gaussian} & $3.5\pm0.3$ & $18.7\pm 1.6$ & \multirow{2}{*}{$(1.7\pm 0.1) \times 10^{-27}$} & \multirow{2}{*}{$(8.88\pm 0.35) \times 10^{22}$} & \multirow{2}{*}{$(4.26\pm 0.17) \times 10^{32}$}\\
 & $0.8\pm0.6$ & $4.3\pm 3.2$ &  &  & \\
\multirow{2}{*}{2\textsuperscript{nd} Gaussian} & $11.3\pm1.2$ & $60.3\pm 6.4$ & \multirow{2}{*}{$(2.8\pm0.3) \times10^{-27}$} & \multirow{2}{*}{$(14.61\pm 1.17) \times 10^{22}$} & \multirow{2}{*}{$(7.01\pm 0.56) \times 10^{32}$}\\
 & $7.9\pm0.9$ & $42.1\pm 4.8$ &  &  & \\
 \hline
\enddata
\tablecomments{
$\rm L_{4.8~GHz}$ and physical sizes calculated using the redshift $z = 0.3760$ (see Table \ref{tab:jwsttable}).
}
\end{deluxetable*}


\subsection{Detection rate}\label{subsec:det}

We observed a targeted sample of 106 VLA sources in the TDF \citep{scuba} with the VLBA at 4.8 GHz. This significantly improved the resolution of the TDF in radio bands from 0\farcs7 with the VLA to 4 mas with the VLBA. We detected 12 of 106 sources, or a VLBA detection fraction of $\sim$11\% ($\sim$ 3.3 $\mu$Jy rms sensitivity). For brighter radio sources with VLA 3 GHz flux densities greater than 50 $\mu$Jy, the detection rate increases to $\sim$35\%. Table \ref{tab:detections} gives the source properties of the detections.
The deconvolved images of the detected sources are shown in Figure \ref{fig:sources}. Table \ref{tab:non detections} gives the properties of the undetected sources. 


\begin{deluxetable*}{ccccccc}[ht]
    \tablecaption{Brightness temperatures ($\rm T_b$) of the VLBA Detections.
    \label{tab:temperatures}}
    \tablewidth{0pt}
    \tablehead{
    \colhead{PC} & \colhead{VLA ID} & \colhead{$\phi_{\rm Maj}$}  & \colhead{$\phi_{\rm min}$} &
    \colhead{$\Omega$}  & 
    \colhead{$\rm T_b$}  
    \\
    \colhead{} & \colhead{} & \colhead{(mas)} &\colhead{(mas)}  &
    \colhead{($10^{-16}$ sr)} &  \colhead{($10^5$ K)} 
    \\
    \colhead{(1)} & \colhead{(2)} & \colhead{(3)} &
    \colhead{(4)} & \colhead{(5)} & \colhead{(6)}  
    }
    \startdata
     3 & 124 & $ 4.5$ & $ 3.8$ &12.4 &   $>4.5$ \\
     7 & 141 & $ 4.2$ & $ 3.8$ &11.8 	& $>5.8$ \\
    14 & 182 &  $ 4.4$ & $ 3.8$ &12.1 	& $>2.5$ \\
    24 & 260 & $ 2.5$ & $ 2.1$ &3.8  	& $>5.4$ \\
    25 & 262 & $ 4.3$ & $ 3.7$ &11.9 	& $>2.2$ \\
    26 & 279 & $ 4.4$ & $ 3.9$ &12.5	& $>4.6	$ \\
    41 & 382 &  $ 3.9$ & $ 3.7$ &10.6	& $>1.5	$ \\
    46 & 403 & $ 4.0$ & $ 3.7$ &11.0 	& $>1.0$ \\
    47 & 434 & $ 3.9$ & $ 3.7$ &10.7	& $>0.9	$ \\
    64 & 528 &  $4.9^{+0.2}_{-0.3}$ & $3.0^{+0.3}_{-0.2}$ &11.1	& $11.3^{+1.4}_{-1.0}$ \\
    67 & 554 & $ 4.2$ & $ 2.7$ &8.5  	& $>12.7$ \\
    71 & 614 &  $ 4.3$ & $ 2.8$ &8.9 	& $>4.8$\\
    \enddata

    \tablecomments{
        {\bf(1)}: Phase Center (PC) number, used as reference for the VLBA sources in this project;  
        {\bf(2)}: VLA ID of the 3 GHz counterparts \citep{scuba}; 
        {\bf(3, 4)}: deconvolved major ($\phi_{\rm Maj}$) and minor($\phi_{\rm min}$) axes of the VLBA detections. For the unresolved sources, the deconvolved maj/min axes are defined as the respective beam sizes, given that this would be their maximum angular size to remain unresolved (thus the $T_{\rm B}$ are lower limits);
        {\bf(5)}: solid angle ($\Omega$) in steradians;
        {\bf(6)}: Brightness temperature ($\rm T_B$) per Equation 1.
        }
\end{deluxetable*}


\subsection{Resolved source PC 64 (VLA ID 528)}\label{subsec:64}

The only significantly resolved source (PC 64 or VLA ID 528) was re-fitted with two Gaussian components with results in Table \ref{tab:SRC64 Properties}. Both Gaussians had very similar centroids, but the physical sizes differ. The first component is elongated with the axis ratio 4.3. The second component has a radio-emitting region approximately 40 pc larger than that of the first. It exhibits a rounder structure, with the axis ratio 1.5. The first component could be a small jet, while the second component could be an AGN core or plasma heated when the small jet is terminated/interacts with the surrounding medium.

\subsection{Brightness temperature}\label{subsec:tb}

The brightness temperature ($T_B$) of each of the 12 detections is calculated according to the following equation,
\begin{equation}
    \rm T_b = \frac{2 ln(2) }{ k \Omega} S_{\rm int} (1+z) \lambda^2
\end{equation}
where $\lambda = 0.0623$ m ($\nu = 4.8$ GHz), Boltzmann's constant $\rm k=1.380649 \times 10^{-23}$ J$\rm K^{-1}$, $z$ is taken to be $0$ for all the sources as we do not have information regarding the redshift of these sources, solid angle $\Omega$ = $\pi$(deconvolved major axis $\times$ deconvolved minor axis). For unresolved sources, we define the deconvolved size $\equiv$ beam sizes, and the size of the radio-emitting region remains indeterminate, allowing only lower limits to be estimated for the brightness temperature. 

We find that the brightness temperatures for all the detected sources exceed 10$^5$ K. The $T_B$ values, listed in Table \ref{tab:temperatures}, are significantly higher than those typically produced by SF processes \citep{1992ARA&A..30..575C}. In nearby galaxies, such high brightness temperatures can arise from AGN, as well as supernovae or their remnants. However, in more distant galaxies ($z>$0.1), brightness temperatures in excess of 10$^5$ K can be reliably linked to accretion-related AGN emission, as the high energies required for such $T_B$ values point to an AGN origin \citep{kewley}.


\begin{figure}[t]
 \epsscale{1.1}
 \plotone{"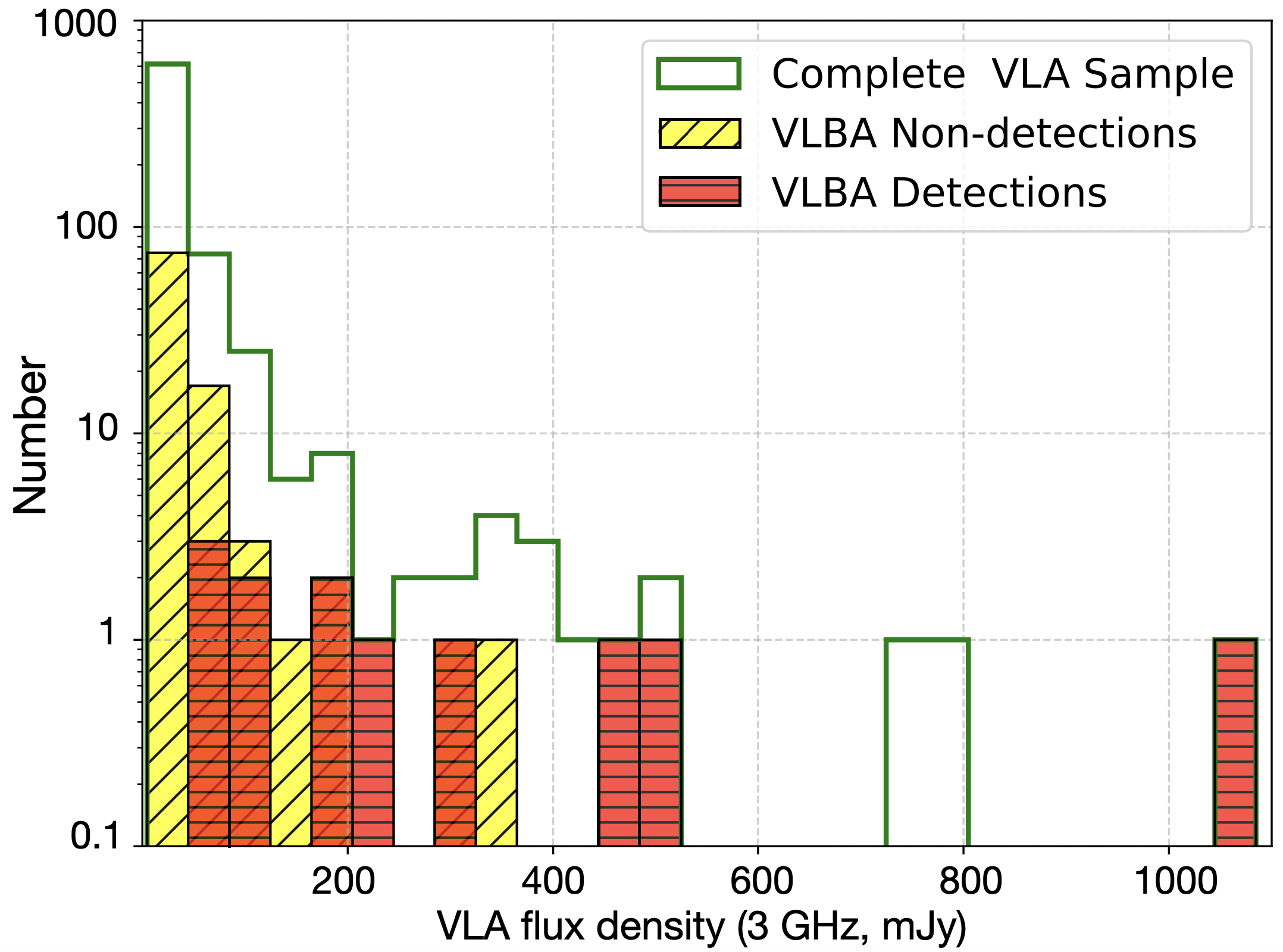"}
 \caption{Distribution of VLA peak flux densities ($S_{\rm peak}$) of the VLBA observed sources. The VLBA detections are depicted in red bins with horizontal dashes, and the non-detections in yellow bins with forward slashes. The complete VLA sample of the TDF \citep{scuba} are shown with empty green steps. The eight brightest VLA sources, with flux density in excess of 1100 mJy, are off scale to the right. 
 }
 \label{fig:Histo_vla}
\end{figure}


\begin{figure}[t!]
 \epsscale{1.1}
 \plotone{"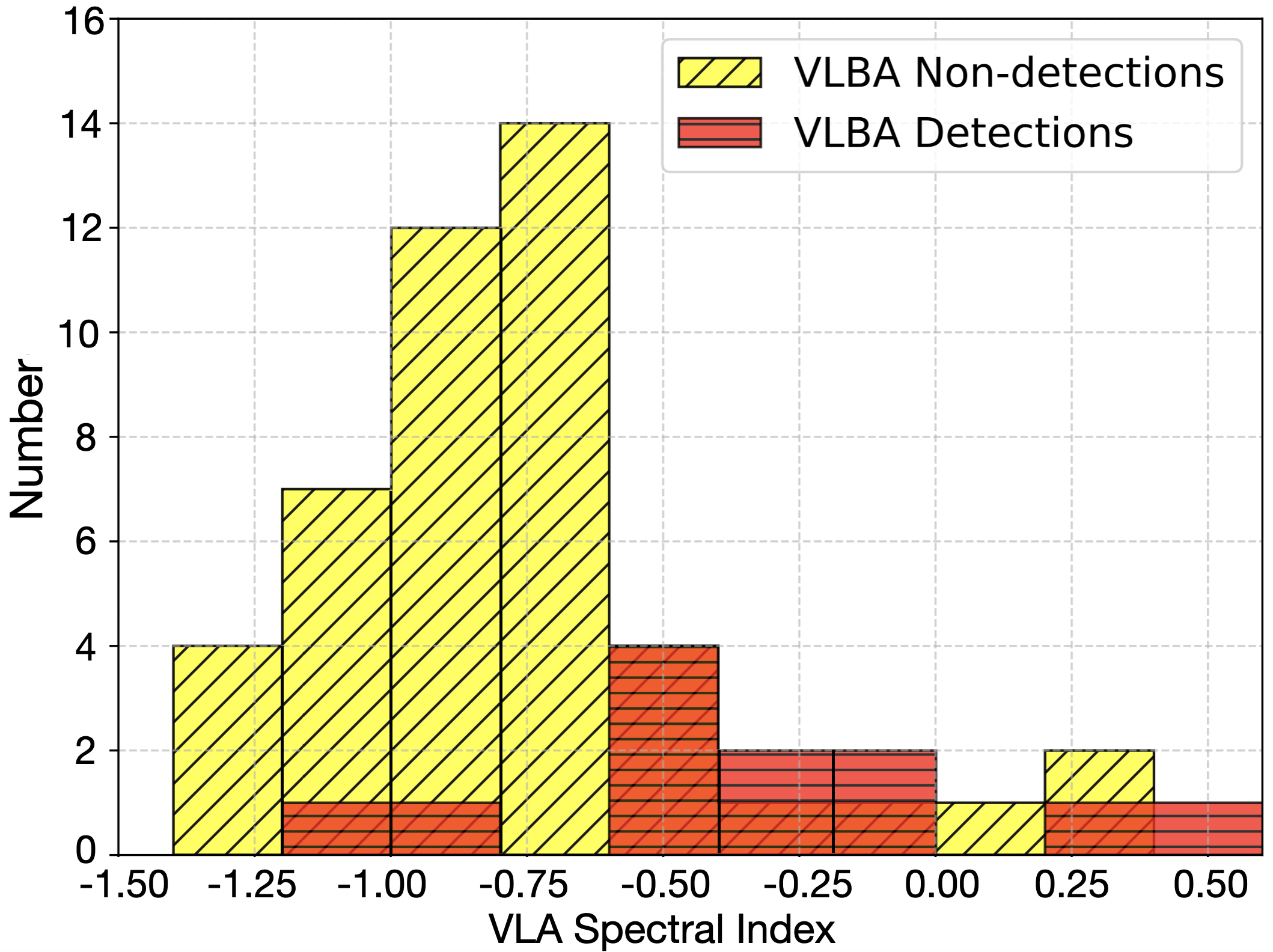"}
 \caption{Distribution of spectral indices \citep[measured from VLA 3 GHz observation;][]{scuba} for the VLBA detections (red bins with horizontal dashes) and non-detections (yellow bins with forward slashes) in the TDF.
 \label{fig:Histo_NEP}
 }
\end{figure}


\begin{figure}[t!]
 \epsscale{1.1}
 \plotone{"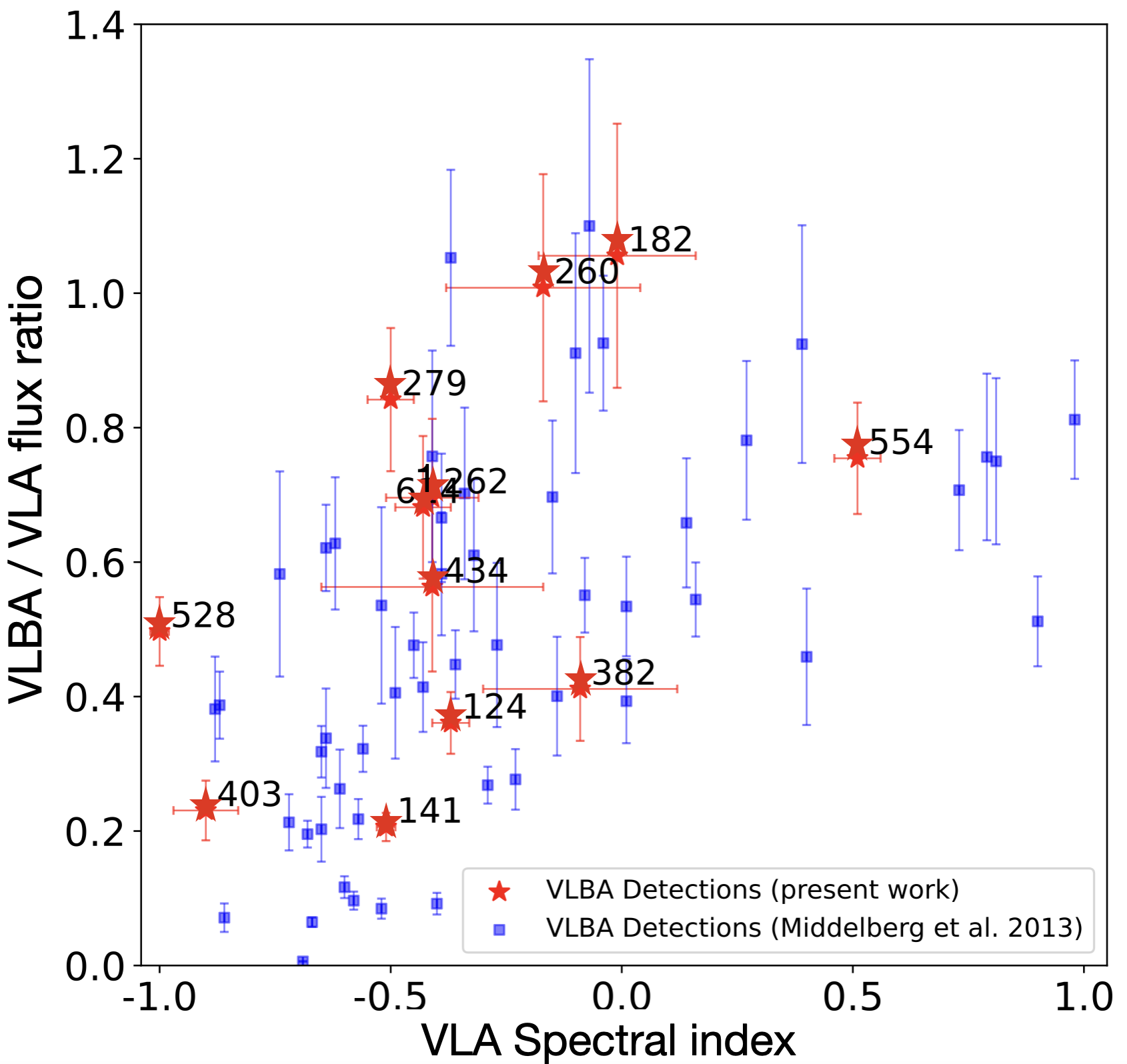"}
 \caption{VLBA/VLA flux density ratio vs. VLA spectral index of the VLBA-detected sources in the TDF (red stars) and the Lockman Hole/XMM field \citep[blue squares;][]{Middelberg}. 
 The TDF fluxes were measured at 4.8 GHz with the VLBA and at 3 GHz with the VLA, with the latter being extrapolated to 4.8 GHz using the source's radio spectral index. In contrast, the Middelberg fluxes (for both VLA and VLBA) were measured at 1.4 GHz.
 \label{fig:NEP_RatiovsSpecindex}
 }
\end{figure}


\subsection{Radio fluxes and spectral indices}\label{subsec:rsi}

Figure \ref{fig:Histo_vla} presents the distribution of 3 GHz peak flux densities for the VLA sample. When observed, nearly all of the brightest VLA sources were detected by the VLBA. All detected sources have $S$(3 GHz) $>$ 50 $\mu$Jy, highlighting a flux threshold for VLBA detection. The VLBA coordinates for all 12 detected sources are within 0\farcs03 of the VLA positions. The VLBA non-detections could just be a matter of sensitivity, though two VLA sources with flux densities above 200 mJy (PC 6 and 33) were undetected in our VLBA survey. These two cases may exhibit intrinsically diffuse radio emission i.e. substantial SF activity creating their VLA flux density. Such extended emission would be resolved out in the VLBA observations. Another possibility is that PC 6 and PC 33 exhibit intrinsic radio variability on long timescales. Although deep field surveys have generally found the $\mu$Jy$-$mJy radio sky to be quiescent \citep{rad}, variability at these flux densities has been reported in other studies \citep{moo,han}, making this a plausible explanation for the observed non-detections.

The radio spectral indices of the VLBA detections are centered around $\alpha \sim -0.4$, and the majority of detections show $\alpha \gtrsim -0.5$ (Figure \ref{fig:Histo_NEP}). In contrast, sources that are not detected with the VLBA, predominantly have $\alpha \lesssim -0.5$. Sources 1, 27, and 40 (VLA IDs 113, 282 and 379), have a flat or inverted spectral index and would be expected to appear in VLBA observations. However, these sources have VLA peak flux densities between 30 and 95 $\mu$Jy, likely rendering them too faint for VLBA detection. Alternatively, their radio emission may be variable, potentially resulting in a diminished flux density during the VLBA observation epochs. In contrast, sources 46 and 64 (VLA IDs 403 and 528), which exhibit very steep spectra, were detected with the VLBA. This could indicate that at least some of their radio emission is from vigorous SF within the host galaxy, with an AGN component giving the VLBA signal. Alternatively, these detections may represent the AGNs jets, observable at VLBA resolutions.\\

\cite{Middelberg} conducted a VLBI survey of 217 radio sources in the Lockman Hole/XMM field and found that most of the detections had a flat spectral index. Despite our sample being approximately five times fainter than and at a higher frequency (4.8 GHz instead of 1.4 GHz) than that of \cite{Middelberg}, the same trend in the spectral index distribution suggests that we are observing a similar population at lower luminosities or greater distances..

Figure \ref{fig:NEP_RatiovsSpecindex} compares the VLBA/VLA flux density ratio with the VLA spectral index.
The VLBA/VLA flux density ratio increases with flatter spectral indices. This trend can be attributed to synchrotron self-absorption, which requires higher densities (indicated by a higher VLBA/VLA ratio) and results in a characteristic flat spectrum. The compactness of these high-density sources makes them more likely to be detected by the VLBA. However, the question remains as to why these compact cores do not produce larger-scale radio emission, despite there being no apparent restriction that prevents it.


\begin{figure}[t]
     \epsscale{1.2}     \plotone{"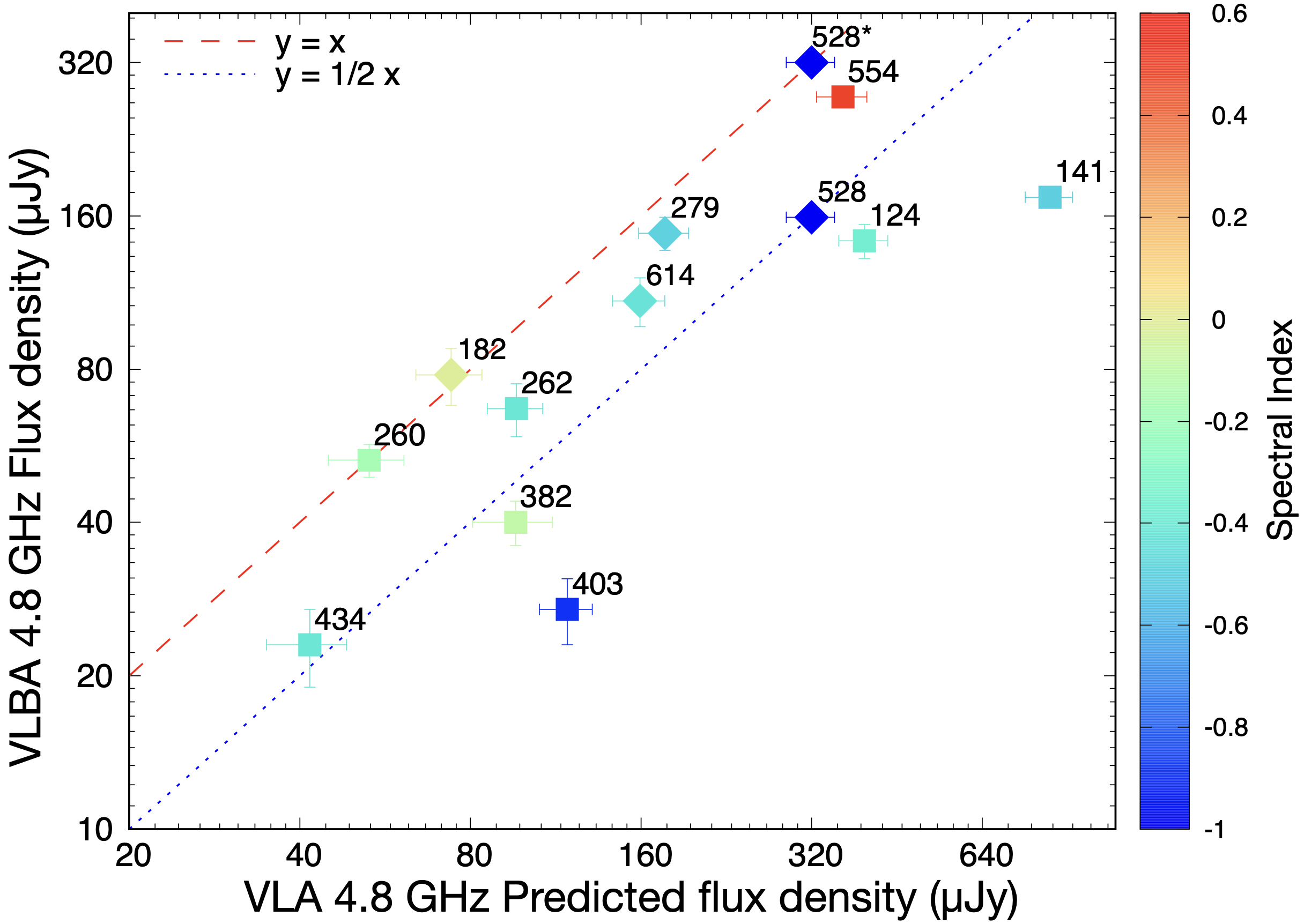"}
     \caption{VLBA 4.8 GHz vs. predicted VLA 4.8 GHz peak flux density ($S_{\rm peak}$), color coded by spectral index $\alpha$. Each source is labelled according to the classification of their WISE counterparts, with the diamonds depicting the AGN-dominated galaxies and the squares depicting the non-AGN dominated, and marked using their respective VLA IDs. The red dashed line represents the location were the observed VLBA 4.8 GHz flux density ($S_{\rm 4.8}$) is equal to the predicted VLA $S_{\rm 4.8}$. The blue dotted line represents the location where the observed VLBA $S_{\rm 4.8}$ is equal to half of the predicted VLA $S_{\rm 4.8}$. ``Flatter spectrum" sources ($\alpha \gtrsim -0.5$) are closer to the 1:1 equality/dashed line. The only resolved detection (VLA ID 528, or PC 64) is presented both with detected VLBA $S_{\rm int}$ (labeled as 528) and $S_{\rm peak}$ (labeled as 528*). 
     }
     \label{fig:predictedVLA+WISE}
\end{figure}


We calculated the expected VLA 4.8 GHz flux densities using spectral indices from the VLA 3 GHz observations. By analyzing which fraction of the predicted VLA emission was detected by the VLBA, we found a strong correlation between this fraction and the VLA spectral index. Figure \ref{fig:predictedVLA+WISE} plots the VLBA 4.8 GHz peak flux density versus the predicted VLA 4.8 GHz peak flux density. The $\alpha \gtrsim -0.5$ sources have a higher fraction (located closer to the equality/dashed line) than steeper-spectrum sources. This suggests that the flatter-spectrum sources are truly compact and produce optically-thick, probably self-absorbed, synchrotron emission. In contrast, the steeper-spectrum sources are physically larger and produce optically-thin synchrotron emission \citep{keller}.\\

\subsection{Mid-infrared colors}
\label{subsec:wisecolor}

\begin{figure}[t]
     \epsscale{1.1}
     \plotone{"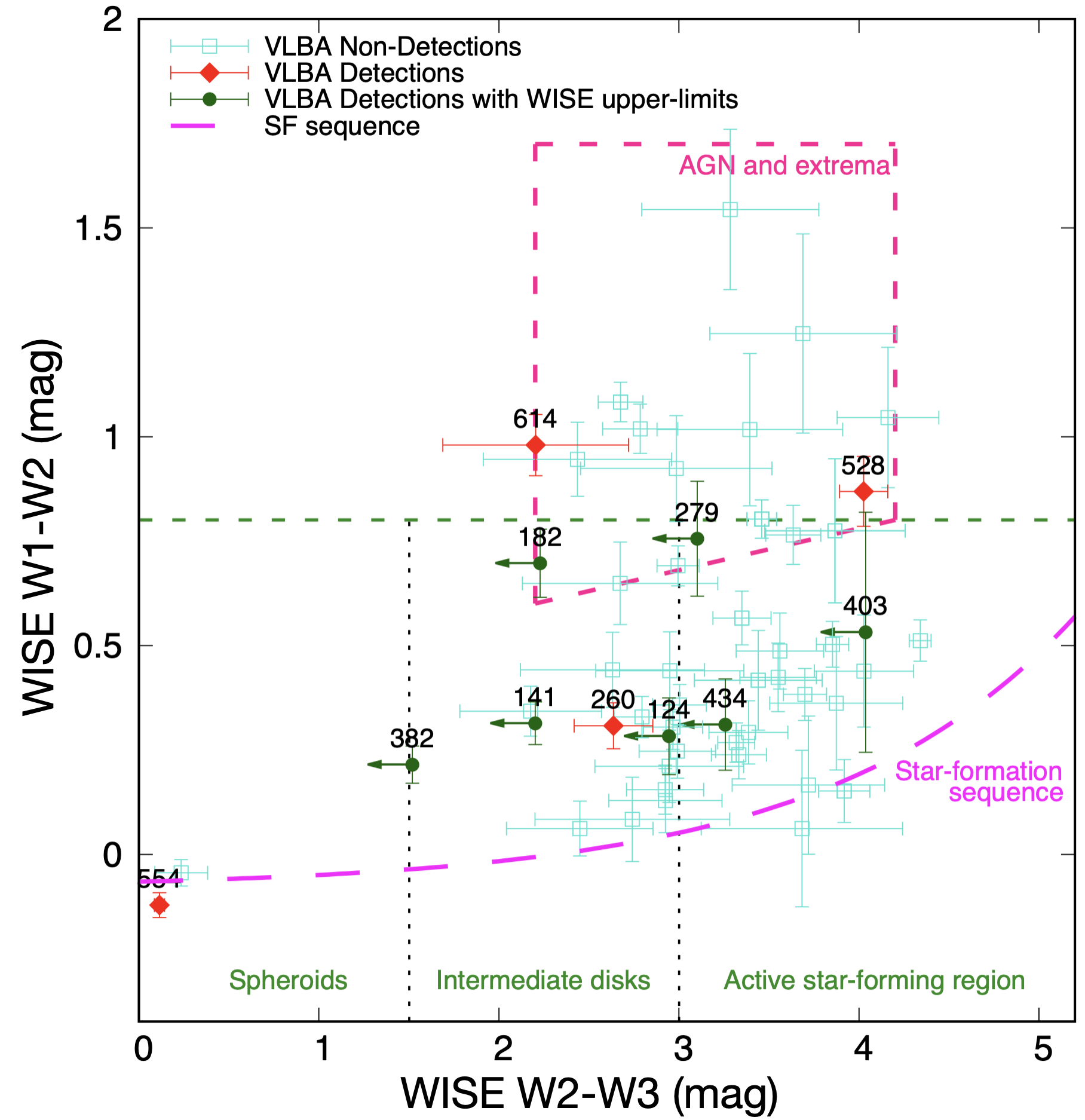"}
     \caption{WISE colour–colour diagram showing W1$-$W2 versus W2$-$W3 colors for the VLBA detections with WISE measurements (red diamonds), VLBA detections with WISE upper limits (green circles), and non-detectons (blue open squares). Lines mark the \cite{jarrett} categories: spheroid-dominated, intermediate-disc, active star-formation region, and AGN-dominated galaxies. The horizontal green dashed line represents the AGN threshold \citep{stern}, above which sources are likely influenced by AGN dust. The magenta dashed curve shows the `star formation sequence' from \cite{Jarrett19}, depicting the typical color transition of galaxies from quiescent to actively forming stars.
     }
     \label{fig:WISE colors W1-W2 vs W2-W3}
\end{figure}

We analyzed the mid-infrared colors of the detected sources' WISE counterparts, and classified the sources following the framework outlined by \cite{jarrett}. This separated the sources into four main categories: spheroid-dominated, intermediate-disk-dominated, star-formation-dominated, and AGN-dominated (see Figure \ref{fig:WISE colors W1-W2 vs W2-W3}). This classification is refined using the AGN selection threshold from \citet{stern}, above which mid-infrared emission is indicative of dust heated by active galactic nuclei. Additionally, we included the `SF sequence' from \cite{Jarrett19}, which illustrates the typical color progression for normal galaxies ranging from quiescent to highly star-forming. The WISE classification was developed for low redshifts and might need to be adjusted for higher redshifts.\\


As shown in Figure \ref{fig:WISE colors W1-W2 vs W2-W3}, four VLBA-detected sources were detected in all the WISE bands. These mostly classified as intermediate-disc or AGN-dominated galaxies, and none fall into the star-formation region. In contrast, many of the VLBA non-detected sources are in the star-formation region, indicating different primary emission sources between detections and non-detections, consistent with VLBA detections being AGN.\\


\begin{figure}[t]
     \epsscale{1.1}
     \plotone{"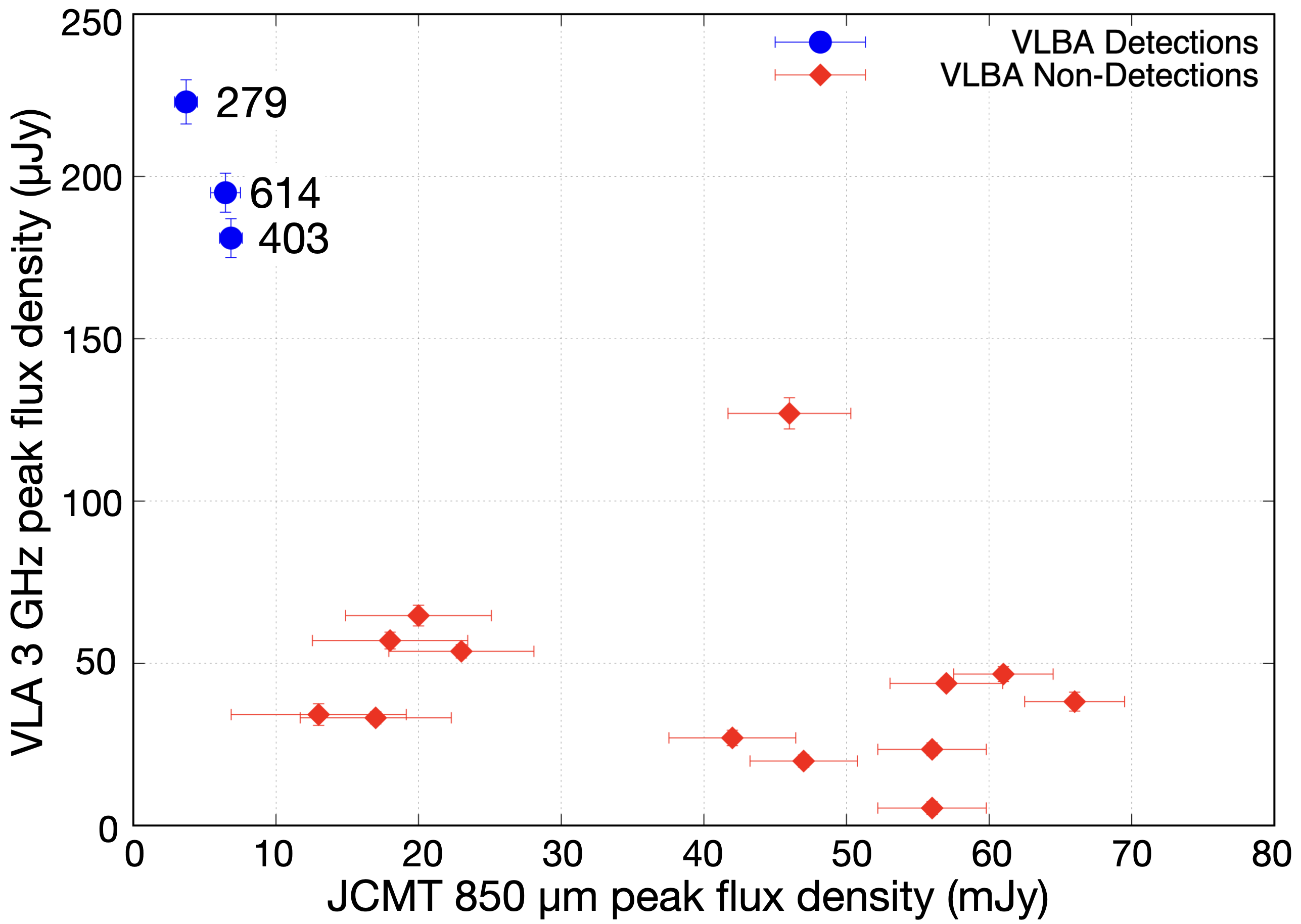"}
     \plotone{"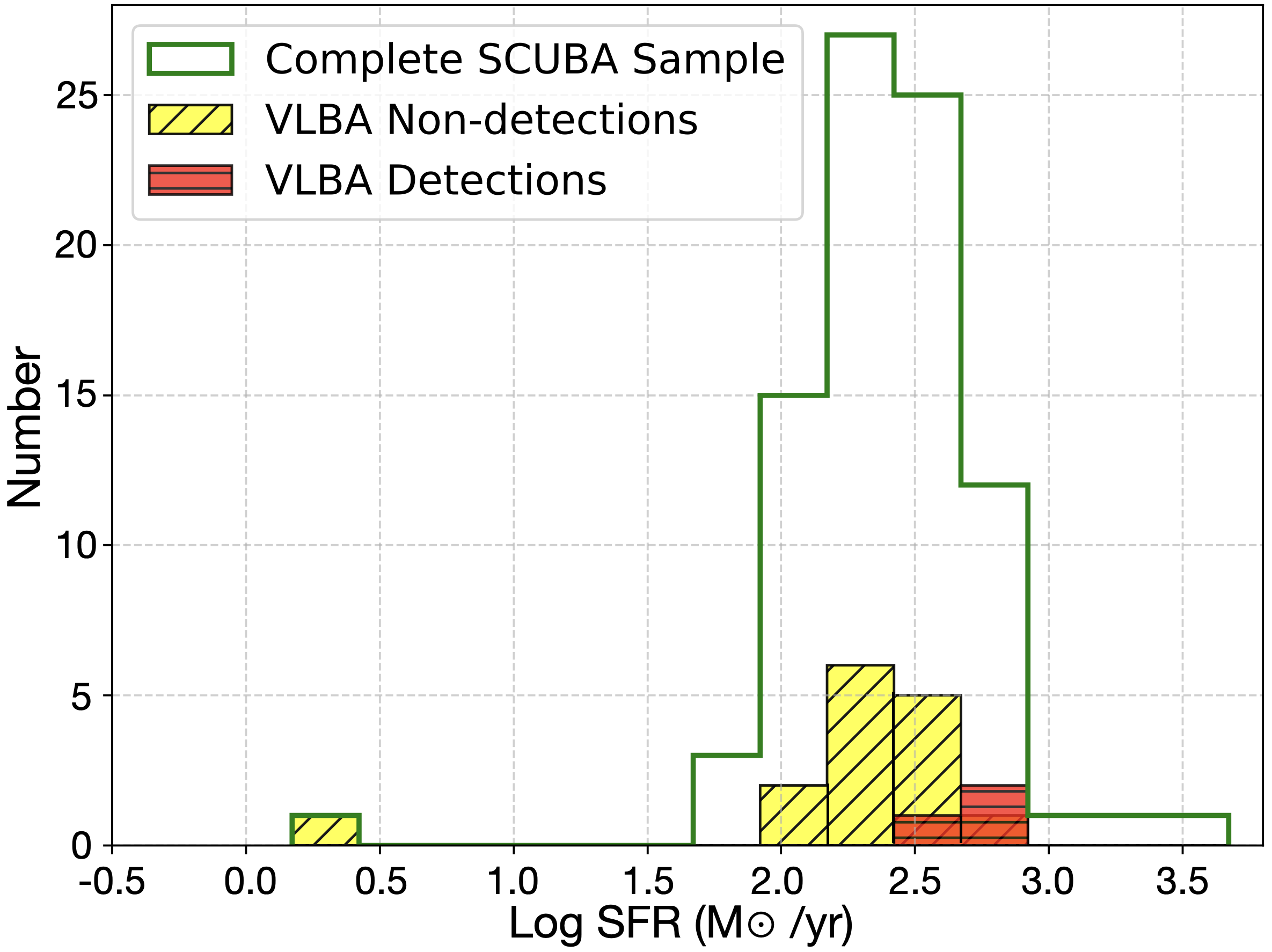"}
     \caption{\textbf{Top :} JCMT flux density vs VLA 3 GHz flux density for VLBA detections in blue circles and non-detections in red diamonds. Detections are identified by their corresponding VLA IDs. \textbf{Bottom :} Histogram depicting the SFR distribution (in logarithmic scale) of the complete SCUBA sample (green empty steps), VLBA detections (red bins filled with horizontal lines) and non-detections (yellow bins filled with forward slashes). 
     }
     \label{fig:scuba}
\end{figure}


\subsection{Physical properties from submillimeter detections}

\cite{scuba} cross-matched the VLA radio sample of the TDF with their 850 $\mu$m SCUBA-2 observations, identifying 85 possible VLA counterparts of SCUBA-2 sources, of which 17 match our VLBA phase centers. Three of these (PC 26, 46, and 71) were detected by VLBA, while the remaining 14 were non-detections. The three VLBA-detected sources in the SCUBA sample are at redshifts of approximately $z \sim$ 1.5 with estimated stellar ages ranging from approximately 80 to 500 Myr. The physical properties derived from SCUBA-2 for the parent galaxies of VLBA detections show no substantial differences from the non-detections in key attributes such as stellar mass, SFR, or other derived galaxy parameters.

All three VLBA detections show low JCMT peak flux densities compared to the non-detections (Figure \ref{fig:scuba}, top panel), which could suggest comparatively lower levels of SF in the host galaxies of these three sources, and/or less dusty star-forming galaxies.

To investigate this further, we examined the SCUBA$-$derived physical properties, focusing on the SFR of both detected and undetected sources \citep{scuba}. Surprisingly, we found that the estimated SFRs were relatively high even for the detected sources (in the range of $\sim$ 450$-$700 $M_{\odot}$/yr), contrary to what might be expected based on their lower JCMT flux densities (see Figure \ref{fig:scuba}, bottom panel). This discrepancy between peak flux density and measured SFR raises interesting questions about the underlying mechanisms in these galaxies, suggesting that while the SCUBA flux density could reflect localized SF activity, the total SFR measurements derived from SCUBA may still capture broader SF processes across the host galaxy (see Section \ref{sec:sfr} for a complete discussion).

Further detailed analysis would be necessary to resolve these contrasting indicators of SF in VLBA-detected sources versus non-detections. However, we also note that because of the limited angular resolution and positional uncertainties of SCUBA observations, we remain cautious in associating the SCUBA counterparts with specific radio sources.\\


\subsection{Properties of the sources from infrared detections}\label{subsec:optir}


\begin{figure}[t]
     \epsscale{1.1}
     \plotone{"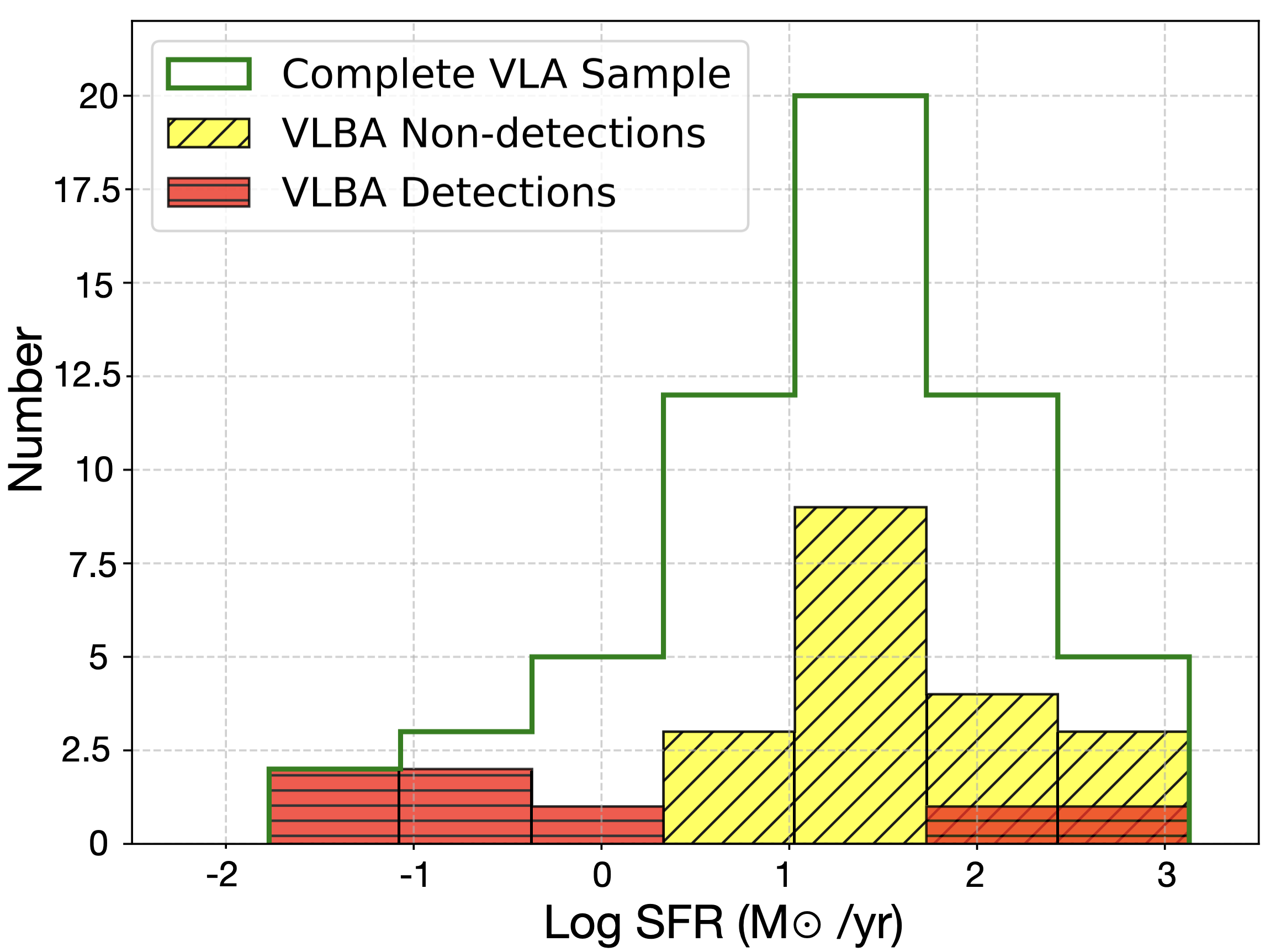"}
     \caption{SFR distribution (in logarithmic scale) of the complete VLA sample with published JWST counterparts (green empty steps) and non-detections (yellow bins filled with forward slashes). We also plot the SFR measured from JWST counterparts of the VLBA detections (red bins filled with horizontal lines) with both published as well as the new SFR measurements reported in this paper.}
     \label{fig:jwstsfr}
\end{figure}


\begin{deluxetable*}{ccLccl}[t]
    \tablecaption{Redshift and SFR ($\rm M_{\odot}/yr$) of the eight VLBA-detected sources within the NIRCam field of view. Radio luminosity ($\rm W/Hz$) and the physical size of the emission region, derived using the measured redshifts. All of the sources listed here are unresolved and we used the beam size to estimate the maximum possible extent of the emission region. An SFR estimate for PC 64 is not available as it falls outside the NIRISS field of view.}
    \label{tab:jwsttable}
    \tablewidth{0pt}
    \tablehead{
    \colhead{PC} & \colhead{VLA ID} & \colhead{Redshift}  & \colhead{4.8 GHz Luminosity} 
    &\colhead{Size}  &\colhead{SFR}  
    \\
    \colhead{} & \colhead{\citep{scuba}} & \colhead{} & \colhead{($\rm W/Hz$)} & \colhead{($\rm pc$)} &\colhead{($\rm  M_{\odot}/yr$)}     
    }
    \startdata   
    3 & 124 & $1.05$ &  $(8.90\pm0.68)\times10^{23}$ & $<37$ & $0.02^{+0.13}_{-0.02}$ \\
    7 & 141 & $1.0175$ &  $(1.00\pm0.03)\times10^{24}$ & $<35$  & $1.8^{+2.1}_{-0.8}$ \\
    14 & 182 & $1.02$ &  $(4.52\pm0.58)\times10^{23}$ & $<36$ & $0.3^{+1.1}_{-0.3}$ \\
    24 & 260 & $0.5445$ &  $(6.63\pm0.50)\times10^{22}$ & $<17$ & $0.1^{+1.2}_{-0.09}$ \\
    41 & 382 & $0.3741$ &  $(2.05\pm0.20)\times10^{22}$ & $<20$ & $0.04^{+0.05}_{-0.01}$ \\
    46 & 403 & $1.42$  &  $(3.54\pm0.52)\times10^{23}$ & $<35$ & $55.6^{+31.0}_{-35.6}$ \\
    47 & 434 & $0.95$ &  $(1.12\pm0.19)\times10^{23}$ & $<32$ & $860.0^{+156.4}_{-278.7}$ \\
    64 & 528 & $ 0.3760 $ &  $(8.26\pm0.26)\times10^{22}$ & $<22$ & $--$ \\
    \enddata
\end{deluxetable*}


To understand the nature of infrared counterparts, we cross-matched our VLBA-detected sources with JWST PEARLS observations. \cite{pearlsjwst} carried out JWST/NIRCam observations covering a 16 arcmin$^2$ region of the TDF and detected 4.4 $\mu$m counterparts for 62 of the 63 VLA radio sources, of which only one overlaps with our VLBA-detected sample. Willner et al. (in prep.) has examined new NIRCam data of the JWST counterparts for the radio sources at the TDF. In these, we have identified a total of eight matches with our VLBA-detected sources, while the remaining four sources were outside the NIRCam coverage (Figure \ref{fig:vlajwst}). Each detected source is located within a galactic nucleus, with some host galaxies showing late-type features. All sources exhibit a compact NIR nucleus, though only some show clear point-source signatures. PC 3 (VLA ID 124) is elongated with a bright nucleus, potentially with a spiral disk where SF could contribute to the radio flux. PC 7 (VLA ID 141) has an elongated disk, possibly with ongoing SF. PC 14 (VLA ID 182) is a bright elliptical with a compact red nucleus but no clear point-source signature and is notable for lensing a background galaxy into an Einstein ring (Adams et al., in prep). PC 24 (VLA ID 260) is another bright elliptical but with a distinct point-source nucleus. PC 41 (VLA ID 382) is a radio source with a large elliptical NIRCam counterpart with a bright but extended nucleus. PC 46 (VLA ID 403) appears as a red disk, potentially spiral, with a compact red nucleus. PC 47 (VLA ID 434) is a very red source with a prominent dust lane, resembling a disk with a bright nucleus at longer wavelengths. PC 64 (VLA ID 528) is a large, smooth face-on disk with a bright point-source nucleus.

\cite{ortiz} identified a sample of 66 galaxies in JWST/NIRCam, with point-source features in their cores. Thirteen of these have VLBA observations, and three were detected (PC 3, 7, and 64). Among the detections, PC 64 (VLA ID 528) was classified by \cite{ortiz} as a point-source galaxy core, indicating that its infrared emission probably originates from the AGN. In contrast, PC 3 (VLA ID 124) and PC 7 (VLA ID 141) were classified as compact stellar bulges. For all three sources, the median fractional AGN contribution to their 0.1–30 $\mu$m flux is approximately 0.21--0.25. Interestingly, among the ten VLBA non-detections, nine were classified as bulges, while four (PC 21, 33, 57, and 95) appeared as point sources, suggesting that the infrared emission in these sources is likely dominated by a central AGN. Hence, their non-detection at VLBA scales is intriguing.\\

For the VLBA non-detections, 49 of 94 have been detected with NIRCam (Figure \ref{fig:jwstnd}). The VLBA-undetected sources with S(3 GHz) $>$ 100 $\mu$Jy show no major differences from the detected sources except for PC 70 (VLA ID 593). For the remaining sources, it is unclear whether they were too resolved for the VLBA to detect or simply fell below the VLBA's sensitivity limit. PC 6 (VLA ID 134) is a face-on, reddened disk with a point-source nucleus, except for one spiral-shaped region; given its S(3 GHz) = 340 $\mu$Jy, its non-detection is surprising. PC 10 (VLA ID 150) appears to be part of a complex major merger with patchy dust and a point-like nucleus. PC 33 (VLA ID 314) is dominated by a point-source nucleus \citep{ortiz}, has faint red irregular patches and is another surprising VLBA non-detection. PC 35 (VLA ID 319) is a nearly face-on spiral with a very bright nucleus. PC 70 (VLA ID 593) is a spiral with multiple bright spots, likely H II regions, with an extended and faint nucleus. The remaining two bright VLA radio sources that were not detected with the VLBA (PC 303 and PC 587) have no NIRCam images yet.\\

\subsection{Redshift and SFR measurements}

We used the NIRISS spectra of JWST for the detected galaxies to obtain a measurement of the associated SFR (see Table \ref{tab:jwsttable}). We note that although PC 64 (VLA ID 528) was covered in JWST NIRCam, it does not have coverage in NIRISS. We constrained the SFR using the NIRISS spectra by simultaneously fitting the photometric data and NIRISS grism spectra using the method outlined by \cite{vin}, the flexible star-formation histories of Dense Basis \citep{iy}, and SED models from the Flexible Stellar Population Synthesis (FSPS) models \citep{con} utilizing the MILeS and BaSeL libraries and Kroupa initial mass function \citep{kr}. 

We also provide the redshifts of the eight detected objects within the NIRCam field of view. The redshifts given to four decimal places are derived from Binospec spectroscopy, whereas those with two decimal places are photometric estimates. The number of decimal digits reflects the associated uncertainty. Further details on how redshifts were measured, including the full set of host galaxy properties based on these new infrared observations, will be presented in Willner et al. (in prep.). The measured redshifts were used to estimate the size of the emission regions and the radio luminosity of the sources (see Table \ref{tab:jwsttable}).


\begin{deluxetable*}{ccccc}
\tablecaption{Comparison of radio and host galaxy properties of high  excitation  radio AGN (HERANG) and low  excitation  radio  AGN (LERAGN) \label{tab:HERAGN}}
\tablewidth{0pt}
\tablehead{
\colhead{Property} & \colhead{PC 64 (VLA ID 528)}& \colhead{LERAGN}& \colhead{HERAGN} & \colhead{Reference}
}
\startdata
$\rm L_{4.8~GHz}$	& $\rm 8.26 \times 10^{23}$ \multirow{2}*{$\rm \frac{W}{Hz}$} & \multirow{2}{*}{$\lesssim \rm 10^{26}$ $\rm \frac{W}{Hz}$} & \multirow{2}*{$\gtrsim \rm 10^{26}$ $\rm \frac{W}{Hz}$}  & \multirow{2}*{\cite{2012Best}}  \\
$\rm L_{1.4~GHz}$ &  $\rm 2.9 \times 10^{23} ~~~~$ & & &  \\
$\rm L_{22~\mu m}$	& $\rm 1.4 \times 10^{44}$ $\rm \frac{erg}{s}$ & $\lesssim \rm 10^{44}$ $\rm \frac{erg}{s}$ & $\gtrsim \rm 10^{44}$ $\rm \frac{erg}{s}$ &  \cite{Gurkan}  \\
\enddata
\tablecomments{$\rm L_{4.8~GHz}$ and $\rm L_{22~\mu m}$ calculated using redshift $z = 0.3760$ (see Table \ref{tab:jwsttable}). $\rm L_{1.4~GHz}$ was extrapolated using its spectral index ($\alpha =-1$).
}
\end{deluxetable*}


\section{Discussion}\label{sec:dis}

We conducted a detailed analysis of 106 radio sources in the North Ecliptic Pole (NEP) field using high-resolution VLBA observations, detecting 12 sources.


\subsection{Origin of the radio emission} \label{sec:ori}

Radio sources are commonly classified as AGN or non-AGN by analyzing their spectral index \citep{pacho,kel}. AGN are typically identified by flat or inverted spectral indices ($\alpha \gtrsim 0.7$), which suggest compact, optically thick synchrotron sources. In contrast, ``steep" indices ($\alpha \sim -$ 1.0) are linked to optically thin emission from radio lobes, often indicating older particle populations. For intermediate indices $\alpha \sim -$0.7, classification becomes ambiguous, as both AGN and starburst galaxies can produce these values through ongoing synchrotron particle injection \citep{pacho}. As illustrated in Figure \ref{fig:predictedVLA+WISE}, compact VLBA detections in our sample typically display flat spectral indices ($\alpha \gtrsim -0.5$), which are characteristic of compact AGN structures. Conversely, VLBA detections with more spatially extended, kpc-scale emission exhibits steeper spectral indices ($\alpha \lesssim -0.5$).\\

The sharp rise in the VLBA/VLA flux density ratio around the spectral index ($\alpha \gtrsim -0.5$, Figure \ref{fig:NEP_RatiovsSpecindex}) likely reflects two distinct accretion regimes (see Table \ref{tab:mechanisms}). These regimes correspond to different Eddington ratios, representing distinct modes of accretion onto SMBHs. Assuming this relationship holds, the origin of radio emission in these two cases could differ significantly. Specifically, AGNs with flat-spectrum indices tend to exhibit compact, pc-scale radio emission (indicated by a high VLBA/VLA flux density ratio), whereas those with steeper spectra are associated with larger-scale emission. 

\begin{figure}[t]
     \epsscale{1.1}
     \plotone{"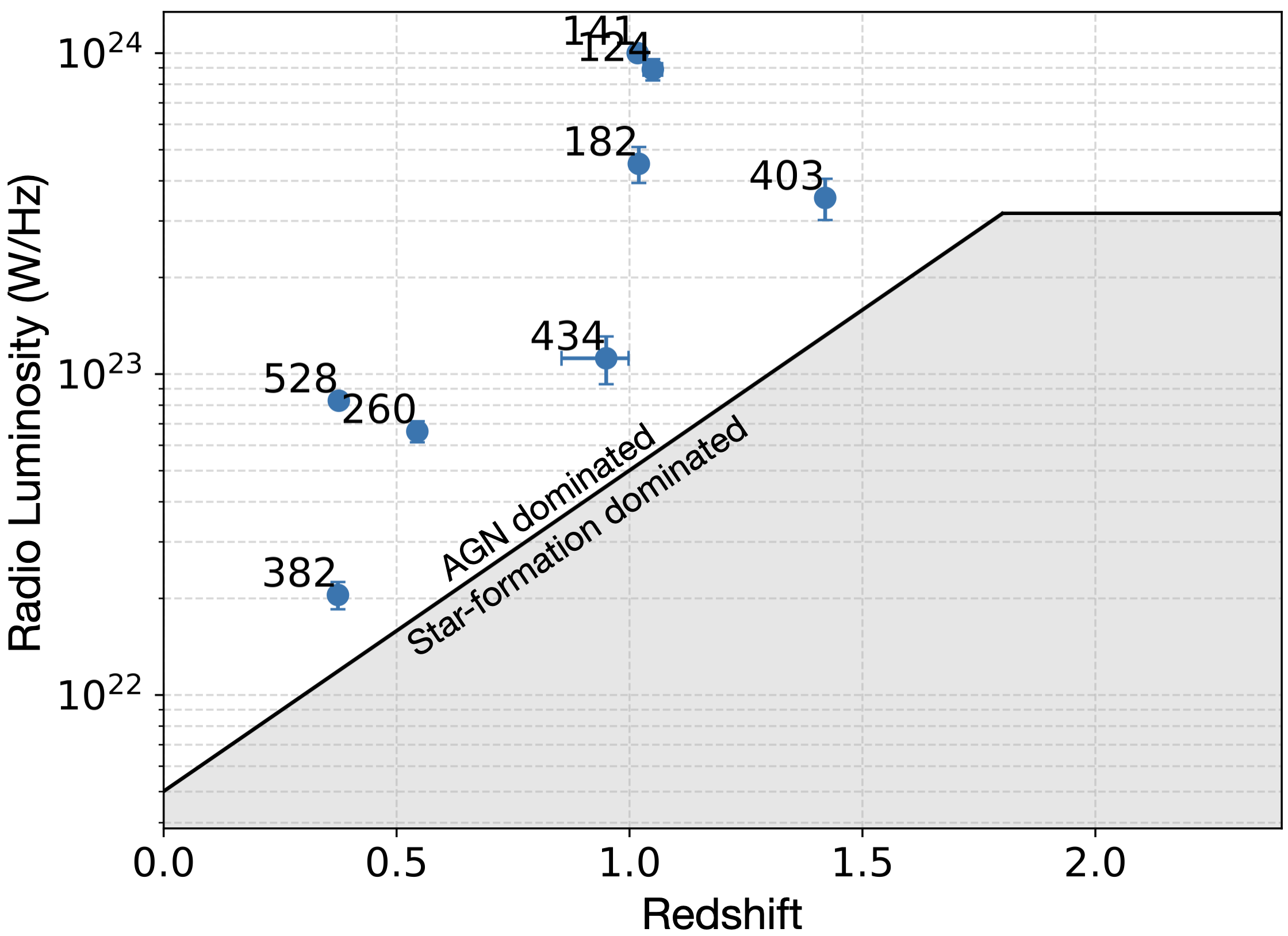"}
     \caption{4.8 GHz radio luminosity (in logarithmic scale) vs redshift for the VLBI detections. The detected sources are labeled using their VLA IDs. The solid black line represents the threshold, $P_{cross}$, which separates AGN-dominated and star-formation-dominated radio emission, based on the selection criteria of \cite{maglio}. The grey-shaded region marks the regime in which SF is expected to be the dominant contributor to radio emission, while the area above the gray-shaded region corresponds to the AGN-dominated regime.}
     \label{fig:radioz}
\end{figure}

Figure \ref{fig:radioz} shows the distribution of the VLBI-detected sources in the redshift–radio luminosity plane. We use the selection criteria of \cite{maglio} to separate AGN-dominated and SF-dominated radio emission, which defines $P_{cross}$ as the luminosity above which AGN-driven radio emission overtakes that from SF in a radio-selected population. At $z<1.8, P_{cross}$ is determined using the radio luminosity functions of \cite{mcap}. Beyond $z>1.8$, the radio luminosity function of SF galaxies declines rapidly, and $P_{cross}$ is fixed at $10^{23.5} \rm W/Hz$, ensuring that contamination from star-forming galaxies remains below 10\% \citep{maglio}. As evident in Figure \ref{fig:radioz}, all our VLBI sources lie above this threshold, confirming that our observations primarily probe AGN-related radio emission.


The high brightness temperatures observed in our detections also imply that the compact radio emission originates from AGNs (see Table \ref{tab:temperatures}), which are typically associated with relatively low SFR. WISE infrared counterparts for the VLBA detections were primarily either AGN or intermediate-disk dominated, as evidenced by mid-infrared colors typical of AGN, while none of the VLBA detections are classified as purely star-forming. Many non-detected sources, however, exhibited WISE classifications pointing to active SF, indicating that our VLBA detections probe AGN driven radio emission.

\subsection{Star formation rate} \label{sec:sfr}

VLBA-detected sources that have a JWST counterpart have SFR significantly lower than non-detected sources with JWST counterparts. This is consistent with AGN-dominated emission. Among the seven VLBA-detected sources with available JWST spectra, five exhibit very low SF activity, with SFR values below 1 $M_{\odot}$ yr$^{-1}$. The only two sources with significantly higher SFRs are PC 46 (VLA ID 403) and PC 47 (VLA ID 434), which may be AGN–starburst composites where AGN activity coexists with substantial SF, similar to NGC 1068 and NGC 4945 \citep[e.g.,][]{agn1,agn2} or a transition phase from SF-dominated to AGN-dominated. Both sources display complex morphologies in their JWST counterparts (Figure \ref{fig:vlajwst}), suggestive of mergers, further supporting the idea that their radio emission arises from a combination of starburst activity and a compact AGN. For PC 46 (VLA ID 403), the low VLBA/VLA flux ratio ($\sim$0.2) suggests that the starburst component is probably resolved out by VLBA. However, an intriguing contrast is seen in PC 47 (VLA ID 434), where the high VLBA/VLA ratio ($\sim$0.6) indicates that the majority of its radio emission originates from an AGN. Hence, the unusually high SFR in this source is particularly interesting. 

For the three VLBA-detected sources with SCUBA counterparts, the SFR values are surprisingly high. We caution that more precise submillimeter positions are necessary to confidently identify the true VLA counterparts within the SCUBA positional uncertainties; moreover, even if the associations are correct, the elevated SFR estimates for SCUBA-detected sources may still be affected by methodological uncertainties. SCUBA’s SFR and redshift estimates rely on multi-wavelength spectral energy distribution (SED) fitting, which integrates data from optical, NIR, and MIR photometry, de-boosted 850 $\mu$m measurements, and 3 GHz VLA data for submillimeter galaxies \citep{scuba}. Including the radio flux in these fits could lead to overestimated SFRs, particularly in sources where the radio emission includes significant AGN contributions, such as those with VLBA detections.

In fact, for PC 26 (VLA ID 279) and PC 71 (VLA ID 614), we find that the VLBA/VLA flux ratio is approximately 0.7--0.8, suggesting that about 70--80\% of the radio flux is AGN-driven rather than due to SF. Consequently, any SFR estimate based on VLA flux density would lead to an overestimated value. The only source in our sample where the SFR derived from SCUBA-2 is likely reliable is PC 46 (VLA ID 403), which has a low VLBA/VLA ratio of $\sim$0.2, implying that most of the VLA flux arises from SF activity. Notably, the JWST spectrum of PC 46 (VLA ID 403) also confirms a higher SFR compared to other sources in our sample.

In contrast, JWST derives host-galaxy properties through SED fitting using photometry from NIRCam, HSC, and MMIRS, which may yield more accurate SFR values by isolating AGN contributions \citep{pearlsjwst}.

\subsection{Host galaxy morphology}

JWST data show that the eight VLBA-detected sources all show compact nuclei in the NIRCam images, and half show noticeable point-source signatures, i.e., resembling the diffraction PSF. However, while diffraction spikes are a reliable indicator of relatively bright point sources, we note that some of the other sources may still host a central point source that is simply too faint to produce visible diffraction spikes. Some of these counterparts have an elliptical morphology and no evidence of dust (although dust can not be ruled out in the nucleus), while others show clear dust reddening. Five of the eight VLBA sources with identified JWST counterparts are situated in the centers of late-type galaxies, two within the centers of spirals with substantial bulge components, and one within the bulge of a merging system (Figure \ref{fig:vlajwst}). This positional trend indicates that the VLBA-detected sources are predominantly located in the centers of spheroidal structures or bulge-dominated components, likely at lower redshifts than higher-redshift star-forming galaxies. Many of these higher-redshift galaxies, still in an active phase of SF, may not yet harbor detectable weak AGNs, explaining their typical non-detection by the VLBA. The complex merger PC 46 (VLA ID 403) appears to present a mix of radio emission sources, likely consisting of a brighter, steeper-spectrum distorted synchrotron disk alongside a faint, flat-spectrum AGN, which we are able to detect with the VLBA observations.

\subsection{VLBA detected sources}

For all the 12 VLBA detections, we attribute the parsec-scale radio emission to a LERAGN at the center of the respective host galaxies. The VLA and VLBA flux densities, along with radio spectral indices, are listed in Table \ref{tab:detections}; brightness temperatures in Table \ref{tab:temperatures}; and redshifts, radio luminosities, and physical sizes in Table \ref{tab:jwsttable}. The properties of the individual sources detected in our VLBA observations are discussed in detail below. \\

\textbf{PC 3:} PC 3 (VLA ID 124) exhibits a possible jet structure in the VLA image (Figure \ref{fig:vlajwst}), with VLBA detection concentrated in its core (within an area of 37 pc), suggesting that AGN activity could be driving the observed radio emission. In our multi-wavelength counterpart search, we identified this source in both the WISE catalog and JWST imaging. \citet{ortiz} classified it as a compact stellar bulge with only a 0.24\% AGN contribution to its bolometric infrared luminosity. The JWST image reveals an elongated galaxy with a bright central point source. If the elongation represents a spiral disk, SF could contribute to radio flux, but JWST spectra indicate a very low SFR.\\

\textbf{PC 7:} PC 7 (VLA ID 141) stands out with the highest VLA integrated and peak flux density values, among the 12 sources detected by VLBA. The VLBA to VLA flux density ratio is one of the lowest in our sample. It displays a prominent two-sided jet structure in its VLA image with radio-bright nucleus. The VLBA detection is centered precisely in the galaxy (Figure \ref{fig:vlajwst}). It has counterparts in the WISE and JWST observations. The NIRCam source is a
bright point-source nucleus with an elongated disk, which could have SF. \cite{ortiz} identify this source as a compact stellar bulge, with AGN emission accounting for only 0.21\% of its total bolometric infrared luminosity.\\

\textbf{PC 14:} PC 14 (VLA ID 182) has the flattest radio spectrum of all 12 detections ($\alpha \sim 0$), indicating compact, optically thick emission typical of AGN cores. The VLBA/VLA flux ratio is almost $\sim$1, suggesting that almost all radio emission comes from a compact AGN. The VLBA emission comes from a central region smaller than 36 pc. The source has counterparts in both WISE and JWST observations. The JWST NIRCam observation shows that it is a bright elliptical galaxy with a compact red nucleus but no obvious point-source signature. The JWST spectra indicate a very low SFR. The source is additionally interesting because it is imaging a background galaxy into an Einstein ring (Adams et al. in prep).\\

\textbf{PC 24:} PC 24 (VLA ID 260) has identified counterparts in the WISE, SDSS, and JWST catalogs. Detected in the first epoch of the JWST data release, it has a spectral energy distribution (SED) consistent with that of a typical radio-loud galaxy \citep{pearlsjwst}. The JWST image reveals a bright elliptical with point-source signature in the nucleus. It has a significantly low SFR compared to the SFR range observed across all TDF field images {pearlsjwst}. The source has an identifiable optical counterpart in SDSS 0\farcs04 away from its VLBA position, with an unusually red color, suggesting significant dust obscuration. The mid-infrared counterpart falls in the ``Intermediate disk" region of the WISE color-color diagram (see Figure \ref{fig:WISE colors W1-W2 vs W2-W3}). \\

\textbf{PC 25:} PC 25 (VLA ID 262) exhibits no detectable counterparts in SDSS, WISE and SCUBA. The source is yet to be observed by the JWST. The nearest optical source found in the SDSS is approximately 10\arcsec away, a substantial angular distance that weakens the likelihood of a physical association between the two. The absence of detections across these various instruments and wavelengths—ranging from optical and infrared to submillimeter—suggests that this object may lie at an exceptionally high redshift, be heavily obscured and faint, or possess an inherently unusual emission profile.\\

\textbf{PC 26:} PC 26 (VLA ID 279) has not been observed with JWST, but has counterparts detected by WISE and SCUBA. The SFR measured from the submillimeter emission with SCUBA-2 suggests high SF activity. The stellar population associated with this source is estimated to have an age of around 80 Myr, indicative of a relatively young burst of SF within the host galaxy \citep{scuba}. \\

\textbf{PC 41:} PC 41 (VLA ID 382) is one of the three fainter VLBA detections. It displays a subtle polar jet-like structure in observations conducted with the VLA (Figure \ref{fig:vlajwst}). The source has counterparts in WISE, SDSS, and JWST observations. The optical counterpart is closely aligned with the VLBA position (offset by 0\farcs07). The low redshift of the source ($z \sim 0.3741$) and absolute $r$-band magnitude ($\rm M_{r}=-22.33$) suggest it is luminous enough to be classified as a luminous AGN, consistent with the observed jet structure. JWST/NIRCam counterpart shows a large elliptical with a very bright nucleus but without any point-source appearance. The JWST spectra reveal a very low SFR.\\

\textbf{PC 46:}  PC 46 (VLA ID 403) shows a faint VLBA detection, with one of the lowest VLBA/VLA flux density ratios and among the steepest radio spectra in the sample. It has several multi-wavelength counterparts, including SCUBA, WISE and JWST. In JWST images, the source appears red in the F200W–F444W bands (Figure \ref{fig:vlajwst}), indicating potential dust obscuration, an older stellar population in the host galaxy, or quenching from merger. It shows a red disk, perhaps with spiral arms, and a red compact nucleus. Looking at the JWST image, the host could be a complex merger, and the radio observations are possibly a mix of radio emission: a brighter steeper spectrum distorted synchrotron disk and a faint flat spectrum AGN that was detected with the VLBA. It has a very high SFR of $\sim$55 M$_{\odot}$ yr$^{-1}$.\\

\textbf{PC 47:} PC 47 (VLA ID 434) is the faintest of the 12 VLBA detections. It has counterparts in WISE and JWST. The WISE colors of this source indicate that its infrared emission primarily originates from SF, a conclusion further supported by JWST, which confirms a high SFR. The JWST images show that it is a very red source with a prominent dust lane across it. At long wavelengths, it looks like a disk, perhaps with spiral arms, and a bright nucleus. The red color index is similar to PC 46, possibly suggesting a mix of star-forming and older stellar populations within the galaxy (Figure \ref{fig:vlajwst}). The high VLBA/VLA flux ratio and relatively flat radio spectrum suggest that radio emission is predominantly AGN-driven. However, the SFR, as measured from the JWST spectra, is the highest in our sample. The lack of kpc-scale radio emission despite significant SF remains an open question and warrants further investigation.\\

\textbf{PC 64:} PC 64 (VLA ID 528) is the only resolved source in our sample of 12 detections (see Table \ref{tab:SRC64 Properties}). It has counterparts in SDSS, WISE and JWST. The SDSS data of the optical counterpart (SDSS J172340.67+654952.7), separated only by 0\farcs07, place the galaxy in the red sequence ($u=22.36,~g=22.88,~r=21.05,~i=20.57,~z=20.12$), with absolute r magnitude $\rm M_{r}=-20.21$. Based on the WISE classification in \cite{Gurkan}, it could be a high-excitation radio galaxy (HERG) with a powerful AGN, or a low-excitation radio galaxy (LERG), as it falls right in between the two distinct areas ($\rm L_{22~\mu m}$ $=\rm 1.2 \times 10^{44} \rm ergs/s$). However, as can be seen in Table \ref{tab:HERAGN}, the radio luminosity of the source clearly falls below $10^{26}$ W Hz$^{-1}$, placing it in the jet-mode accretion-driven Low Excitation Radio AGN (LERAGN) classification of \cite{2012Best}. Given that LERGs release their accretion energy in the form of jets or winds \citep{Gurkan}, this might support the component distinction mentioned above of a small, elongated jet and its interaction with the surroundings. The JWST counterpart with NIRCam shows a large, smooth face-on disk with bright point-source nucleus. \cite{ortiz} classify this source as having a central red point-source component, indicating significant infrared emission from hot dust surrounding the AGN.\\

\textbf{PC 67:} PC 67 (VLA ID 554) is the brightest of all VLBA detections and is the only one with an inverted radio spectrum ($\alpha \sim 0.5$, Table \ref{tab:detections}). JWST has not observed the field, but it has a WISE counterpart. It is the only source in our sample to be classified in the ``Spheroids" region of the WISE color-color diagram (Figure \ref{fig:WISE colors W1-W2 vs W2-W3}), indicating a galaxy morphology typically associated with older, elliptical, or spheroidal galaxies. This classification suggests that the host galaxy may have low levels of SF with radio emissions likely dominated by AGN activity. \\

\textbf{PC 71:} PC 71 (VLA ID 614) has counterparts in WISE and SCUBA, but it has not been observed with JWST. This source exhibits the highest W1–W2 color in the WISE diagram, along with a high enough W2$–$W3 color that places it within the AGN region (Figure \ref{fig:WISE colors W1-W2 vs W2-W3}) as classified by \cite{jarrett}, although a dusty SF could also contribute to high W1$–$W2 color. The SCUBA counterpart suggests significant SF activity with a high SFR \citep{scuba}.

\section{Conclusion}\label{sec:con}

The VLBA observations of the JWST North Ecliptic Pole Time-Domain Field make up one of the most sensitive VLBI deep surveys conducted to date. The observing frequency is also higher than most, which reduces the impact of synchrotron self-absorption on the detected sources.

The survey detected 12 of 106 known radio sources, with a detection rate of $\sim$11$\%$, and achieved a mas resolution suitable for pinpointing compact AGN emissions primarily concentrated within the central parsec-scale regions. For sources with S(3 GHz) brighter than 50 $\mu$Jy, the detection rate increases to $\sim$35\%. By design, this study probes AGN that are intrinsically radio-faint. This could correspond to either luminous, distant sources or faint, nearby ones. The radio luminosities of sources with known redshifts are $0.2-16.7 \times10^{23}$ W Hz$^{-1}$, consistent with the characteristics of low-excitation radio AGN. For most of these galaxies, the radio emission is unresolved and confined to the central $<$40 pc, suggesting an origin closely linked to the AGN. This compact radio emission likely traces the base of the jet, that is, the region where the jet is launched, which may coincide spatially with the accretion-disk corona. The radio emission is likely either optically thin synchrotron radiation from an outflow originating near the center or synchrotron self-absorbed emission from an optically thick region.

Our detections have high brightness temperatures T$_B \gtrsim 10^{5}$ K, suggesting non-thermal, AGN-driven emission, with 2 GHz spectral indices ($\alpha \gtrsim -$0.5) consistent with SMBH origins. Non-detections showed steeper indices, indicative of star-forming regions. Comparatively, VLBA detections had higher VLBA/VLA flux density ratios, reinforcing the likelihood of compact, AGN-dominated sources, whereas non-detected sources implying more extended radio emissions. WISE infrared data also showed that most VLBA detections were either AGN or intermediate-disk dominated, while non-detections often indicated active SF. A more careful comparison with sub-millimeter sources will be highly beneficial, once higher-precision sub-millimeter positions become available.

Compared to earlier deep VLBI surveys at 1.4--1.6 GHz (see Table \ref{tab:surveys}), our 4.8 GHz observations probe fainter sources with higher angular resolution and, due to its higher frequency, less sensitive to synchrotron self-absorption. Despite the differences in frequency and depth, our detection fraction and spectral index trends remain broadly consistent with previous findings, confirming that compact, high-brightness temperature AGN cores are a persistent feature of faint radio-quiet AGN populations. Although radio emission in radio-quiet AGN was traditionally thought to be primarily driven by star formation, our higher-frequency, higher-sensitivity VLBA observations reveal that in most of our sources, star formation contributes less than 50\% of the total VLBA-scale radio emission, with several cases where the emission is clearly AGN-dominated.

Eight of the twelve VLBA sources observed with JWST/NIRCam revealed infrared counterparts predominantly located in the nuclei of early-type, bulge-dominated galaxies. This association suggests that these AGNs are often embedded within dense galactic centers, contrasting with higher redshift, star-forming galaxies, where weak AGNs remain undetected. The JWST-based SFRs for VLBA-detected sources were lower than the estimates derived from SCUBA-2, underscoring the role of JWST in differentiating AGN activity from extended SF processes. This combination of VLBA, VLA, JWST, WISE, SCUBA-2 and SDSS data highlights the intricate relationship between AGN and SF in the TDF, showing how AGN impact their host galaxies in different ways across various wavelengths.

For future work, to better disentangle SF and AGN-related radio emission, higher-resolution radio observations of the TDF are needed. Spatially resolving the compact nuclear emission will allow us to study the morphology of the radio sources and distinguish AGN cores from extended star-forming regions. Additionally, multi-frequency radio observations will enable spectral index measurements, helping to differentiate between thermal and non-thermal emission components and providing insights into the underlying physical processes driving the radio emission. Finally, further VLBA observations of the field will enhance sensitivity and enable the study of source variability.\\


\noindent \textbf{ACKNOWLEDGEMENT}\\

\noindent We thank the anonymous referee for their constructive comments, which improved the quality of this manuscript. RAW acknowledges support from NASA JWST Interdisciplinary Scientist grants NAG5-12460, NNX14AN10G and 80NSSC18K0200 from GSFC. SNM and HBH acknowledge support from NASA JWST 600 Interdisciplinary Scientist grant 21-SMDSS21-0013. This work is based in part on observations made with the NASA/ESA/CSA James Webb Space Telescope. Some of the data presented in this article were obtained from the Mikulski Archive for Space Telescopes (MAST) at the Space Telescope Science Institute, which is operated by the Association of Universities for Research in Astronomy, Inc., under NASA contract NAS 5-03127 for JWST. The specific observations analyzed can be accessed via \dataset[doi:10.17909/4kwv-d132]{https://doi.org/10.17909/4kwv-d132}. This publication also uses data products from the Wide-field Infrared Survey Explorer, which is a joint project of the University of California, Los Angeles, and the Jet Propulsion Laboratory/California Institute of Technology, funded by the National Aeronautics and Space Administration. The National Radio Astronomy Observatory is a facility of the National Science Foundation operated under a cooperative agreement by Associated Universities, Inc. The Legacy Surveys consist of three individual and complementary projects: the Dark Energy Camera Legacy Survey (DECaLS; Proposal ID 2014B-0404; PIs: David Schlegel and Arjun Dey), the Beijing-Arizona Sky Survey (BASS; NOAO Prop. ID 2015A-0801; PIs: Zhou Xu and Xiaohui Fan), and the Mayall z-band Legacy Survey (MzLS; Prop. ID 2016A-0453; PI: Arjun Dey). Pipeline processing and analyses of the data were supported by NOIRLab and the Lawrence Berkeley National Laboratory (LBNL). The Legacy Surveys project is honored to be permitted to conduct astronomical research on Iolkam Du’ag (Kitt Peak), a mountain with particular significance to the Tohono O’odham Nation. NOIRLab is operated by the Association of Universities for Research in Astronomy (AURA) under a cooperative agreement with the National Science Foundation. LBNL is managed by the Regents of the University of California under contract to the U.S. Department of Energy.

\clearpage
\restartappendixnumbering
\appendix

\section{Appendix information}

\startlongtable
\begin{deluxetable*}{cc|ccc|cc}
\tablecaption{VLBA Non-detections \label{tab:non detections}}
\tablewidth{0pt}
\tablehead{ 
\multicolumn{2}{c|}{Source}    & \multicolumn{3}{c|}{VLBA} & \multicolumn{2}{c}{VLA} \\
\hline
\colhead{PC} & \multicolumn{1}{c|}{VLA ID} & \colhead{$S_{\rm peak}$\tablenotemark{a}} & \colhead{rms} & \multicolumn{1}{c|}{PB Gain} & \colhead{3 GHz flux} & \colhead{Spectral Index } \\
\colhead{} &  \multicolumn{1}{c|}{} & \colhead{($\mu \rm{Jy/beam}$)}  & \colhead{($\mu \rm{Jy/beam}$)} & \multicolumn{1}{c|}{} & \colhead{($\mu \rm{Jy}$)} & \colhead{($\alpha$)}\\
\colhead{(1)} & \multicolumn{1}{c|}{2} & \colhead{(3)} & \colhead{(4)} & \multicolumn{1}{c|}{5} & \colhead{(6)} & \colhead{(7)}
}
\startdata
1	& 113 & 48.47 & 3.5 & 0.361 &31.3$\pm$1.7 &0.39$\pm$0.49\\
2	& 120 & 47.14 & 3.5 & 0.371 &14.1$\pm$1.5 &-1.01$\pm$0.26\\
4	& 128 & 51.35 & 3.5 & 0.341 &38.2$\pm$2.9 &-0.78$\pm$0.42\\
5	& 130 & 56.93 & 3.5 & 0.307 &37.5$\pm$1.9 &-1.00$\pm$0.37\\
6	& 134 & 46.53 & 3.5 & 0.376 &340.0$\pm$10.3 &-1.15$\pm$0.20\\
8	& 142 & 45.40 & 3.5 & 0.385 &36.2$\pm$1.8 &-1.12$\pm$0.39\\
9	& 149 & 55.93 & 3.5 & 0.313 &35.5$\pm$2.6 &-1.32$\pm$0.43\\
10	& 150 & 51.63 & 3.5 & 0.339 &112.0$\pm$3.7 &-0.98$\pm$0.23\\
11	& 155 & 39.02 & 3.4 & 0.436 &39.6$\pm$1.8 &-1.40$\pm$0.35\\
12	& 173 & 40.89 & 3.4 & 0.416 &52.8$\pm$3.0 &-1.05$\pm$0.39\\
13	& 181 & 54.57 & 3.6 & 0.330 &33.1$\pm$2.6 &-0.68$\pm$0.45\\
15	& 197 & 38.75 & 3.4 & 0.439 &32.5$\pm$1.7 &-0.68$\pm$0.43\\
16	& 201 & 26.04 & 3.3 & 0.634 &50.4$\pm$3.9 & na\\
17	& 207 & 51.81 & 3.5 & 0.338 &64.7$\pm$3.2 &-1.29$\pm$0.29\\
18	& 218 & 35.78 & 3.4 & 0.475 &29.9$\pm$1.6 &-0.93$\pm$0.44\\
19	& 222 & 40.40 & 3.4 & 0.421 &40.8$\pm$2.5 &-0.71$\pm$0.39\\
20	& 230 & 22.61 & 3.2 & 0.708 &23.0$\pm$1.3 & na\\
21	& 232 & 21.48 & 3.2 & 0.746 &50.3$\pm$1.9 &-0.61$\pm$0.32\\
22	& 246 & 24.93 & 3.3 & 0.656 &69.0$\pm$3.0 &-0.93$\pm$0.34\\
23	& 256 & 27.98 & 3.3 & 0.590 &20.2$\pm$1.5 & na\\
27	& 282 & 69.70 & 3.6 & 0.255 &94.0$\pm$3.2 &0.11$\pm$0.26\\
28	& 283 & 17.13 & 3.1 & 0.890 &55.3$\pm$2.7 &-0.67$\pm$0.38\\
29	& 289 & 18.31 & 3.1 & 0.846 &13.5$\pm$1.1 & na  \\
30	& 303 & 43.85 & 3.4 & 0.391 &196.0$\pm$7.0 & -1.27$\pm$0.22\\
31	& 305 & 16.99 & 3.0 & 0.882 &46.7$\pm$2.3 &  -0.99$\pm$0.35 \\
32	& 306 & 18.05 & 3.1 & 0.858 &17.0$\pm$2.2 & na  \\
33	& 314 & 46.40 & 3.5 & 0.376 &324.0$\pm$9.8 &-1.07$\pm$0.20\\
34	& 317 & 30.43 & 3.3 & 0.545 &34.2$\pm$3.3 & na\\
35	& 319 & 15.84 & 3.0 & 0.947 &103.0$\pm$3.8 &-0.80$\pm$0.27\\
36	& 320 & 20.38 & 3.1 & 0.760 &30.8$\pm$1.5 &-0.67$\pm$0.44\\
37	& 337 & 16.32 & 3.0 & 0.919 &33.2$\pm$1.9 &-0.90$\pm$0.43\\
38	& 338 & 16.07 & 3.0 & 0.933 &19.1$\pm$1.2 & na  \\
39	& 368 & 40.44 & 3.5 & 0.427 &30.4$\pm$1.5 &-0.93$\pm$0.45\\
40	& 379 & 52.02 & 3.5 & 0.336 &39.2$\pm$2.0 &0.35$\pm$0.40\\
42	& 390 & 16.60 & 3.0 & 0.903 &32.8$\pm$3.2 & na   \\
43	& 392 & 18.60 & 3.1 & 0.833 &41.1$\pm$1.6 &-0.55$\pm$0.37 \\
45	& 398 & 41.25 & 3.6 & 0.436  &60.7$\pm$3.8 &-0.09$\pm$0.37\\
48	& 441 & 15.48 & 3.0 & 0.969 &47.5$\pm$2.3 &-0.51$\pm$0.36 \\
49	& 459 & 16.00 & 3.0 & 0.937 &40.2$\pm$1.6 &-0.75$\pm$0.36\\
50	& 466 & 17.38 & 3.1 & 0.891 &36.1$\pm$2.0 &-0.68$\pm$0.42\\
51	& 470 & 17.49 & 3.1 & 0.886 &32.7$\pm$1.4 &-1.10$\pm$0.41\\
52	& 478 & 17.37 & 3.1 & 0.892 &21.7$\pm$2.1 & na   \\
53	& 481 & 17.45 & 3.1 & 0.888 &19.4$\pm$2.0 & na   \\
54	& 482 & 17.63 & 3.1 & 0.879 &23.1$\pm$1.3 & na  \\
55	& 493 & 22.92 & 3.2 & 0.698 &27.8$\pm$2.9 & na   \\
56	& 495 & 26.68 & 3.3 & 0.618 &27.0$\pm$2.4 & na   \\
57	& 497 & 19.28 & 3.1 & 0.804 &30.3$\pm$2.3 & na   \\
58	& 499 & 21.45 & 3.2 & 0.746 &24.5$\pm$2.2 & na   \\
59	& 501 & 19.25 & 3.1 & 0.805 &46.6$\pm$2.4 &-0.60$\pm$0.38 \\
60	& 502 & 21.37 & 3.2 & 0.749 &41.8$\pm$2.4 &-0.68$\pm$0.40\\
61	& 505 & 27.35 & 3.3 & 0.603 &57.0$\pm$2.6 &-0.80$\pm$0.31\\
62	& 506 & 18.52 & 3.1 & 0.837 &36.4$\pm$2.3 &-0.81$\pm$0.47\\
63	& 507 & 44.38 & 3.5 & 0.391 &53.7$\pm$2.1 &-0.93$\pm$0.31\\
65	& 542 & 34.07 & 4.5 & 0.660 &54.4$\pm$2.0 &-0.36$\pm$0.31\\
66	& 547 & 32.19 & 4.5 & 0.699 &43.8$\pm$1.8 &-0.84$\pm$0.34\\
68	& 587 & 95.21 & 5.2 & 0.273 &188.8$\pm$7.6 &-0.49$\pm$0.32\\
69	& 592 & 44.90 & 4.7 & 0.523 &60.3$\pm$2.2 &-0.99$\pm$0.30\\
70	& 593 & 46.06 & 4.7 & 0.510 &127.0$\pm$4.8 &-1.03$\pm$0.24\\
72	& 627 & 58.95 & 4.8 & 0.407 &52.1$\pm$3.0 &-1.04$\pm$0.37\\
73	& 640 & 68.68 & 4.9 & 0.357  &53.4$\pm$4.3 & na\\
74	& 647 & 93.94 & 4.9 & 0.261 &58.3$\pm$2.4 &-0.73$\pm$0.30\\
75	& 351 & 22.16 & 4.1 & 0.925   &21.9$\pm$1.2 & na\\
76	& 374 & 21.51 & 4.1 & 0.953 &24.7$\pm$1.3 &-0.70$\pm$0.51\\
77	& 384 & 20.41 & 4.0 & 0.980 &22.6$\pm$1.8 & na\\
78	& 402 & 20.66 & 4.0 & 0.968 &12.1$\pm$1.1 & na\\
79	& 477 & 53.72 & 4.7 & 0.437 &33.8$\pm$3.8 & na\\
80	& 127 & 49.30 & 4.8 & 0.487 &18.6$\pm$2.4 & na\\
81	& 152 & 45.04 & 4.7 & 0.522 &14.1$\pm$1.4 & na\\
82	& 159 & 39.93 & 4.6 & 0.576 &22.4$\pm$2.5 & na\\
83	& 162 & 39.23 & 4.6 & 0.586 &17.6$\pm$2.1 & na\\
85	& 200 & 40.37 & 4.5 & 0.557 &20.4$\pm$2.6 & na\\
86	& 219 & 40.66 & 4.7 & 0.578 &15.2$\pm$1.6 & na\\
91	& 248 & 40.94 & 4.7 & 0.574 &11.1$\pm$1.4 & na\\
93	& 258 & 29.72 & 4.4 & 0.740 &15.8$\pm$1.2 & na\\
95	& 300 & 22.67 & 4.1 & 0.904 &18.1$\pm$1.3 & na\\
97	& 312 & 31.81 & 4.4 & 0.692 &16.8$\pm$2.8 & na\\
99	& 358 & 34.32 & 4.5 & 0.656 &18.5$\pm$2.7 & na\\
100	& 372 & 19.66 & 3.9 & 0.992 &14.3$\pm$2.3 & na\\
103	& 386 & 40.26 & 4.6 & 0.570 &18.3$\pm$1.3 & na\\
105	& 409 & 19.82 & 3.9 & 0.989 &6.8$\pm$1.3 & na\\
107	& 440 & 38.35 & 4.5 & 0.588 &43.1$\pm$1.7 &-0.40$\pm$0.36\\
108	& 445 & 24.22 & 4.2 & 0.867 &19.9$\pm$1.9 & na\\
109	& 446 & 20.69 & 4.0 & 0.962 &18.6$\pm$1.8 & na\\
110	& 449 & 21.68 & 4.0 & 0.925 &20.4$\pm$1.6 & na\\
111	& 456 & 34.11 & 4.4 & 0.648 &19.4$\pm$1.3 & na\\
112	& 471 & 23.16 & 4.2 & 0.898 &12.4$\pm$1.2 & na\\
116	& 511 & 28.73 & 4.4 & 0.766 &16.9$\pm$1.4 & na\\
117	& 533 & 29.59 & 4.4 & 0.743 &15.4$\pm$3.2 & na\\
118	& 535 & 29.17 & 4.4 & 0.754 &23.5$\pm$2.0 & na\\
120	& 544 & 42.89 & 4.7 & 0.548 &24.4$\pm$2.4 & na\\
123	& 552 & 36.96 & 4.6 & 0.622 &14.4$\pm$2.7 & na\\
124	& 558 & 38.98 & 4.7 & 0.603 &9.3$\pm$1.4 & na\\
125	& 574 & 44.77 & 4.6 & 0.514 &33.4$\pm$4.0 & na\\
126	& 580 & 45.75 & 4.6 & 0.503 &24.5$\pm$1.5 & na\\
\enddata
\tablenotetext{a}{$S_{\rm peak} = \frac{5*rms}{PBC}$}. The typical sensitivity loss from delay and time smearing is $48\%$ at 6\arcsec offset.
\tablecomments{
{\bf(1)}: Phase Center (PC) number, used as reference for VLBA sources in this project;\\
{\bf(2)}: VLA ID from the parent sample \citep{scuba};\\ 
{\bf(3)}: Estimated peak flux density ($S_{\rm peak}$) of the source (4.8 GHz) after primary beam gain correction. The $S_{\rm peak}$ was estimated from the corresponding VLBA image and can be used as an upper limit on the VLBA flux density;\\ 
{\bf(4)}: Local VLBA noise rms measured with natural weighting;\\
{\bf(5)}: Primary beam Gain.\\
{\bf(6)}: VLA 3 GHz flux density ($\mu \rm{Jy}$) from \cite{scuba};\\
{\bf(7)}: Spectral index $\alpha$ of the VLA 3 GHz counterpart.
}
\end{deluxetable*}

\begin{figure*}[ht]
     \epsscale{1.2}     \plotone{"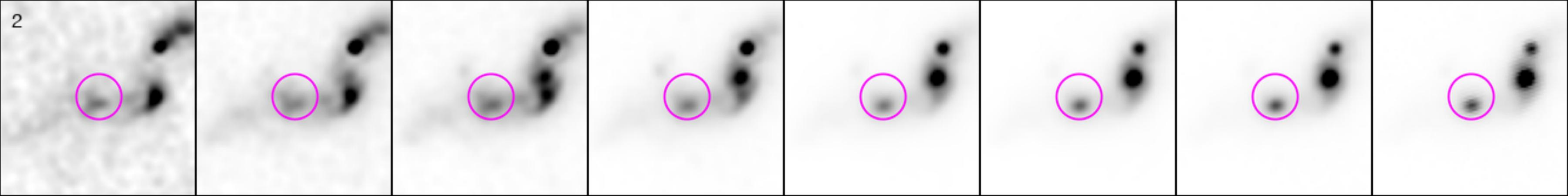"}     \plotone{"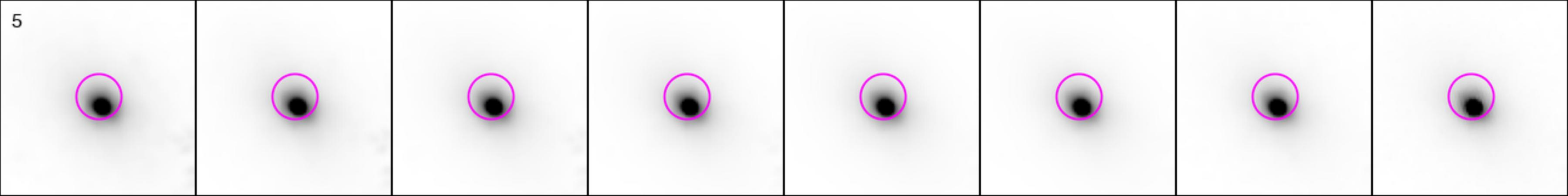"}    \plotone{"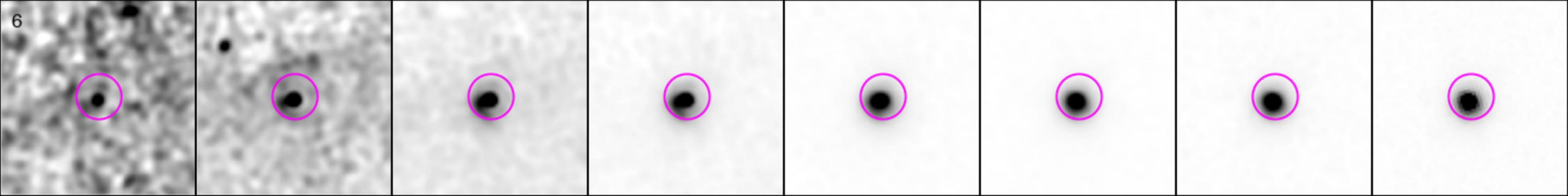"}     \plotone{"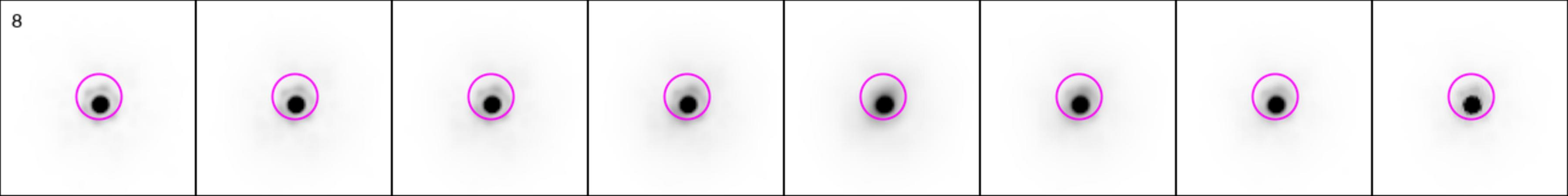"}     \plotone{"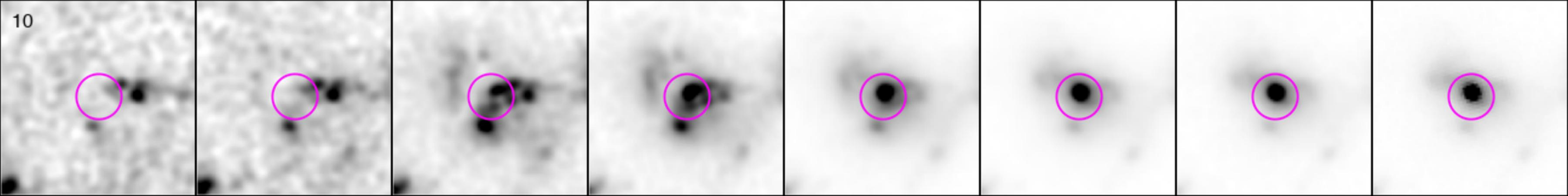"}    \plotone{"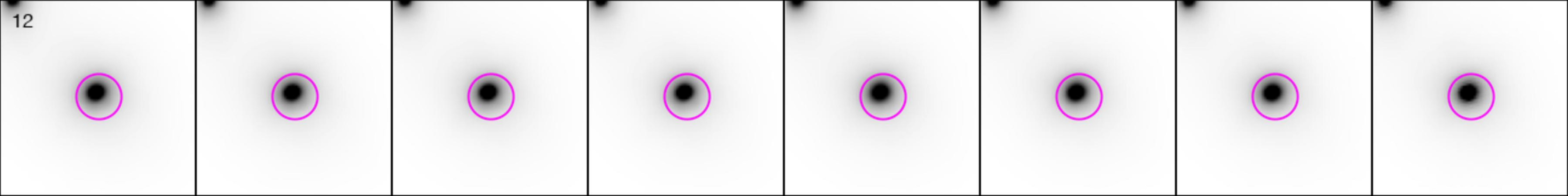"}     \plotone{"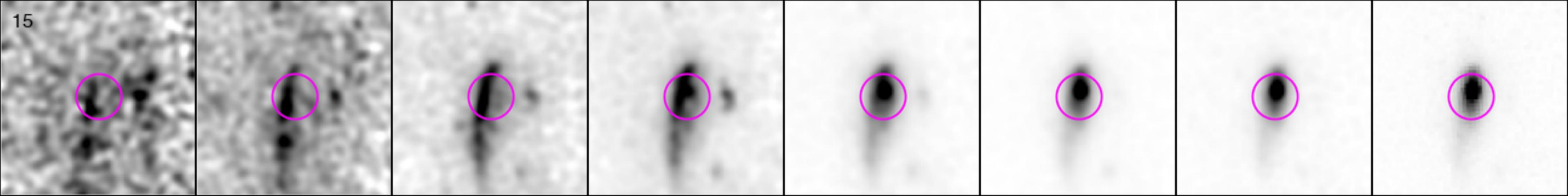"}    \plotone{"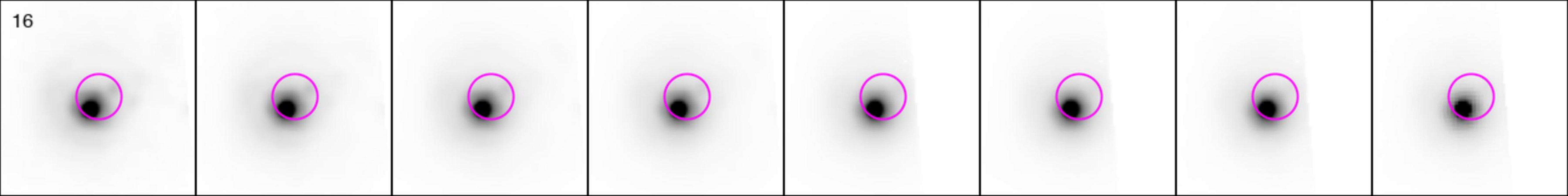"}
     \plotone{"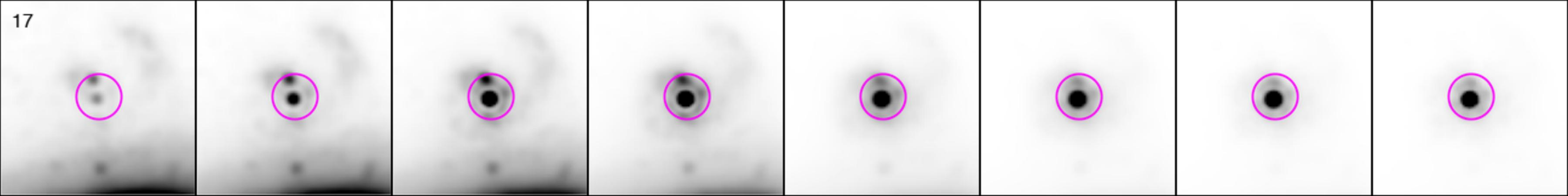"}     \plotone{"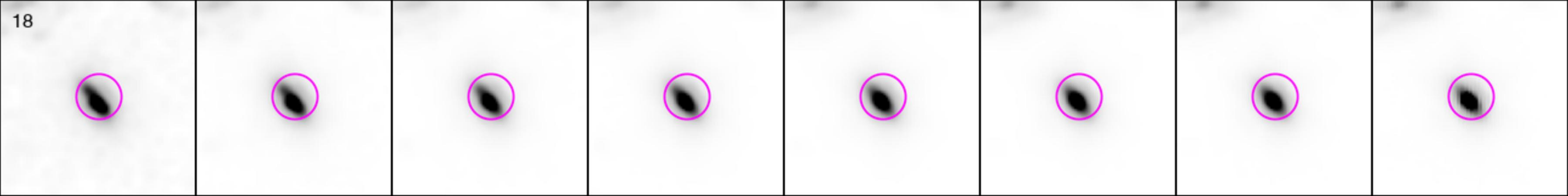"}  

\end{figure*}
\begin{figure*}[ht]
     \epsscale{1.2}    \plotone{"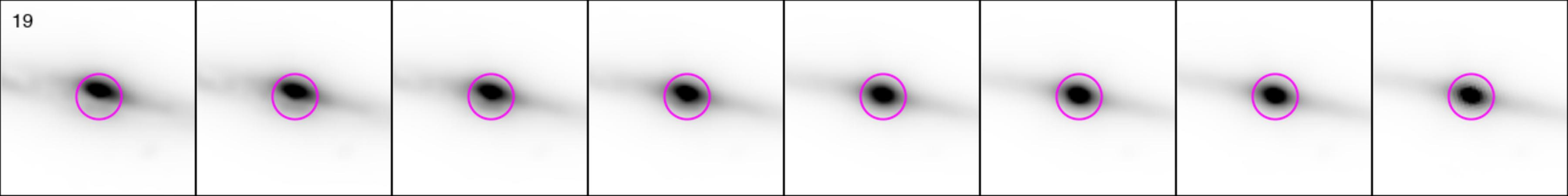"}     \plotone{"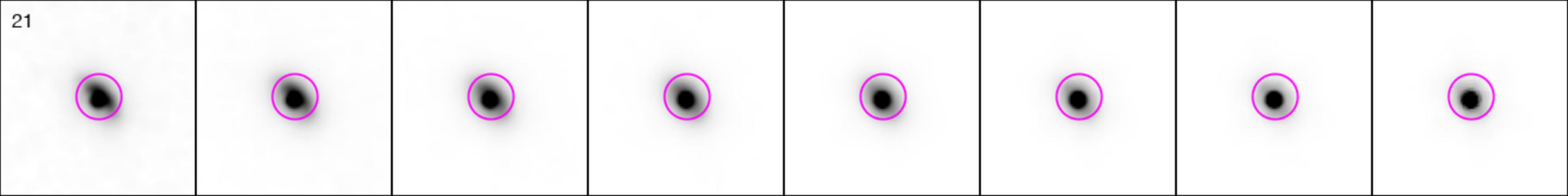"}     \plotone{"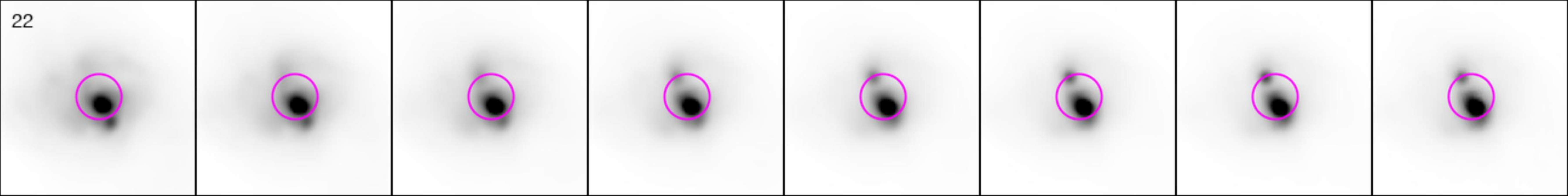"}    \plotone{"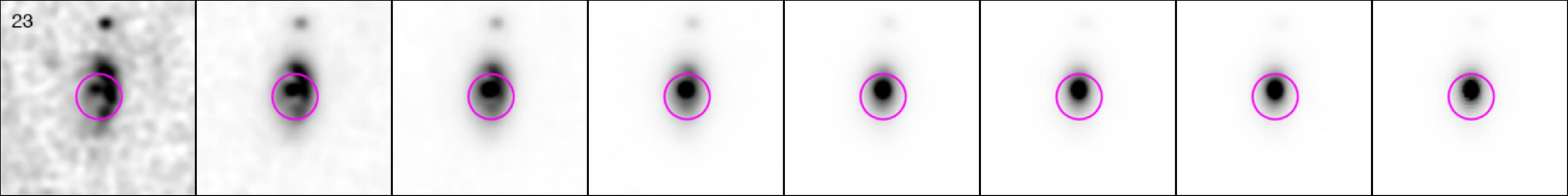"}     \plotone{"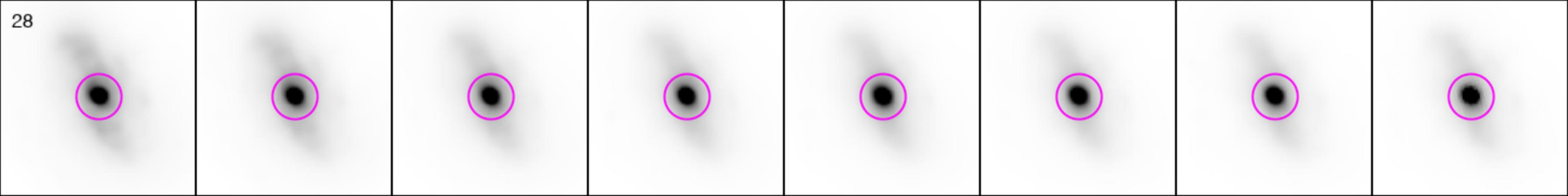"}    \plotone{"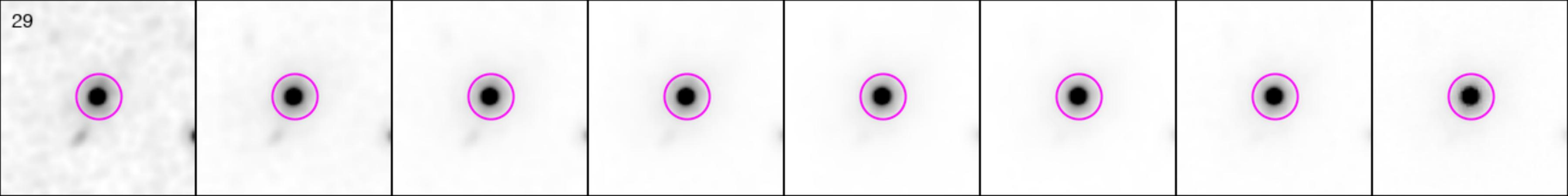"}
     \plotone{"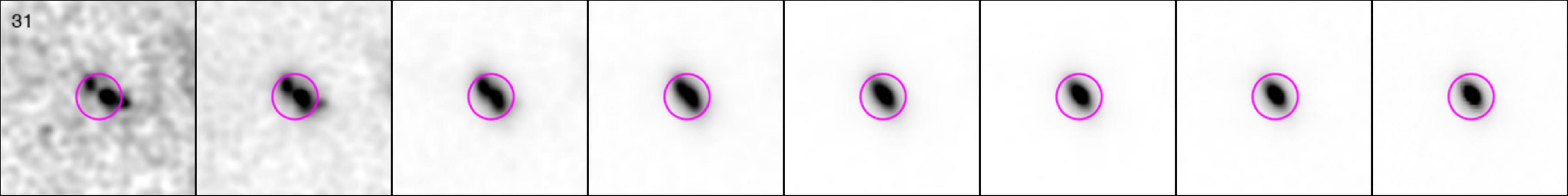"}     \plotone{"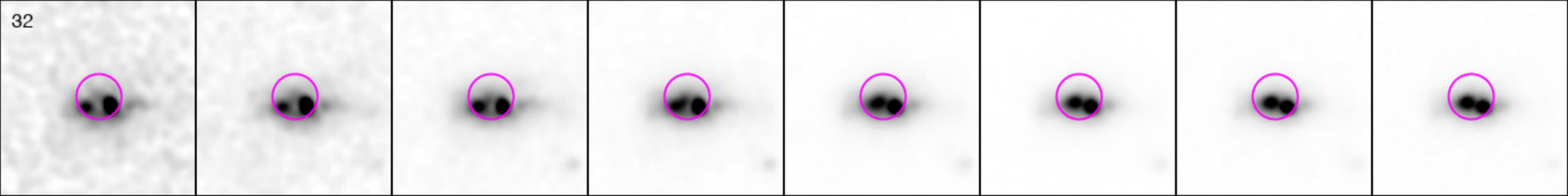"}    \plotone{"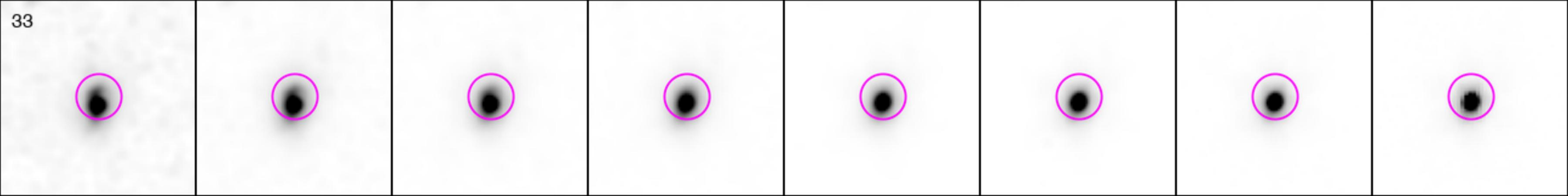"}     \plotone{"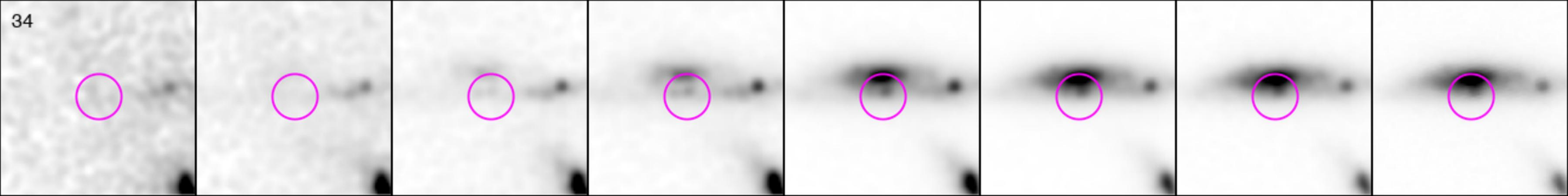"} 
\end{figure*}     
\begin{figure*}[ht]
\epsscale{1.2}
     \plotone{"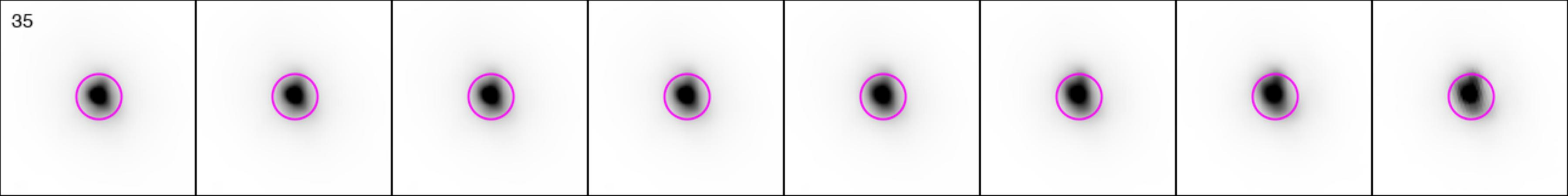"}    \plotone{"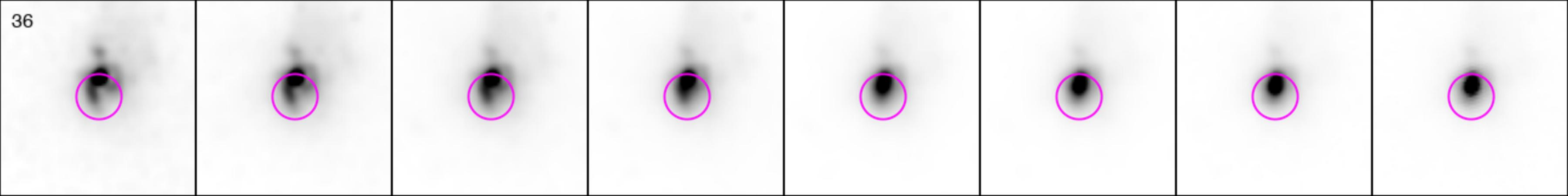"}     \plotone{"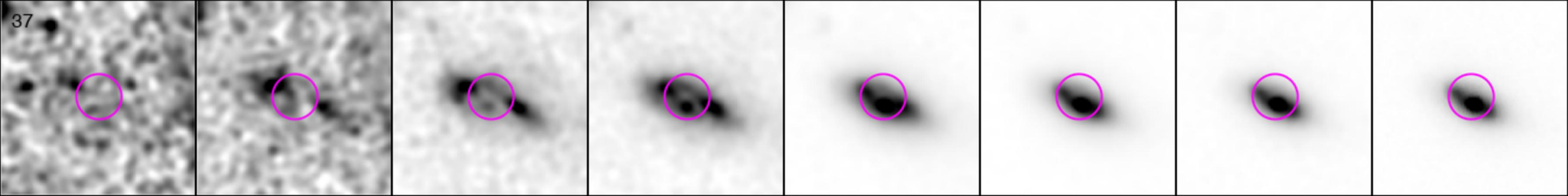"}    \plotone{"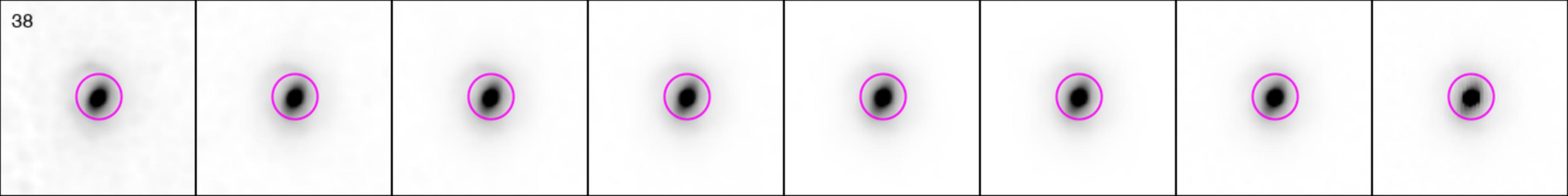"}
     \plotone{"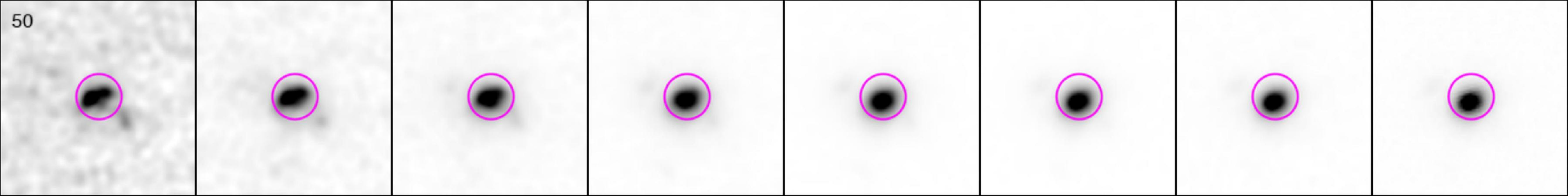"}     \plotone{"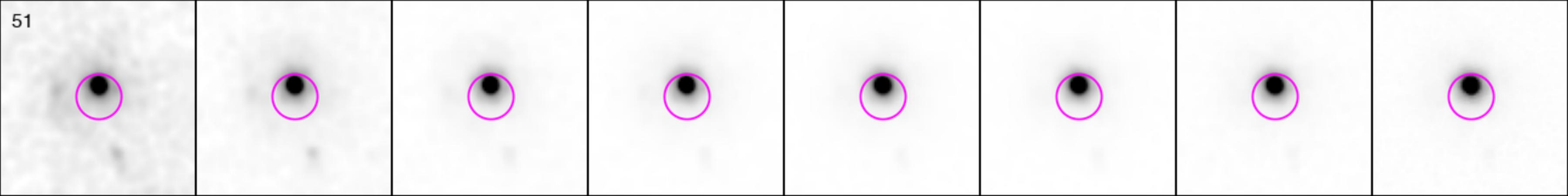"}    \plotone{"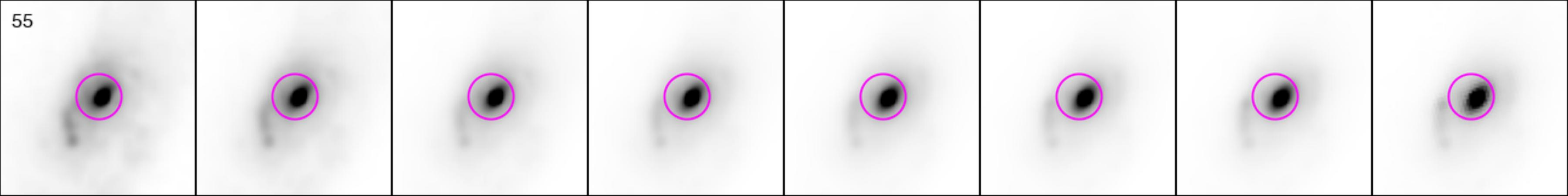"}     \plotone{"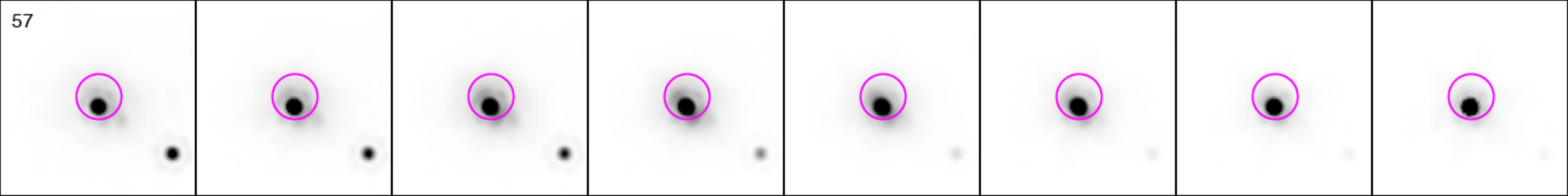"}     \plotone{"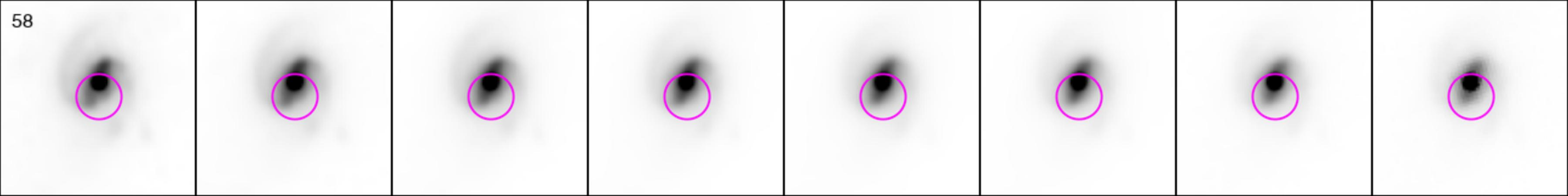"}    \plotone{"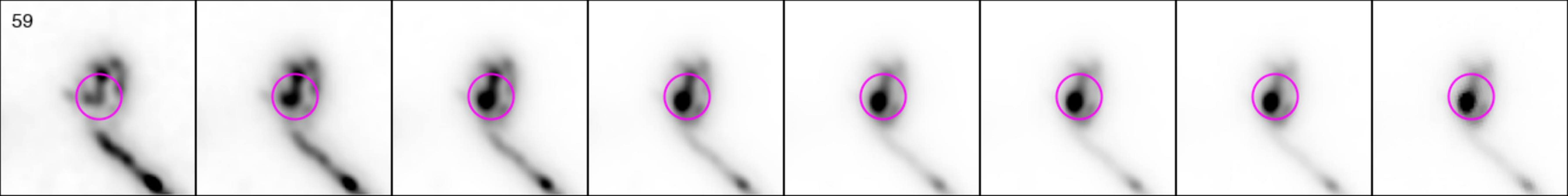"}  
\end{figure*}     
\begin{figure*}[ht]
\epsscale{1.2}

    \plotone{"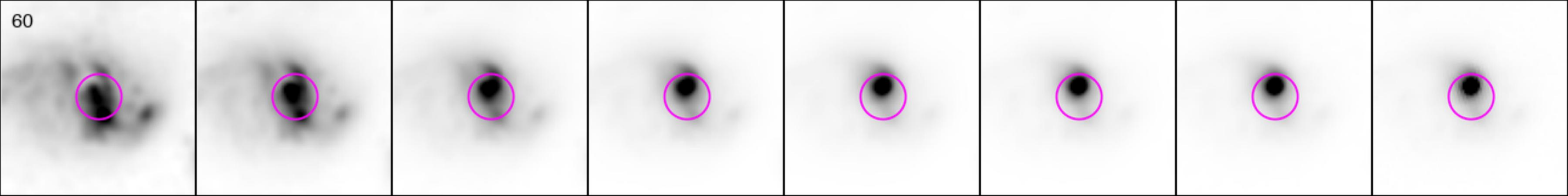"}    \plotone{"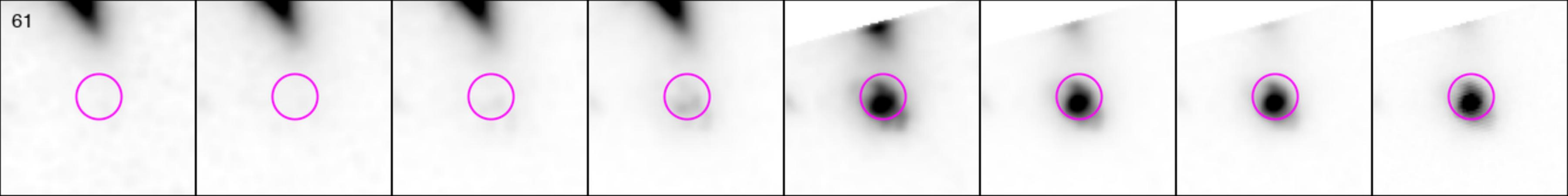"}
     \plotone{"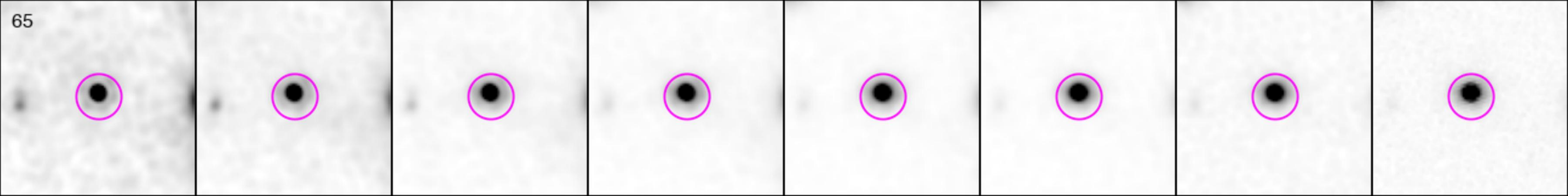"}     \plotone{"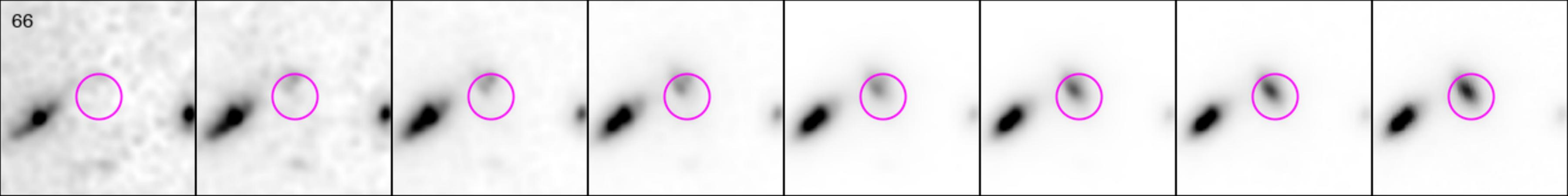"}    \plotone{"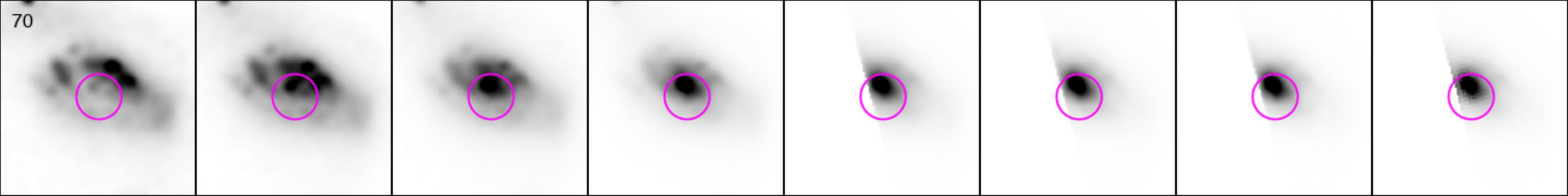"}     \plotone{"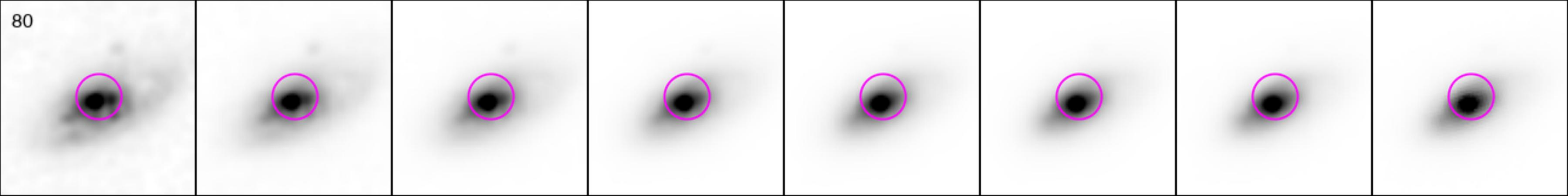"}     \plotone{"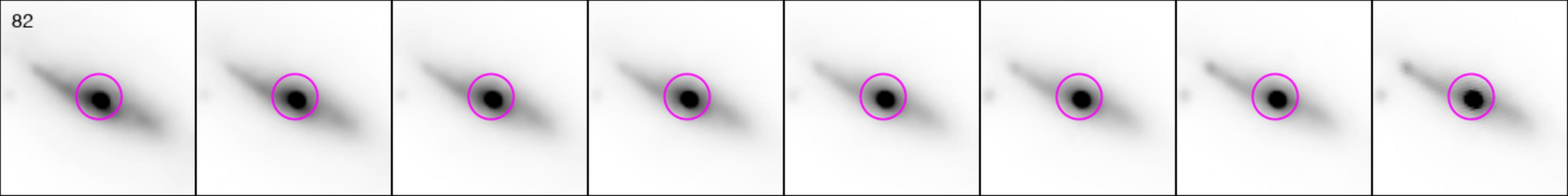"}    \plotone{"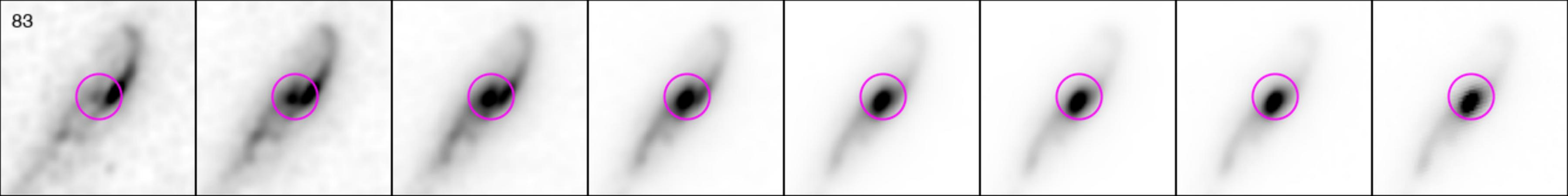"}         \plotone{"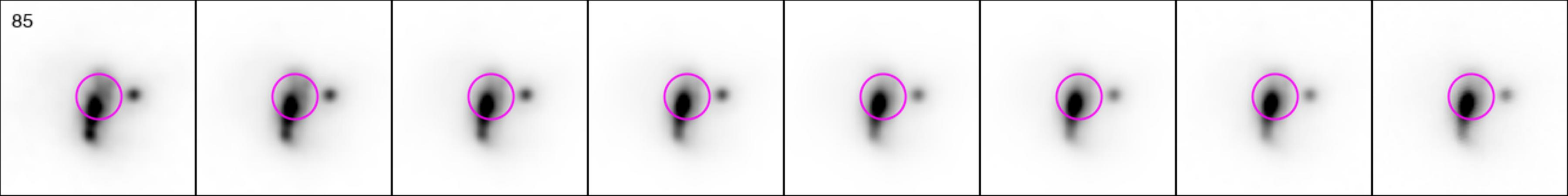"}
     \plotone{"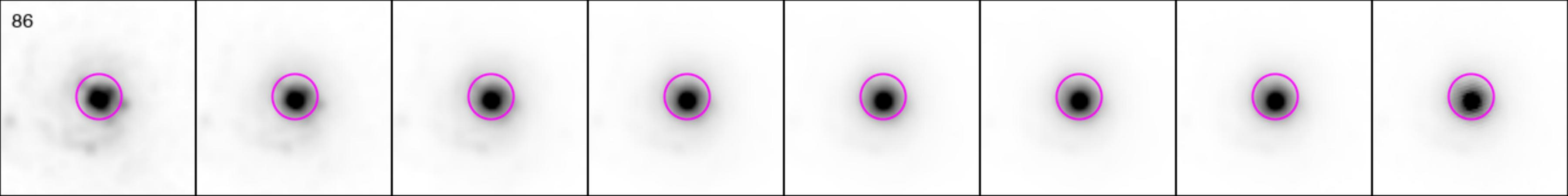"} 
\end{figure*}     
\begin{figure*}[ht]
\epsscale{1.2}
        \plotone{"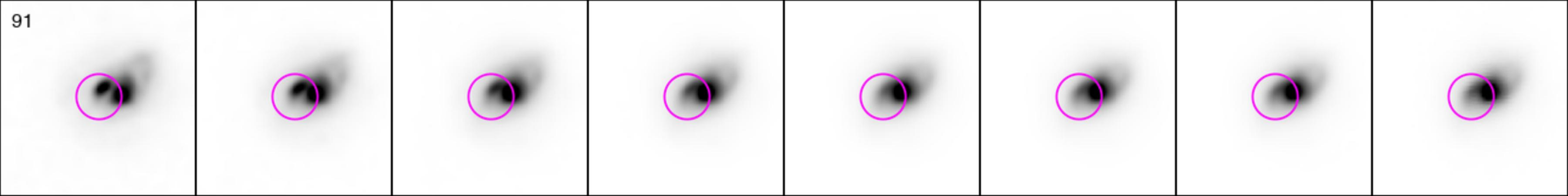"}     \plotone{"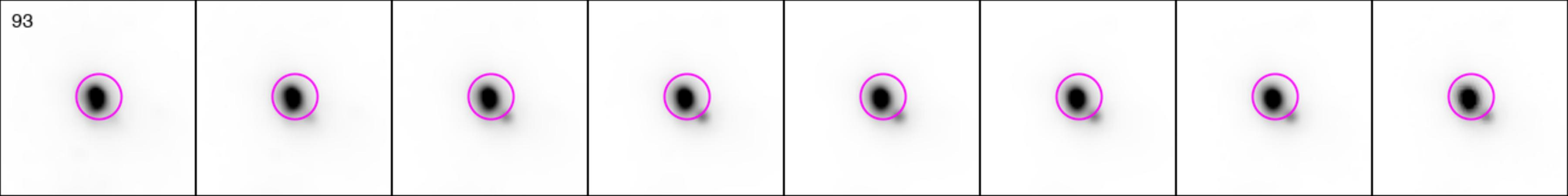"}     \plotone{"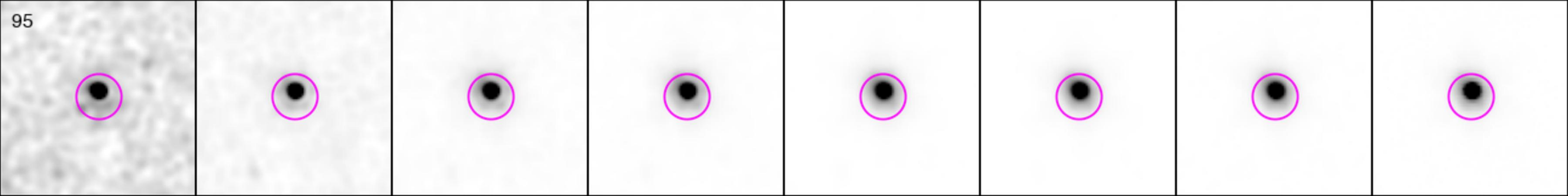"}         \plotone{"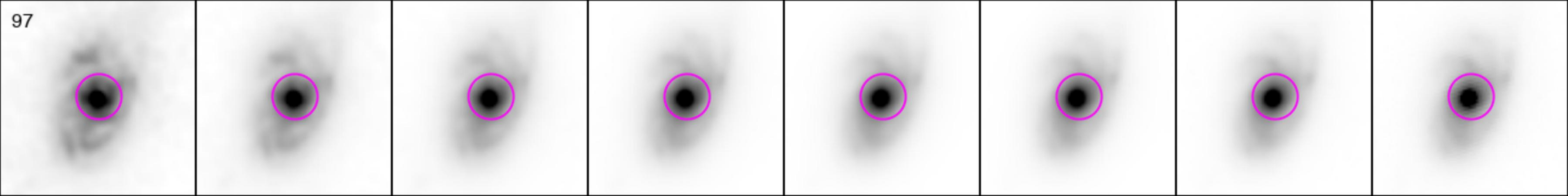"}    \plotone{"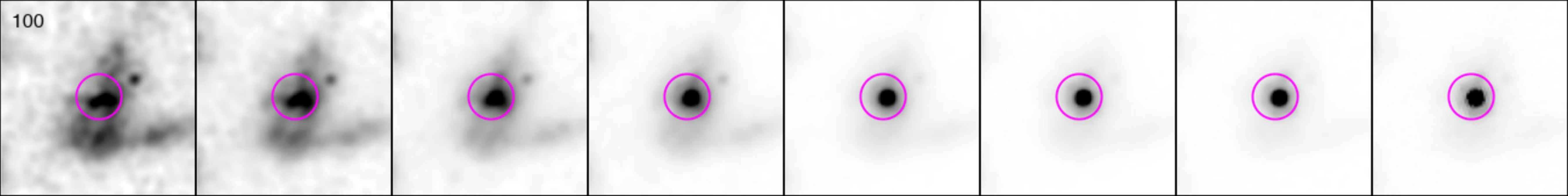"}
         \plotone{"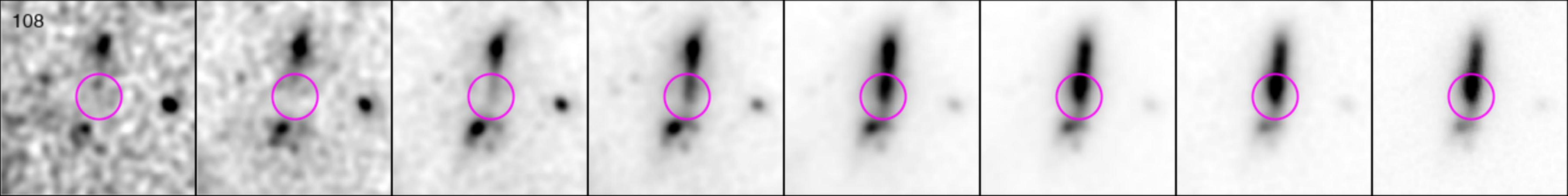"}  
         \plotone{"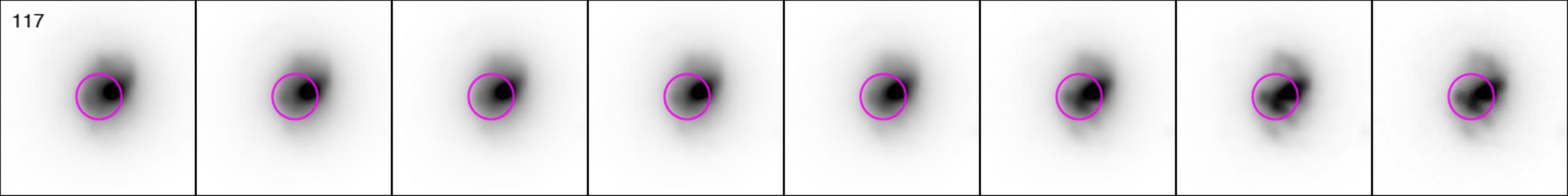"}        \plotone{"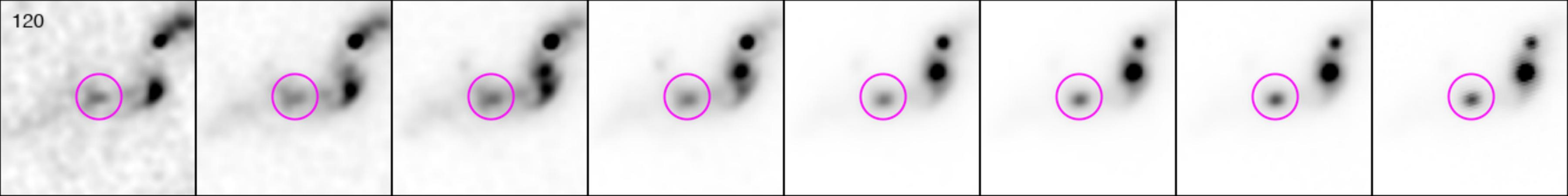"}    \plotone{"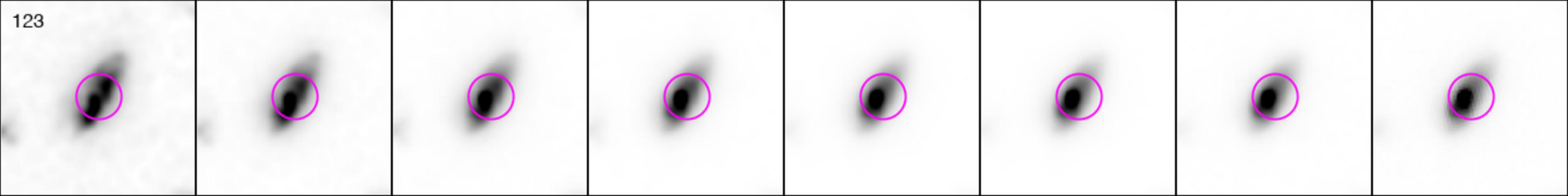"}
         
     \caption{The postage stamps of the 49 VLBA non-detections with JWST counterparts. The eight columns correspond to the eight filters of F090W, F115W, F150W, F200W, F277W, F356W, F410M, and F444W. Each panel is 3\arcsec x 3\arcsec. The magenta circle has a diameter of 0\farcs7, and indicates the radio source position from the VLA parent sample \citep{scuba}. The corresponding VLA IDs for each of the non-detections are tabulated against the VLBA PC numbers in Table \ref{tab:non detections}\\  
     }
     \label{fig:jwstnd}
\end{figure*}

\bibliography{references}{}
\bibliographystyle{aasjournal}

\end{document}